\documentclass[10pt]{article}
\usepackage{amsmath,amsfonts,amssymb,amsthm,bbm}
\usepackage{pdfsync}
\usepackage{float}
\usepackage{subfigure}
\usepackage{caption}
\usepackage{graphicx}
\usepackage[latin1]{inputenc}
\usepackage{hyperref}
\usepackage[all]{xy}
\usepackage{stmaryrd}
\usepackage{psfrag}
\usepackage{rotating}
\usepackage{braket}
\usepackage{amsthm}
\usepackage{enumerate}
\usepackage[normalem]{ulem}
\allowdisplaybreaks

\newcommand{\bit}{\begin{itemize}}
\newcommand{\eit}{\end{itemize}}
\newcommand{\bd}{\begin{description}}
\newcommand{\ed}{\end{description}}

\newcommand{\bc}{\begin{center}}
\newcommand{\ec}{\end{center}}
\newcommand{\Ref}[1]{(\ref{#1})}

\newcommand{\C}{{\mathbb C}}
\newcommand{\N}{{\mathbb N}}
\newcommand{\R}{{\mathbb R}}
\newcommand{\Z}{{\mathbb Z}}

\newcommand{\SL}{\mathrm{SL}}

\newcommand{\be}{\begin{equation}}
\newcommand{\ee}{\end{equation}}
\newcommand{\bea}{\begin{eqnarray}}
\newcommand{\eea}{\end{eqnarray}}
\newcommand{\bs}{\begin{subequations}}
\newcommand{\es}{\end{subequations}}
\newcommand{\nn}{\nonumber}

\newcommand{\f}{\frac}

\newcommand{\ra}{\rangle}

\newcommand{\mean}[1]{\langle{#1}\rangle}

  \newcommand{\g}{\gamma}
\renewcommand{\d}{\delta}  \newcommand{\eps}{\epsilon}  
 \renewcommand{\th}{\theta}      \renewcommand{\l}{\lambda}
\let\m=\mu    \let\r=\rho \newcommand{\s}{\sigma}       
\let\G=\Gamma \let\D=\Delta   \let\L=\Lambda  

\newcommand{\Wthree}[6]{\left(\begin{array}{ccc} #1 & #2 & #3 \\ #4 & #5 & #6 \end{array}\right)}
\newcommand{\Wfour}[9]{\left(\begin{array}{cccc} #1 & #2 & #3 & #4 \\ #5 & #6 & #7 & #8 \end{array}\right)^{(#9)}}

\begin{document}

\title{\bf 2-vertex Lorentzian Spin Foam Amplitudes for Dipole Transitions}
\author{\Large{Giorgio Sarno$^1$, Simone Speziale$^1$ and Gabriele V. Stagno$^{1,2}$}
\\
\small{$^1$ Aix Marseille Univ., Univ. de Toulon, CNRS, CPT, UMR 7332, 13288 Marseille, France} \\
\small{ $^2$Sapienza University of Rome, P.le Aldo Moro 5, (00185) Roma, Italy }
}
\date{\today}

\maketitle

\begin{abstract}
\noindent 
We compute transition amplitudes between two spin networks with dipole graphs, using the Lorentzian EPRL model with up to two (non-simplicial) vertices. We find power-law decreasing amplitudes in the large spin limit, decreasing faster as the complexity of the foam increases. There are no oscillations nor asymptotic Regge actions at the order considered, nonetheless the amplitudes still induce non-trivial correlations. Spin correlations between the two dipoles appear only when one internal face is present in the foam. We compute them within a mini-superspace description, finding positive correlations, decreasing in value with the Immirzi parameter.
The paper also provides an explicit guide to computing Lorentzian amplitudes using the factorisation property of SL(2,C) Clebsch-Gordan coefficients in terms of SU(2) ones. We discuss some of the difficulties of non-simplicial foams, and provide a specific criterion to partially limit the proliferation of diagrams. We systematically compare the results with the simplified EPRLs model, much faster to evaluate, to learn evidence on when it provides reliable approximations of the full amplitudes. Finally, we comment on  implications of our results for the physics of non-simplicial spin foams and their resummation.
\end{abstract}

\tableofcontents

\section{Introduction}
The spin foam formalism offers a covariant approach to the dynamics of loop quantum gravity, see \cite{PerezLR} for an introduction. 
The state of the art is well described by the EPRL model \cite{EPR,EPRL,FK,LS2}, which 
includes the Immirzi parameter, can be extended to provide transition amplitudes to all spin network states \cite{KKL,CarloGenSF}, and admits a quantum group deformation conjectured to describe the case of non-vanishing cosmological constant \cite{Haggard:2015yda}; Most importantly, the large spin asymptotics of the 4-simplex vertex amplitude contains exponentials of the Regge action \cite{BarrettEPRasymp,BarrettLorAsymp}. 
On the other hand, a systematic evaluation of spin foam transition amplitudes for given boundary data is hindered by their sheer complexity. 
For instance, numerical calculations of spin correlations have been performed so far only with a single 4-simplex and with the old Barrett-Crane model \cite{IoDan,IoDan2}.
The situation is even worse for Lorentzian signature, because of the unbounded group integrations in the vertex amplitude. Testing the EPRL Regge asymptotics is numerically very hard \cite{IoSU2asympt,noiLor}, and spin correlations are out of reach for the time being.\footnote{There is on the other hand a growing literature on results using approximate numerical methods for toy models or symmetry reduced models, e.g. \cite{Delcamp:2016dqo,Bahr:2016hwc,Bahr:2017eyi}.}
In this respect, the extension of the EPRL model to arbitrary vertices \cite{KKL,CarloGenSF} can be used to simplify the problem of explicit evaluations, by considering foams which are combinatorially simpler than the simplicial ones. For instance, one can consider boundary graphs in the form of a `flower' (a single node with links starting and ending on it), or of a `dipole' (two nodes and all links connecting one node to the other): the lowest-order spin foams are then much simpler than the simplicial ones, and one can hope to evaluate them explicitly with analytic and numerical methods. 
This is what we do in this paper: we choose a simple boundary given by two dipoles with 4-links, and study a dozen different spin foam amplitudes with up to two vertices and one internal face. 

Schematically, we compute amplitudes for the following spin foam expansion:
\be\label{sf}
 \parbox[2cm]{1cm}{\begin{picture}(0,70) (0,0) \includegraphics[width=1.1cm]{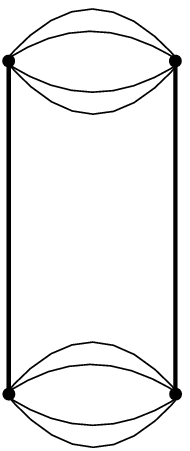} \end{picture}}\hspace{0.3cm}\,\, 
 +\l\quad\parbox[2cm]{1cm}{\begin{picture}(0,70) (0,0) \includegraphics[width=1.1cm]{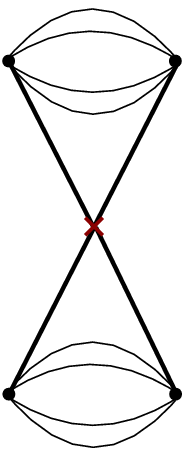} \end{picture}}\hspace{0.3cm}\,\,
 +\l^2\quad\parbox[2cm]{1cm}{\begin{picture}(0,70) (0,0) \includegraphics[width=1.1cm]{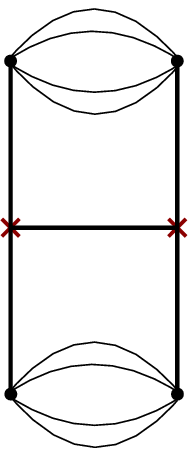} \end{picture}}\hspace{0.3cm}\,\, +\l^2\quad\parbox[2cm]{1cm}{\begin{picture}(0,70) (0,0) \includegraphics[width=1.1cm]{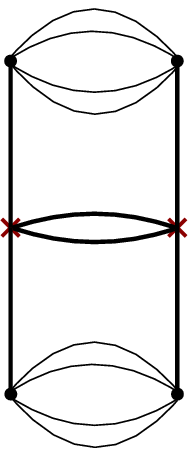} \end{picture}}\hspace{0.3cm}
 \ee
 Here $\l$ is a bookkeeping `coupling constant' of the vertex expansion. 
The method we use to evaluate the Lorentzian amplitudes is the factorization introduced in \cite{Boosting}, together with Wolfram's Mathematica and C++ codes to compute the boost integrals and SU(2) Clebsch-Gordan coefficients.
The explicit evaluations are very costly, and we limited detailed computations to those foams that we identified as the most interesting ones.

Non-simplicial transition amplitudes have been used in preliminary work in spin foam quantum cosmology based on a dipole boundary graph \cite{Rovelli:2008aa,Vidotto,Borja:2011ha,Bianchi:2011ym,VidottoLor,Kisielowski:2012yv} or on a flower one \cite{Rennert:2013pfa}. Similar simple graphs appear also in symmetry-reduced models of quantum cosmology using the canonical framework \cite{Battisti:2010he,Borja:2010gn,Livine:2011up}.\footnote{These canonical models are different from the main loop quantum cosmology approach (see e.g. \cite{Ashtekar:2011ni} for a review), in which the graph structure effectively disappears; and from \cite{Alesci:2013xd,Alesci:2017kzc}, where a regular 6-valent graph is used.}
Apart from mini-superspace applications, simpler graphs also provide a case study to gain control over the structure of the spin foam expansion. Being able to compute the explicit value of the amplitudes, or even just the large spin scaling or the divergence structure, are crucial ingredients to understand how to organize them, be it for resumming them or taking their continuum limit. 
To begin with, we are interested in questions of the following type: \emph{(i)} at fixed boundary graph, how do the different foams scale in a large spin expansion? \emph{(ii)} which foams dominate at large spins? 
\emph{(iii)}
what correlations between initial and final states are introduced by the foam?
In the case of \Ref{sf} we found the following answers: The foams become quickly sub-dominant as their complexity increases; The large spin scaling is not simply a function of the number of vertices, edges and faces of the foam. There are non-trivial correlations even in the absence of the Regge asymptotic behaviour; Only in the presence of an internal face (but not always) the correlations couple spins on the two dipoles, otherwise the amplitudes factorize and only spins and intertwiners within each connected part of the graph are correlated. 

The paper is organized as follows. First, we provide a summary of our main results and the consequences that can be drawn from them in Section~\ref{results} at the end of the Introduction, for the benefit of the reader already familiar with spin foams, and who wants to quickly identify the methods and results closest to her interests.
In the main body of the paper we provide an explicit guide on how to compute the EPRL Lorentzian amplitudes for generalized (non-simplicial) foams.
In Section~\ref{extention} we briefly recall the definition of the  vertex amplitude; We discuss the existence of non-integrable graphs, and introduce a face criterion to partially limit the proliferation of diagrams, the hardest issue in using non-simplicial spin foams. The criterion we use, of faces as minimal cycles only, basically makes the 2-complex rigidly determined by its 1-skeleton. 
The exclusion of non-integrable graphs and non-minimal faces significantly reduce the number of foams to be computed.
For instance at one vertex, graph integrability eliminates 16 of the 20 possible vertex graphs listed in \cite{Kisielowski:2012yv} for the Euclidean theory;  our face criterion eliminates 3 more, leaving only one, which is the one considered in \cite{Vidotto}. With two vertices and one edge we have again a single admissible type of vertex graph. With two edges and an internal face we have seven topologically distinct vertex graphs. The graphs and associated foams are presented in Section~\ref{graphstype}.  A complementary Appendix~\ref{AppDVD} contains the complete list relaxing our face criterion.
In Section~\ref{BoosterFact} we review the factorization property introduced in \cite{Boosting}, and explain how we use it to provide analytical and numerical studies of the amplitudes. It provides us also with a simplified version of the amplitude (denoted EPRLs in \cite{Boosting}), which provides a good approximation in certain cases, and can be used to compute some analytic estimates and much faster numerical evaluations. One of the results of this paper is that we tested that for most foams of \Ref{sf} the simplified model gives correct estimates of both scaling and correlations.
This result and all exact numerical evaluations and approximate analytic estimates we performed of the EPRL amplitudes are reported in Section~\ref{DDFoams}. 
In Section~\ref{SecDLD6} we focus on the dominant non-factorized foam with one internal face, and discuss and compute in some details the spin correlations it induces.
The final Section contains our conclusions with the implications of our results, and a perspective on future work, explaining the limits we face at the moment at the numerical level, and what could be improved.
Two additional Appendices contain the recoupling theory rules used for the evaluations of the spin foams, a summary of scaling properties of the booster functions, and the results of evaluating the 3 one-vertex foams left out by our face criterion.

We follow the conventions of \cite{Varshalovich} for SU(2) and its recoupling theory, and those summarised in \cite{Boosting} for $\SL(2,\C)$. We refer to the nomenclature of \cite{RovelliVidotto} for spin foams: the boundary graph is characterised by nodes connected by links, and as the boundary of a 2-complex, each node is the boundary of an edge (and one only), each link the boundary of a face (and one only). 

\subsection{Summary of the main results}\label{results}
The notation for \Ref{sf} is as follows: we use thin lines for the links of the boundary dipole graphs, and round dots for their nodes; thick lines for the edges of the foams and (red) crosses for their vertices; we call $j_a$ the spins of the lower dipole ($a=1\ldots 4$) and $(i,t)$ its left and right intertwiners, and similarly but with primed letters for the upper dipole. Then \Ref{sf} is a function $W(j_{a},j_{a}';i,t;i',t')$ (ignoring the dependence on $\l$) determined by the Lorentzian EPRL model. 

\begin{itemize}

\item Aware of the risk of an uncontrollable proliferation of non-simplicial spin foams, we introduced a strong criterion to select the 2-complexes:
we admit only faces corresponding to minimal cycles. With this choice, to be motivated and detailed below, there is only one admissible vertex graph at one vertex; only one at two vertices and one edge; seven topologically distinct vertex graphs for foams with one internal face. 

 \item As shown already in \cite{VidottoLor}, the one-vertex foam factorizes in two contributions each depending only on lower or upper spins. There are thus no correlations between lower and upper graphs, but only within each connected component. We find that also the two-vertex-one-edge foam factorizes in the same manner. Non-factorized amplitudes appear only in the presence of an internal face, and thus lower-upper spin correlations. 
 
\item The spin correlations we found do not have an immediate geometric interpretation. In fact, most of the foams considered do not contain non-trivial $nj$-symbols, but only (generalized) $\th$-graphs, and Regge actions do not appear in the asymptotics. 
We numerically computed the spin correlations, using for cost sustainability the simplest non-factorized amplitude and a mini-superspace approach with all spins fixed to be equal. We found positive correlations between lower and upper spins, with a monotonical decrease in $\g$. We also considered spin correlations in the general boundary framework with a unit-width Gaussian state peaked on a background geometry $j_0$. The resulting correlations peak near the Planck scale at $\g$-dependent value, and have a $\g$-dependent power-law tail in $j_0$.

\item The large-spin leading order of each foam decreases as the complexity of the foam increases. 
The power of the large spin scaling depends on the explicit combinatorics of the faces, and it is not simply a function of the number of vertices $V$, edges $E$ and faces $F$ of the 2-complex. This is unsurprising given the non-topological nature of the EPRL model, however one could have still hoped for a dependence on the three numbers $(V,E,F)$ alone. We discuss this point and what we can learn for power counting of scalings from the foams here considered. 

\item Numerical estimates show that the foam with one vertex ($DVD$, for dipole-vertex-dipole) scales like $N^{-3}$ in the homogeneously large spin limit; the foam with two vertices and one edge ($DED$, for dipole-edge-dipole) scales like $N^{-6}$; the simplest non-factorized foam with an internal face ($DLD$, for dipole-loop-dipole) scales like $N^{-9}$.
Using both analytical and numerical methods, we found that the large spin behaviour of \Ref{sf} has the following structure:
\begin{align}
W(Nj_a,Nj'_a;i,i';t,t') &=\delta_{j_a,j'_a}\d_{i,i'}\d_{t,t'}+\l\f{f(j_a) f(j'_a)}{N^3} \d_{it}\d_{i't'} 
\nonumber\\ & \quad + \l^2\f{g^2(j_a) g^2(j_a')}{N^6} \d_{it}\d_{i't'} + \l^2\f{h(j_a,j'_a,i,t,i',t')}{N^9}, \label{Wseries}
\end{align}
where the functions  $f, g$ and $h$ will be presented below. 

\item We systematically compared the EPRL evaluations with those of the simplified EPRLs model introduced in \cite{Boosting}, corresponding to additional impositions of the primary simplicity constraints (via the $Y$-map): this is much faster to evaluate, and furthermore analytic estimates of the scalings are known. 

For the first two foams ($DVD$ and $DED$) the EPRLs provides an exact evaluation of the full model; for the simplest foam with an internal face it does not, but it still captures the right large spin scaling and the spin correlations, hence providing a valid approximation for most questions. There are 3 additional foams of the 7 with an internal face that have non-factorized amplitudes. For these the simplified model amplitudes differ significantly from the complete EPRL model. They could have slower decay behaviour than $N^{-9}$ (by one or two powers, a priori) because of the presence of unbounded summations, and may thus dominate the correlations in the large spin limit; but we do not know at present.

\item Our results can also be immediately applied to a nothing-to-dipole expansion, 

\be
W_{D} =\,
 \l \parbox[2cm]{1cm}{\begin{picture}(0,30) (0,0) \includegraphics[width=1.1cm]{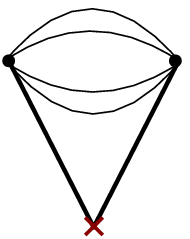} \end{picture}}\hspace{0.3cm}\,\, 
 +\l^2 \quad\parbox[2cm]{1cm}{\begin{picture}(0,30) (0,0) \includegraphics[width=1.1cm]{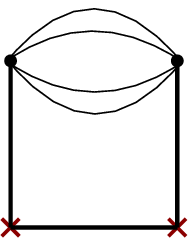} \end{picture}}\quad\cdots\label{seriesHH}
\ee

\end{itemize}
There are three principal lessons that can be drawn from these results. The first is that Lorentzian amplitudes are explicitly computable, in spite of their complexity. We were limited to small foams, but with improved codes for the booster functions and the guidance from the simplified model here learned to estimate the scalings, the analysis can be pushed significantly further. The second is that large spin decays speed up with the complexity of the foam, allowing a naive hierarchical organization of the expansion. The power of the decay depends explicitly on the routing of the faces and not just on their number. This complicates power-counting arguments even with our criterion to restrict the faces, and show the dominance of disconnected vertex graphs. 
The third is that there are non-trivial spin correlations even in absence of asymptotic Regge actions. Such correlations do not show a squared-inverse power law as in the 4-simplex graviton calculations (see e.g. \cite{Bianchi:2006uf,IoDan,IoDan2,Bianchi:2011hp}), but a non-trivial dependence on the Immirzi parameter. 
It is a priori possible to compute spin correlations on larger graphs and study whether they fall off with the graph distance or background distance, but that would definitively require much more numerical power than currently available to us, or find better approximation schemes. 

\section{Lorentzian EPRL model for general complexes}\
\label{extention}
In this Section we briefly review the definition of the Lorentzian EPRL model \cite{EPRL}, its factorization introduced in \cite{Boosting}, and discuss its extension to non-simplicial complexes. The two aspects that require attention are the presence of non-integrable graphs and the proliferation of foams. We assume the reader to be familiar with the basics of loop quantum gravity and the spin foam model, and refer to the cited literature  \cite{EPR,EPRL,FK,LS2} and to the monographs \cite{PerezLR,RovelliVidotto} for details. Conventions and notations follow \cite{Boosting}.

\subsection{Boundary states, spin foam amplitudes and correlations}
We work with abstract oriented graphs colored by spins $j_l$ on every link $l$, and a set of intertwiner labels $\vec\imath_n$ on every node $n$: these are $v_n-3$ half-integers for a node of valence $v_n$, corresponding to the spins along virtual links once a recoupling scheme is chosen. We define an SU(2) spin network state by
\be\label{sDef}
\bra{g_l}\G,j_l,\vec \imath_n\ra = \prod_{l} D^{(j_l)}_{m_{s(l)}m_{t(l)}}(g_l)\prod_n  \left(\begin{array}{cccc} j_l \\ m_l \end{array}\right)^{(\vec\imath_n)},
\ee
with $s(l)$ and $t(l)$ the source and target nodes of the link $l$, and summations over magnetic indices $m$ implicitly assumed. Here $D^{(j)}(g)$ are Wigner matrices, and the right-most symbol a short-hand notation for generalized Clebsch-Gordan coefficients, see Appendix \ref{AppSU2} for definitions.
These functions provide a basis of the Hilbert space of square-integrable gauge-invariant functions with respect to the SU(2) Haar measure, schematically
\be
\mathcal{H}_{\G} =L^2\left[SU(2)^L/SU(2)^N,d\m_{\rm Haar}\right],\nonumber
\ee
where we denoted by $L$ and $N$ the total number of links and nodes of the graph.
Because of standard conventions in the definition of the Wigner matrices and $\{nj\}$-symbols, the basis \Ref{sDef} is not normalized. That requires multiplying each state by $(\prod_l d_{j_l} \prod_n d_{\vec\imath_n})^{1/2}$, where $d_{j}:= 2j+1$ is the dimension of the SU(2) irrep $j$, and $d_{\vec\imath}=(2i^{(1)}+1)...(2i^{(v_e-3)}+1)$ is the dimension of the virtual links irreps.

Consider next an open 2-complex $\cal C$ whose boundary is $\G$: 
it consists of vertices, edges and faces, such that to each node and link of $\G$ there correspond a unique edge and face of the boundary of $\cal C$.
We color it with spins $j_f$ on the faces and intertwiners $\vec\imath_e$ on the edges, such that  $j_f=j_l$ and $\vec\imath_e=\vec\imath_n$ for each face and edge on the boundary. 
The spin foam formalism \cite{CarloReisenbergerSum,BaezIntro,PerezLR,RovelliVidotto} assigns an amplitude to each pair $(\ket{s}=\ket{\G,j_l,\vec\imath_n},{\cal C})$  in the form of a state sum model,
\be 
\bra{W_{\cal C}} {\G,j_l,\vec\imath_n}\ra = \sum_{\{j_f,\vec \imath_e\}} \prod_f d_{j_f} \prod_e d_{\vec\imath_e} \prod_v A_v(j_f,\vec\imath_e)\,,\label{Z}
\ee
where the summations are at fixed boundary values, and face and edge weights are internal only.\footnote{In other words, the weight for a boundary face and edge is 1. This is a consequence of our choice of basis \Ref{sDef}. If we work instead with normalized boundary spin networks, the boundary face and edge weights would be the square roots of the internal ones. Notice also that in the literature the edge weights are often absorbed in the definition of the vertex amplitude. We choose not to do so, so that the vertex amplitudes are given by the conventional $\{nj\}$-symbols.} The vertex amplitude $A_v$ is model-dependent and carries the core of the dynamics of the theory. 

The physical interpretation of $W_{\cal C}$ is to provide a dynamical amplitude for the (quantum) 3-geometry $\ket{s}$. 
The total transition amplitude was historically defined as the sum over all possible 2-complexes compatible with the boundary graph, and conjectured to implement the projector over physical states in the kernel of the Hamiltonian constraint \cite{CarloReisenbergerSum,BaezIntro}.
Hopes to control the sum and make sense of it mathematically were put on the group field theory approach \cite{DePietri:1999bx,DePietri:2000ii}, and work of the last few years (see e.g. \cite{Bonzom:2012hw,Carrozza:2013wda,Oriti:2014uga}) is starting to bear fruits: for simplicial complexes and simpler models at least (typically topological BF theory) it is now possible to establish convergence or renormalizability of the sum.
Extending these results to the Lorentizan EPRL is a key open question. The study presented in this paper contributes by evaluating individual foams and assessing their relative scaling weights in the non-simplicial case.\footnote{An alternative definition of the dynamics is to view spin foams as amplitudes cut-off at a finite number of degrees of freedom, to be studied in the continuum limit instead of being summed over. See e.g. \cite{Rovelli:2011mf} and \cite{Delcamp:2016dqo,Bahr:2016hwc} for recent results in this directions. }

To give physical content to the amplitudes at fixed foam, one can define dynamical expectation values following the general boundary framework \cite{Bianchi:2006uf}. For an observable $\hat{\cal O}$ and a state $\ket{\Psi}\in{\cal H}_\G$, we consider the quantity
\be\label{EV}
\f{\bra{W_{\cal C}}\hat{\cal O} \ket{\Psi}}{\bra{W_{\cal C}}\Psi\ra}
\ee
as the `$\cal C$-representative' of the dynamical projection on $\ket{\Psi}$.
In this construction, $\ket{\Psi}$ is required to be a semiclassical state peaked on a classical discrete geometry $q_\G$ associated to the boundary. One often works with Gaussian states (see however \cite{Livine:2006it}), with width kept as a free parameter or fixed by dynamical requirements \cite{Dittrich:2007wm}. In special circumstances, the linear map $\bra{W_{\cal C}}$ is normalizable; or alternatively it can be suitably regularized. Then it is also possible to look at the spin foam expectation values
\be\label{PEV}
\f{\bra{W_{\cal C}}\hat{\cal O} \ket{W_{\cal C}}}{\bra{W_{\cal C}}W_{\cal C}\ra}.
\ee

In both cases, a simple choice of observables are those diagonalized by the spin network basis \Ref{sDef}, for which
\be
\bra{W_{\cal C}}\hat{\cal O} \ket{\Psi} = \sum_{j_l,\vec\imath_n} W_{\cal C}(j_l,\vec\imath_n){\cal O}(j_l,\vec\imath_n) \Psi(j_l,\vec\imath_n).
\ee
In particular, we will compute below correlations among spins, defined as
\be\label{corr}
\mean{j_l j_{l'}}_{({\cal C},\Psi)} := \f{ \bra{W_{\cal C}}\hat{J}_{l} \hat{J}_{l'} \ket{\Psi} } {\bra{W_{\cal C}}\Psi\ra} -  \f{ \bra{W_{\cal C}}\hat{J}_{l} \ket{\Psi} } {\bra{W_{\cal C}}\Psi\ra} \f{ \bra{W_{\cal C}}\hat{J}_{l'} \ket{\Psi} } {\bra{W_{\cal C}}\Psi\ra}.
\ee

\subsection{EPRL vertex amplitude}
\label{EPRLva}
For the EPRL model in Lorentzian signature \cite{EPRL}, the vertex amplitude $A_v$ is built from $\SL(2,\C)$ unitary irreducible representations (irreps) of the principal series, see  \cite{Ruhl}. These are labelled by a pair $(\r \in \R,k\in \Z/2)$, and Naimark's canonical basis is chosen diagonalizing the operators $L^2$ and $L_z$ of the matrix subgroup SU(2), with eigenvalues $j(j+1)$ and $m$. The group elements in this basis are represented by infinite-dimensional unitary matrices
\be\label{Dh}
D^{(\r,k)}_{jmln}(g), \qquad (j,l)\geq k, \quad -j\leq m \leq j, \quad -l\leq n \leq l, \qquad g\in\SL(2,\C).
\ee
Only a certain subset of irreps is used, that we refer to as \emph{$\g$-simple} representations; they satisfy
\be\label{simple}
\r = \g k, \qquad k=j.
\ee
This restriction defines an embedding of SU(2) irreps in unitary $\SL(2,\C)$ ones called $Y$-map.

\begin{figure}[h]
\centering
\includegraphics[scale=0.45]{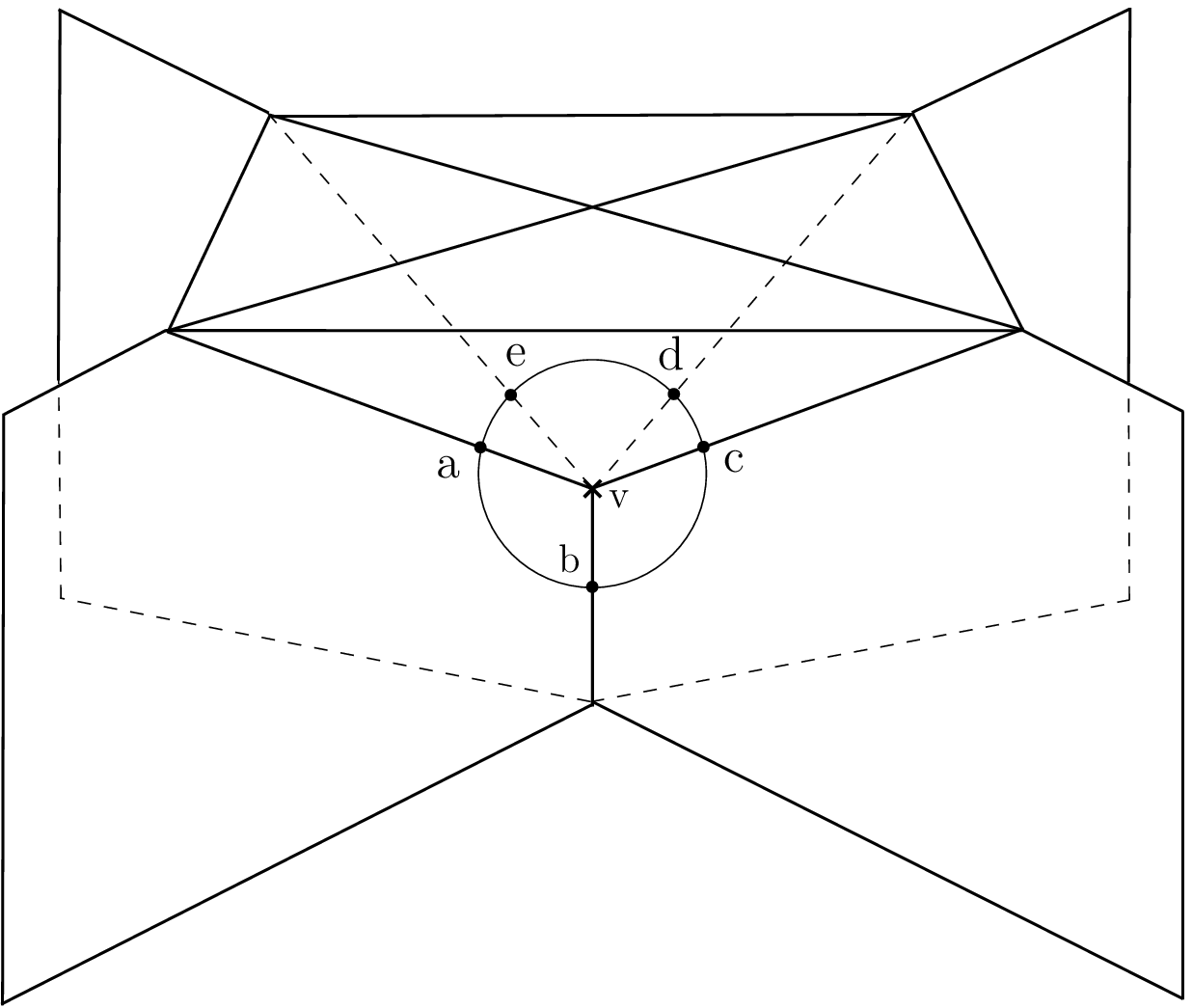}\hspace{1cm}
\includegraphics[scale=0.6]{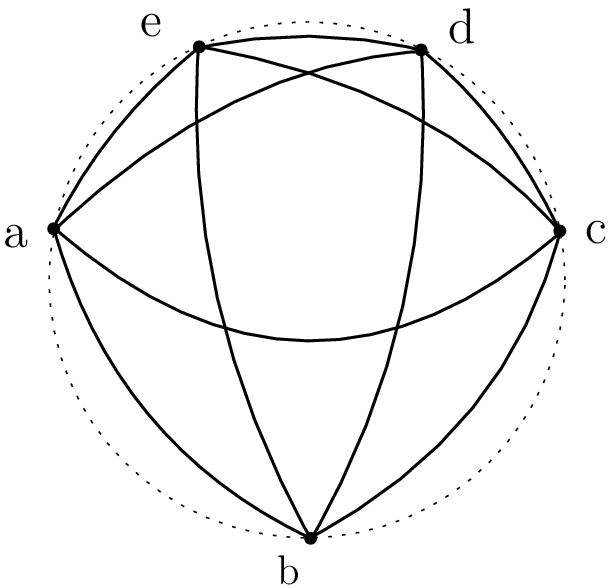}
\caption{\emph{\small{On the left the complex for a single $4$-simplex together with a little sphere surrounding the vertex $v$. On the right the vertex graph associated to it (the pentagon graph) as explained in the text.}}\label{4simplex}}
\end{figure}

The explicit form of the vertex amplitude $A_v$ depends on both the valence of the vertex (i.e. the number of edges attached to it) and the combinatorics of the faces. The combinatorics can be most easily visualized if we draw a sphere around the vertex: each edge projects to a point on the sphere, each face to a line connecting two points. The spherical graph so obtained is referred to as the \emph{vertex graph}, see Fig.~\ref{4simplex} for an example. 
Points and lines of the vertex graph are in 1-to-1 correspondence with edges and faces of the spin foams, and we adapt the labelling accordingly.

\begin{figure}[ht]   
\centering\begin{picture}(100,100)
\put(-150,0){\includegraphics[width=4cm]{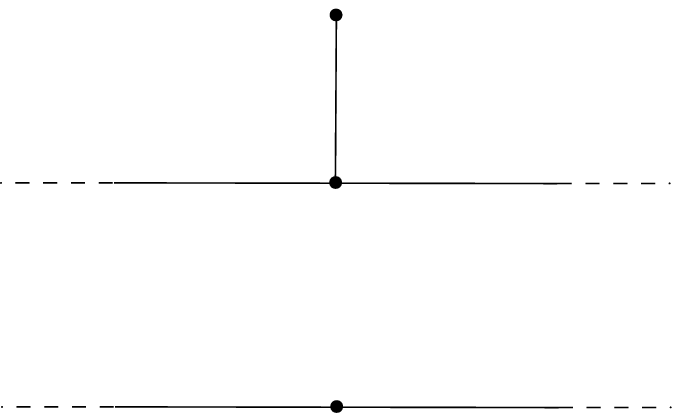} }
\put(0,25){\includegraphics[width=4cm]{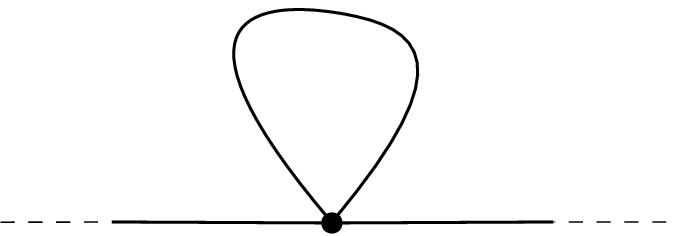} }
\put(150,0){\includegraphics[width=2cm]{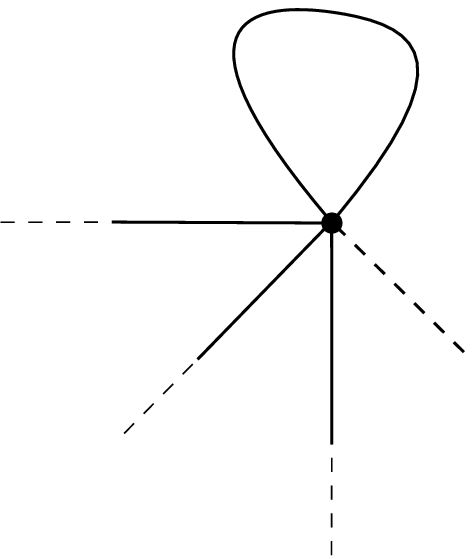} }
\end{picture}
\caption{\label{Figtadpole} {\small{Left: \emph{Vertex graph with uni-valent or bi-valent nodes: divergent amplitudes.} Right: \emph{Vertex graph with a petal: the associated amplitude diverges in the case of two legs, and converges for three or more legs.} }}}
\end{figure}

The amplitude is then defined by the evaluation of a $\g$-simple $\SL(2,\C)$ spin network on the vertex graph, namely
\be\label{Av1}
A_v(j_f, \vec\imath_e) := 
\int_{\SL(2,\C)^{E_v-1}}  \prod_{e=1}^{E_v-\#} dg_{e} \, \prod_{f} D^{(\g j_f,j_f)}_{j_f m_{s(e)} j_f m_{t(e)}}(g^{-1}_{s(e)} g_{t(e)}) 
\, \prod_e \left(\begin{array}{cccc} j_f \\ m_f \end{array}\right)^{(\vec\imath_e)},
 \ee
with the summation over magnetic indices implicitly assumed as in \Ref{sDef}.
Here $E_v$ is the valence of the vertex graph and $\#$ the number of connected components of the vertex graph.
Although in the definition we assign a group element to each node of the vertex graph, we need to remove one integration per connected part of the graph: left-right invariance of the Haar measure makes it redundant, and it would lead to a diverging amplitude because of the non-compactness of the group. This is an important difference from Euclidean models where removing or not the redundant integration does not affect the amplitude. 
However, even once the redundancy has been taken care of, not all vertex graphs are integrable. Examples of graphs for which \Ref{Av1} is not well-defined are reported in Fig.~\ref{Figtadpole}. 
The amplitude of a graph with a uni-valent node diverges trivially because it contains the integration
\be
\int_0^\infty dr \, (\sinh r)^2 \, d^{\r,k}_{jlm}(r) = \infty.
\ee
For a graph with a bi-valent node, the amplitude contains the integration
\begin{align}\label{Dortho}
\int dg\,
D^{(\r_1,k_1)}_{j_1m_1l_1n_1}(g)D^{(\r_2,k_2)}_{j_2m_2l_2n_2}(\bar g) &= 
\f{\d(\r_1-\r_2) \d_{k_1k_2}}{4(\r_1^2 +k_1^2)} \, \d_{j_1j_2}\d_{l_1l_2} \d_{m_1m_2}\d_{n_1n_2}.
\end{align}
The right-hand side, a consequence of the orthogonality of the matrix representations, gives a distributional divergence for general irreps; 
and gives always infinite for the simple ones \Ref{simple} used in the EPRL model.
A related example is a vertex graph containing a closed loop, or `petal'. 
Using the following `sliding' identity from graphical calculus (see Appendix~\ref{AppSU2}), 
\[
\centering      
\includegraphics[width=4.5cm]{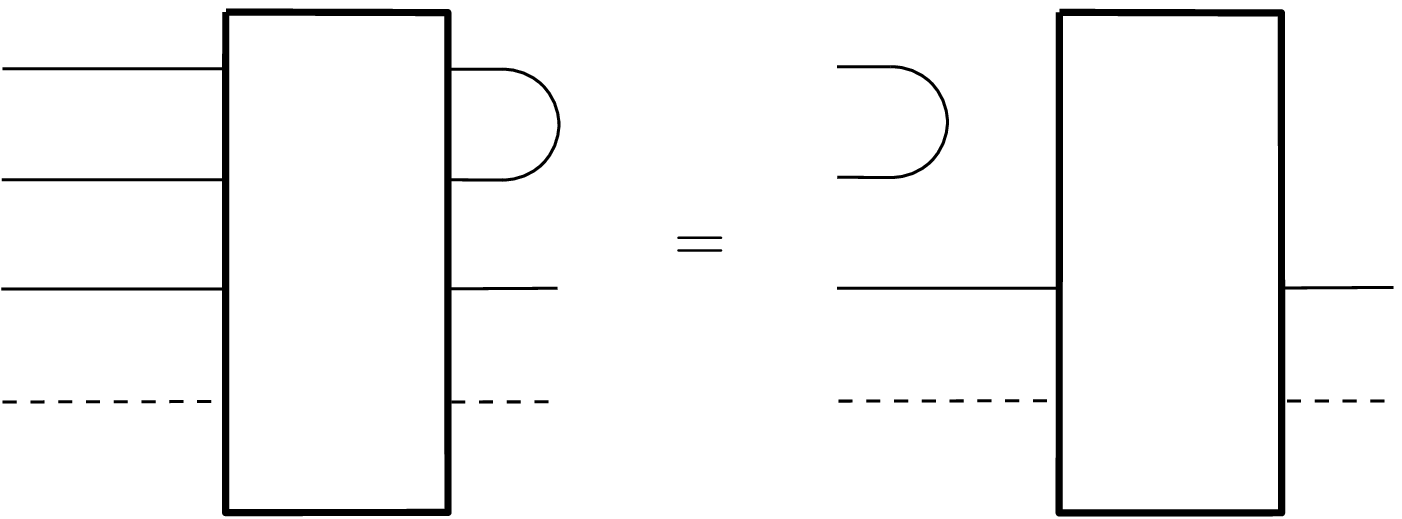}
\]
petals factorize in the evaluation of the amplitude, leaving behind an intertwiner with two lesser links. For a petal with two legs, or tadpole, this gives \Ref{Dortho} and the same divergence as before; whereas for higher-valence petals with three or more legs the amplitude is finite.

The existence of divergent graph introduces a necessary restriction on the admissible foams unlike in the Euclidean case.
We are not aware of a necessary condition for graph integrability. 
A \emph{sufficient} condition
is 3-link-connectivity, meaning any bi-partition of the nodes cannot be disjointed by cutting only two links  \cite{Baez:2001fh,Kaminski:2010qb}. One could then decide to restrict attention to such vertex graphs, for which convergence of \Ref{Av1} is guaranteed. The criterion is however based on very loose bounds, and one can find many vertex graphs that are not 3-link-connected and yet finite: We will encounter some examples below. For this reason, we will remove in our analysis only vertex graphs with uni- and bi-valent nodes, and two-legged petals, and keep non-3-link-connected ones -- at least a priori.\footnote{A related open question is whether the simplified model, see below, allows one to sharpen this sufficient condition. At first sight it looks like removing uni- and bi-valent nodes may be enough for this model, which sits at an intermediate place between the complete EPRL and a pure SU(2) theory. We leave it as an open question for future work.}

The fact that not all vertex graphs are integrable reduces the number of admissible foams with respect to Euclidean amplitudes. 
For instance, out of the 20 possible vertex graphs considered in \cite{Kisielowski:2012yv}, only 4 do not contain tadpoles, and only 3 satisfy the triple-connectivity criterion. See Appendix~\ref{AppDVD} for details.
Nonetheless, the proliferation of generalized spin foams remains a severe problem of this approach, as compared for instance to the use of simplicial ones. 
To have any hope of taming it, it is quite likely that one needs to supplement the definition of the theory with some notion of class of complexes to be considered.\footnote{A radically opposite philosophy has been proposed in \cite{Kisielowski:2011vu,Kisielowski:2012yv}, whereby one tries to accommodate \emph{all possible vertex graphs} to all boundaries. This removes combinatorial constraints on the vertex graph, thus simplifying the problem of listing all admissible foams with a given boundary graph; but at the price of largely increasing their number.} 
For the purposes of this paper, we work with a single but strong restriction on the assignment of the faces in the 2-complex:

\begin{description}
\item{\bf Definition:} The faces of the 2-complex are defined as all minimal cycles of its 1-skeleton, namely minimal closed sequences of edges (for internal faces), or edges and links (for external faces). 
\end{description}

\noindent This definition means that the 2-complex is uniquely specified by its 1-skeleton and the boundary graph, thus imposing a strong restriction on the foams to be considered.
Two important consequences are the following:

\begin{itemize}
\item There are no minimal cycles without a face assigned; nor multiple faces can be assigned to the same minimal cycle.
The latter forbids trivial infinite proliferations like adding ad libitum rolled-up faces or `pillow-like' faces to the bulk edges, 
which corresponds to adding petals or additional links between the same two nodes to vertex graphs.

\item It also forbids the `intermediate static foam' construction used in \cite{Kisielowski:2011vu} to freely associate arbitrary vertex graphs to a given boundary graph. It implies in particular that petals in vertex graphs can only appear if induced by petals in the boundary graph.
\end{itemize}

This definition appears to us natural in order not to be irremediably swamped by the problem of proliferation of non-simplicial foams. Whether it is sufficient to truly tame the combinatorics, and whether it can be relaxed is an interesting but also very difficult question.
With the ensuing restriction imposed, there are still two additional configurations unusual from the simplicial setting worth to be pointed out:
\begin{enumerate}
\item Disconnected vertex graphs: these can lead to factorizations of the transition amplitudes that limit their relevance.
\item Spin foam faces with two or more boundary links in them: these force the boundary spins on the shared links to match, leading to trivial pieces in the amplitudes. Such `face-rigidity' limits as well the relevance of such amplitudes.
\end{enumerate}

With these considerations in mind, we  discuss next our choice of boundary graphs.

\section{Minimal boundary graphs and lowest-order interpolating foams} \label{graphstype}
\begin{figure}[h]
\centering
\includegraphics[scale=1.8]{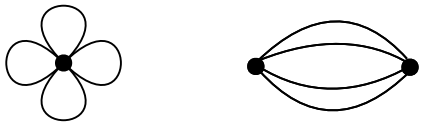}
\caption{\small{\emph{The 4-petal graphs and the 4-link dipole.}} \label{fig:dg}}
\end{figure}

In its original formulation, the EPRL model was defined on simplicial foams, i.e. 2-complexes dual to simplicial decompositions of spacetime. Since the boundary of a simplicial manifold is again a simplicial manifold of co-dimension one, these can be used to compute transition amplitudes only for spin networks with 4-valent graphs dual to a 3d simplicial manifold. A reason to be interested in non-simplicial foams is that one can then look at simpler boundary graphs, and interpolate them with simpler foams. The computational advantage of simpler foams is immediate from \Ref{Avnj}: we get simpler $\{nj\}$ symbols whose evaluation is faster. In the context of abstract spin networks, the simplest possible boundary graphs are flower graphs with a single node, or a dipole graph with two nodes, see Fig.~\ref{fig:dg}. 
Between these two types, dipole graphs have simpler interesting lowest-order spin foams, and have a more intuitive geometric interpretation that makes them preferable for applications.

\subsection{Flower Graphs}
Flower graphs appear often in the discussion of gauge fixing, see e.g. \cite{Charles:2016xwc}. By gauge-invariance, the algebra of observables on any abstract closed graph is isomorphic to that associated with a flower graph, were the number of `petals' matches the number $L-N+1$ of independent loops of the initial graph. The isomorphism depends on a choice of maximal tree on the initial graph, along which all holonomies can be gauge-fixed to the identity. Although flower graphs are very simple, they are not so natural as boundaries of spin foams: each face touching the boundary is rolled-up on itself, meaning it touches twice the same edge. This does not forbid the definition of (non-simplicial) spin foam amplitudes, however it leads to peculiar combinatorics: either a rolled-up face shares both boundaries, or one of the spin foam vertices must have at least one petal in its vertex graph, to absorb the `rolling-up'. This is for instance the case of \cite{Rennert:2013qsa}: there the authors considered a three-petal flower boundary graph, and a single-vertex foam in the bulk. The vertex graph is identical to the boundary graph, and the contribution of the spin foam amplitude is trivial. 
Considering two disconnected flowers for transitions, it is easy to see that our face criterion leads to factorized amplitudes for simple 2-complexes. Using flower boundary graphs leads to subtle combinatorics and not particularly rewarding from the perspective of simplifying the analysis. In the exploratory spirit of this paper, we find it more suitable to work with less peculiar foam structures.

\subsection{Dipole Graphs}
The next simplest graph is a `dipole', namely two nodes connected by a set of $L$ links. It can be associated to a partially gauge-fixed abstract extended graph, where one fixes to the identity all but one of the holonomies on a maximal tree, and it admits a more intuitive geometric interpretation as two atoms of space connected by all faces: the dual graph to a (degenerate) tessellation of the 3-sphere by two $L$-faced polyhedra. 
For the sake of concreteness and simplicity, we restrict attention to 4-valent dipoles, as in Fig.~\ref{fig:dg}. We can consider two different types of transition amplitudes: `nothing-to-dipole' or `dipole-to-dipole'. If $\G$ has a single closed connected component, it is a `nothing-to-$\G$' 
amplitude similar in spirit to the Hartle-Hawking `no-boundary' proposal \cite{HH,Hartle:2008ng}.
If the boundary graph has two closed disconnected components, $W_{\cal C}$ provides a  transition amplitude between two 3-geometries associated with the two $S^3$ boundaries of a 4d hyper-spherical shell. This  set-up was proposed in \cite{Vidotto}, and it is the one we focus on.
It shows very neatly the advantage of working with non-simplicial spin foams: were we to use a simplicial discretization, the minimal configuration would be five tetrahedra for each boundary $S^3$ and a 2-complex with 30 vertices! Whereas with the dipole, one can get transitions with as little as a single vertex.
We list here all possible vertex graphs associated with the expansion \Ref{sf} and the restrictions given by finiteness and our face criterion. 
For completeness, we report in Appendix~\ref{AppDVD} the complete list of vertex graphs relaxing our face criterion. 
With reference to \Ref{sf}, we refer to foams with one vertex as $DVD$ (for dipole-vertex-dipole), two vertices and one edge as $DED$ (for dipole-edge-dipole), two vertices and two edges as $DLD$ (for dipole-loop-dipole). 

\subsubsection*{$DD$ complex}
The first term in \Ref{sf} is a foam with no vertices, and corresponds to the trivial transition amplitude.

\subsubsection*{$DVD$ complex}
\begin{figure}[ht]   
\centering      
\includegraphics[width=6cm]{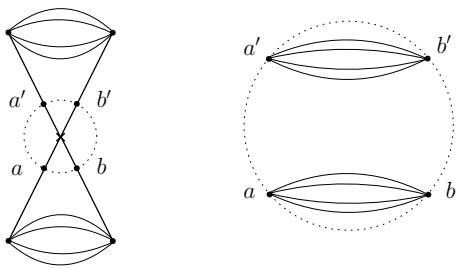}
\caption{\small{\emph{$DVD$ complex and the admissible vertex graph.}} \label{FigDVD}}
\end{figure}

Using our definition of faces, we only have 1 admissible vertex graph, see Fig.\ref{FigDVD}, which is the one originally considered in \cite{Vidotto,VidottoLor}. 
It is disconnected, and thus manifestly leads to a factorized amplitude between initial and final states. 

\subsubsection*{Vertex graph for $DED$}
\begin{figure}[H]   
\centering      
\includegraphics[width=8.5cm]{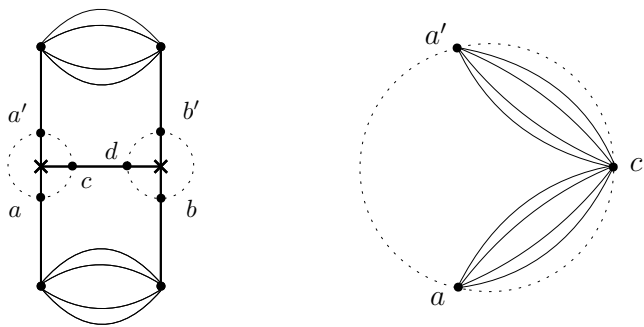}
\caption{\small{\emph{DED complex and the admissible vertex graph.}} \label{FigDED}}
\end{figure}

The smallest foam with two vertices has a single edge connecting them, and the two vertex graphs are necessarily equal and specular from our definition of faces. There is again a single admissible choice, shown in Fig.~\ref{FigDED}: all other routings of the strands would produce faces corresponding to non-minimal cycles.
Although the vertex graph is not disconnected, it still leads to a factorized amplitude. As we will show explicitly below, this simply comes from the fact that we can remove the $\SL(2,\C)$ integration at the middle vertex, thus effectively `disconnecting' the evaluation in two pieces.

\subsubsection*{Vertex graphs for $DLD$}
\begin{figure}[ht]   
\centering      
\includegraphics[width=10cm]{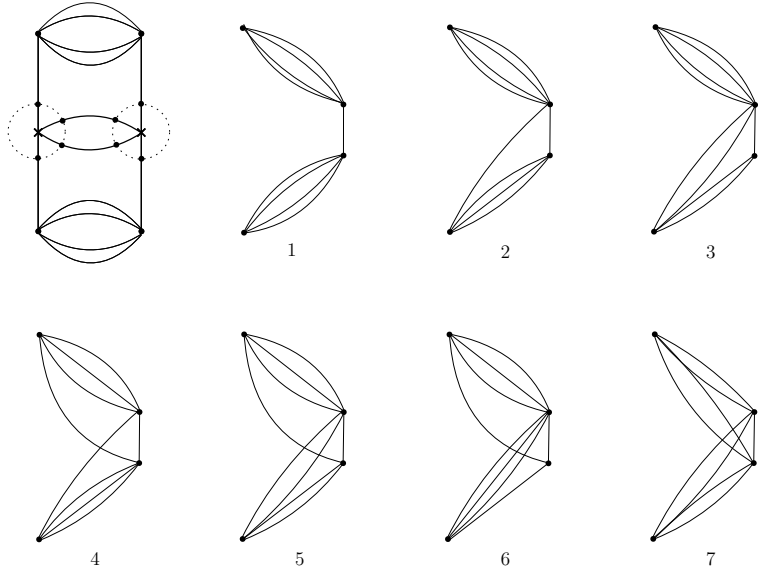}
\caption{\small{\emph{ $DLD$ complex with the seven topologically distinct admissible vertex graphs. Each can give different foams by permutations of the spin labels or by flipping the two internal edges; The graphs 2, 3, 5, and 6 are not symmetric if we flip the internal edges, thus both configurations give two  independent spin foam amplitudes. Graphs 1, 2 and 3 lead to factorized amplitudes, altough 1 and 2 are not 3-link-connected, so their amplitude are not guaranteed to be finite. Graph 6 is the non-factorized amplitude easiest to evaluate, and the one we will study in more details below.\label{FigDLD}}}}
\end{figure}

With two internal edges we have more than just one admissible foam. By our criterion, we can have one internal face only; but the boundary links can be routed in topologically distinct ways without violating the minimal-cycle rule for the faces. This leads to seven different possibilities compatible with our definition of faces, shown in Fig.~\ref{FigDLD}. Again the two vertex graphs must be equal and specular, so we show only one of them. 

Anticipating on the results presented below, the first three diagrams factorize, but not the remaining four. 
These are thus the first ones with non-trivial bulk dynamics, and correlations between initial and final spins.

\bigskip

If we were to allow for faces corresponding to non-minimal cycles, we would have 4 topologically distinct vertex graphs for $DVD$, 3 for $DED$, and 13 for $DLD$, see Appendix~\ref{AppDVD}.

\section{Evaluation of Lorentzian diagrams}

\subsection{Amplitude factorization and booster functions}\label{BoosterFact}
Having listed the foams we are interested in, we now move to their explicit evaluation. The key to do so is the factorization of the amplitudes introduced in \cite{Boosting}. We observe that the vertex amplitude defined in \Ref{Av1} is an $\SL(2,\C)$ tensor invariant, and as such can be represented in terms of Clebsch-Gordan coefficients. For the unitary irreps of the principal series, these can be written as (infinite) sums of Clebsch-Gordan coefficients for the canonical SU(2) matrix subgroup weighted by 1-dimensional boost integrals.
See \cite{Boosting} for details. The result gives
\be
\label{Avnj}
A_v(j_{f}, i_e) = \sum_{l_{f},k_e} \{nj\}(l_{f},k_e)\, \prod_{e=1}^{E_v-\#} B^\g_{v_{e}}(j_{f};l_{f};\vec\imath_e,\vec k_e), 
\ee
where $\{nj\}(l_{f},k_e)$ is the SU(2) $nj$-symbol associated to the vertex graph, labelled by spins $l_{f}$ and $k_e$;
$v_e$ is the valence of the edge $e$ (namely the number of faces it bounds), and 
\begin{align}\label{Bsn}
B^\g_n(j_a,l_a;\vec \imath,\vec k) &= \f1{4\pi}\int_0^\infty dr\sinh^2r \, \sum_{p_a}
 \left(\begin{array}{cccc} j_a \\ p_a \end{array}\right)^{(\vec \imath)}
 \left(\begin{array}{cccc} l_a \\ p_a \end{array}\right)^{(\vec k)}
  \prod_{a=1}^n d^{(\g j_a,j_a)}_{j_al_ap_a}(r).
\end{align}
The boost matrix elements  $d^{(\r,k)}(r)$ for simple irreps are reported in Appendix~\ref{AppSU2}.
The summations over $k_e$ in \Ref{Avnj} are bounded by Clebsch-Gordan conditions, and those over $l_f$ by $l_{f}\geq j_{f}$. 

The factorization result \Ref{Avnj} has the neat effect of minimizing the number of unbounded integrations over boost directions, and reduce the evaluation of the Lorentzian amplitude to an exercise in recoupling theory of SU(2), with the boost contributions as weights localised on the edges, see Fig.~\ref{decaffeinato}. 
The fact that only $E_v-\#$ boosters appear per vertex reflects the redundancy of one group integration in \Ref{Av1}. The freedom to arbitrarily choose the redundant integration can be used at one's advantage to eliminate the booster function with the highest valence $n_e$, a simplification that we will use systematically below. 

For the purposes of using graphical calculus for the recoupling theory, we follow the conventions of \cite{Varshalovich} for the SU(2) part (briefly summarized in Appendix~\ref{AppSU2}), and the following graphical representations for \Ref{Bsn}, 
\be \label{Bshort}
B^{\gamma}_3(j_a;l_a)=
\begin{array}{c}
\psfrag{a}{$j_1$}
\psfrag{b}{$j_2$}
\psfrag{c}{$j_3$}
\psfrag{d}{$l_1$}
\psfrag{e}{$l_2$}
\psfrag{f}{$l_3$}
\includegraphics[width=1.7cm]{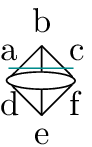}
\end{array} \qquad B^{\gamma}_4(j_a;l_a;i,k) 
= 
\begin{array}{c}
\psfrag{a}{$j_1$}
\psfrag{b}{$j_2$}
\psfrag{c}{$j_3$}
\psfrag{d}{$j_4$}
\psfrag{e}{$l_1$}
\psfrag{f}{$l_2$}
\psfrag{g}{$l_3$}
\psfrag{h}{$l_4$}
\psfrag{i}{$i$}
\psfrag{k}{$k$}
\includegraphics[width=2.2cm]{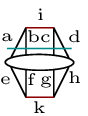}
\end{array}
\ee
\be
B^{\gamma}_5(j_a;l_a;i_1,i_2,k_1,k_2) 
= 
\begin{array}{c}
\psfrag{a}{$j_1$}
\psfrag{b}{$j_2$}
\psfrag{c}{$j_3$}
\psfrag{d}{$j_4$}
\psfrag{e}{$j_5$}
\psfrag{f}{$l_1$}
\psfrag{g}{$l_2$}
\psfrag{h}{$l_3$}
\psfrag{i}{$l_4$}
\psfrag{l}{$l_5$}
\psfrag{u}{$i_1$}
\psfrag{v}{$i_2$}
\psfrag{s}{$k_1$}
\psfrag{t}{$k_2$}
\includegraphics[width=2.9cm]{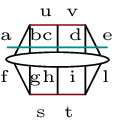}.
\end{array}
\ee
and so on for higher $n$. Here the convention is that both lower and upper sets of spins are assigned from left to right in the $4jm$ symbols.\footnote{This means that all lower (upper) nodes have clockwise (anticlockwise) orientation, see Appendix~\ref{AppSU2}.  As shown in Appendix~\ref{AppBooster}, the booster functions are independent of orientation of the strands. For this reason we omit the arrows on the strands in the picture.}
Because of this diagrammatic expression, the boost integrals 
were referred to as `dipole amplitudes' in \cite{Boosting}. We refrain from using this expression here to avoid confusion with the dipole spin network graph, and call them \emph{booster functions} instead.

In spite of the simplicity achieved by \Ref{Avnj}, the explicit evaluation is still a very complex task: first of all, only for $n=3$ there exist a closed expression for \Ref{Bsn} in terms of finite sums of $\G$ functions, see \cite{Kerimov,Boosting}; for $n\geq 4$ one has to rely on numerical methods to perform the integrals.\footnote{Also for $n\geq 4$ it is possible to analytically perform the $r$ integrals, see \cite{Boosting}; however the results involves new integrations which at present are still of comparable numerical complexity.} The difference is important, since using Mathematica we are able to reliably evaluate $B_3^\g$ for spins up to 100 within few seconds, whereas for $B_4$ we are limited to spins of order 30 by numerical instabilities on hypergeometric functions, and at those spins each value of $B_4$ can take hours to compute. Some numerical calculation times are reported in Fig.~\ref{FigboosterTiming} in the Appendix, based on a standard laptop computer with core speed of 2.5 GHz and 8 Gb of RAM. To perform the calculations of spin foam amplitudes in this paper we used the server maintained by our lab, featuring 32 processors with 3.4 GHz CPU and 24 Gb of RAM. 

Secondly, the factorization has introduced infinite sums over the $l$'s, which play the role of `magnetic numbers' from the $\SL(2,\C)$ viewpoint. These can be very slow to converge and hinder the numerical efficiency. This is for instance the reason why it is difficult to test Barrett's asymptotic formula \cite{BarrettLorAsymp} with Lorentzian boundary data \cite{noiLor}. For some of the foams considered in this paper, the sums turn out to be quickly converging, in which case they can be handled with a controllable cutoff. 

Hence, exact evaluations of Lorentzian EPRL amplitudes are possible but limited, and very costly. One important information that we are interested in is the large spin scaling of the amplitudes. For such estimates the factorization \Ref{Avnj} is very useful because we can use known results from the asymptotics of SU(2) $nj$-symbols \cite{Bonzom:2008xd,BarrettSU2} and combine them with scalings of the booster functions and estimates for the sums. To that end, let us briefly review some scaling properties of the booster functions. Plots showing the behaviours and claims of this section, as well as more details and expressions for the booster functions and their numerical evaluations can be found in Appendix~\ref{AppBooster}.

\subsection{Boosters' scalings}
\begin{enumerate}
\item Minimal spins, homogeneous large spin asymptotics (see \cite{Boosting} and Fig.~\ref{B4asymp}.):
\be
B_{3}^\g(N j_a;N j_a) \sim \left\{ \begin{array}{ll} N^{-3/2} & {\rm if} \quad \sum_aj_a \ {\rm odd}
\\ \label{BnLO} N^{-1} & {\rm if} \quad \sum_aj_a \ {\rm even} \end{array} \right. , \qquad B_{n\geq 4}^\g(N j_a;N j_a;\vec\imath, \vec k) \sim N^{-3/2}. 
\ee
In the second expression, the leading order scaling is the same irrespectively of the intertwiners. However, seen as a function of the intertwiners at fixed spins, it is strongly peaked at equal values. The precise shape and width of the peak depends on the spins considered, but the drop is measured in orders of magnitude already one step away from equal values. Hence, it is qualitatively fair to complete the estimate with a Kronecker delta $ \d_{ \vec\imath \vec k}$. 
There exist also an analytic estimate for the leading order, given by  \cite{Puchta:2013lza}:
\be\label{Puchta}
B_n^\g(N j_a;N j_a;i,k) \sim \f1{(4\pi)^2}\left[\f{6\pi}{(1+\g^2)N\sum_{a}j_a}\right]^{3/2} \,  \f{ \d_{ \vec\imath \vec k}}{d_{\vec\imath}}.
\ee
This correctly reproduce the power law decay, and numerical tests (see e.g. \cite{Boosting} or App.~\Ref{AppBooster}) confirm also the $\g$ and $\sum_a j_a$ dependence. On the other hand, the numerical factor is not very accurate (the error is of order 1), nor the dependence on the intertwiners. Indeed as mentioned above configurations with non-equal intertwiners also decay like $N^{-3/2}$ and not faster as \Ref{Puchta} would suggest.

\item Minimal spins, inhomogeneous large spin asymptotics with one spin small (see Fig.~\ref{B3InHomo}):
\begin{subequations} \label{BInhomo}
\begin{align}\label{B3InhomLim}
& B^\g_3(Nj_{\bar a},j_3;Nj_b,j_3) \sim N^{-1},
\qquad 
B^\g_4(Nj_{\bar a},j_4;Nj_b,j_4;Ni,Nk) \sim N^{-5/2},
\\ & B^\g_5(Nj_{\bar a},j_5;Nj_b,j_5;Ni_1,Nk_1,Ni_2,Nk_2) \sim N^{-7/2},
\end{align}\end{subequations}
where in the above $\bar{a}=1,\ldots, n-1$. 
 
\item Non-minimal spin decay, homogeneous (see \cite{Boosting} and Fig.~\ref{FigB4NonMinimal}): 
\be\label{BnDl}
B_3^\g(j_a;j_a+\D l) \sim \f1{\sqrt{\D l}}, \qquad 
B_4^\g(j_a;j_a+\D l;i,k) \sim \f1{\D l}
\ee

\item Non-minimal spin decay, inhomogeneous: an exponential decay if we rescale a single $l$ label, see Fig.~\ref{FigDecB}. 

\end{enumerate}

\subsection{Simplified model and scaling estimates}
The non-minimal spin decays described in points 3 and 4 just above suggest that for some configurations good estimates of the EPRL model can be obtained looking at the minimal configurations only, $l_a=j_a$. This restriction defines a simplified version of the EPRL model, called EPRLs in \cite{Boosting}, s for simplified. Since the EPRLs amplitudes are much faster to evaluate, it is useful to get a hand on when the simplification gives a good approximation of the full EPRL model. There are three possibilities, depending on the foam:

\begin{enumerate}[(i)]
\item The summations over $l$'s are truncated to the minimal value by Clebsch-Gordan inequalities: this can happen for particular configurations, and if it happens for all $l$s, we refer to such amplitudes as \emph{EPRLs-exact}.
\item The summations are bounded, or unbounded but quickly converging: EPRLs gives different numerical values than EPRL, nonetheless it provides precise estimates of the large spin scalings and of correlations' properties.
\item The summations are unbounded and slowly converging: the EPRLs differs strongly from the EPRL, and does not provide good estimates. 
\end{enumerate}

Because of the intricate structure of the amplitudes, it is not easy to give a general criterion to identify a priori which ones are well approximated by the simplified model. One of the results of this paper is to show for which of the considered foams it happens. The interest in this is that the simplified model is much faster to evaluate, and furthermore, it allows us to use the analytic estimate \Ref{Puchta}.

\section{Dipole-to-dipole spin foam amplitudes}\label{DDFoams}
In this Section we present the explicit form of the amplitudes for the foams listed in Section~\ref{graphstype}. 
For an SU(2) BF theory, the evaluation of the various spin foam amplitudes associated to the above diagrams is a straightforward, if somewhat tedious, exercise in recoupling theory. It is very similar for the EPRL model, using the factorization reviewed above. 
We use graphical calculus for SU(2) with the conventions of \cite{Varshalovich}, briefly reported in Appendix~\ref{AppSU2}, and the notation \Ref{Bshort} for the booster functions. In the diagrams below, every strand represents a face and every box an integration over $\SL(2,\C)$ assigned for each half-edge. 
One has to choose a face orientation, but the result is independent of this choice (not of the orientation of the boundary links, on the other hand). 

\subsection{$DD$ foam}
The first term of \Ref{sf} is a trivial foam, with no internal vertices and faces bounded by a lower and an upper link. The corresponding `face-rigid' amplitude is the identity,
\begin{eqnarray}
W^{DD}(j_a,j'_a;i,t,i',t')=\delta_{j_a,j'_a} \delta_{i,i'}\,\delta_{t,t'}.
\end{eqnarray}

\subsection{$DVD$ foam}
The vertex graph is disconnected (see Fig.~\ref {FigDVD}), leading to an exact factorization of the amplitude into lower and upper contributions.
Each connected component has a single $\SL(2,\C)$ integration after gauge-fixing, and the amplitude reads
\be \nn
\begin{split}
\label{DVD1}
W^{DVD}(j_a,j'_a;i,t,i',t')&= d_id_{i'}d_td_{t'} \int dg_1 dg_2 \prod_{a=1}^4 \, D^{\gamma j_a,j_a}_{j_a m_a j_a n_a}(g_1)\,D^{\gamma j'_a,j'_a}_{j'_a m'_i j'_a n'_i}(g_2) \left(\begin{matrix} j_{a} \\ m_{a} \end{matrix}\right)^{(i)}\left(\begin{matrix} j_{a} \\ n_{a} \end{matrix}\right)^{(t)}\left(\begin{matrix} j'_{a} \\ m'_{a} \end{matrix}\right)^{(i')}\left(\begin{matrix} j'_{a} \\ n'_{a} \end{matrix}\right)^{(t')} \\
& = d_id_{i'}d_td_{t'}
\begin{array}{c}
\psfrag{e}{$j^\prime_1$}
\psfrag{f}{$j^\prime_2$}
\psfrag{g}{$j^\prime_3$}
\psfrag{h}{$j^\prime_4$}
\psfrag{a}{$j_1$}
\psfrag{b}{$j_2$}
\psfrag{c}{$j_3$}
\psfrag{d}{$j_4$}
\psfrag{s}{$i^\prime$}
\psfrag{t}{$t^\prime$}
\psfrag{u}{$i$}
\psfrag{v}{$t$}
\psfrag{1}{$g_1$}
\psfrag{2}{$g_2$}
\psfrag{l}{$I$}
\includegraphics[scale=2]{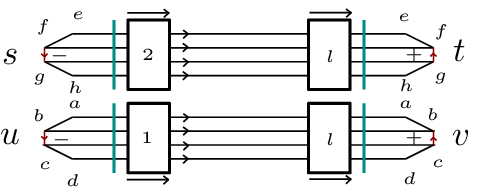}
\end{array} 
\end{split}
\ee
where in the graphical notation we used a box with an identity $I$ to flag the group integrations we removed. The arrows on the box keep track of the edge orientations, and the signs of the spin ordering chosen for the $4jm$ symbols, see Appendix~\ref{AppSU2}.\footnote{In the literature the signs in the graphical calculus are often discarder under the assumption of fixing a convention once and for all, say anticlockwise. We prefer to keep track of them so that we can use an unique algebraic $4jm$ symbol and have a easier mnemonic rule for the boundary spins ($j_a$ goes to $j'_a$ with the same $a$)}.

To proceed with the graphical evaluation we  use \Ref{decaffeinato} to split the boxes in terms of intertwiners. This results in one booster function and one SU(2) 4-valent $\th$ graphs per edge. The $\th$ graph gives a dimensional factor, see \Ref{thetaeq}, and we get
\be
W^{DVD}(j_a,j'_a;i,t,i',t') = d_id_{i'}d_td_{t'}
\begin{array}{c}\psfrag{a}{$j_1$}
\psfrag{b}{$j_2$}\psfrag{c}{$j_3$}
\psfrag{d}{$j_4$}\psfrag{e}{$j_1$}
\psfrag{f}{$j_2$}
\psfrag{g}{$j_3$}
\psfrag{h}{$j_4$}
\psfrag{i}{$i$}
\psfrag{k}{$t$}
\includegraphics[width=2.2cm]{_images/B4.eps}
\end{array} 
\begin{array}{c}
\psfrag{a}{$j'_1$}
\psfrag{b}{$j'_2$}
\psfrag{c}{$j'_3$}
\psfrag{d}{$j'_4$}
\psfrag{e}{$j'_1$}
\psfrag{f}{$j'_1$}
\psfrag{g}{$j'_3$}
\psfrag{h}{$j'_4$}
\psfrag{i}{$i'$}
\psfrag{k}{$t'$}
\includegraphics[width=2.2cm]{_images/B4.eps}
\end{array}.
\ee
The amplitude factorizes as anticipated, and we see that there are no summations over the magnetic spins $l$: In the $DVD$ case the simplified model coincides with the complete model, or in other words, this diagram is EPRLs-exact.
We can then use the estimate \Ref{Puchta} for the large spin scaling, finding
\be\label{DVD1LO}
W_{DVD}(N j_a,N j_a';i,t,i',t') \sim \f1{N^3}\f{27}{32\pi}\f1{(1+\g^2)^3}\f {d_{i}d_{i'} \d_{i t}\d_{i' t'}}{(\sum_aj_a)^{3/2}(\sum_aj'_a)^{3/2}}.
\ee
We have an $N^{-3}$ power law, peaked on configurations with equal intertwiners for each connected component of the boundary graph. 
The only correlations introduced by the amplitude are within spins and intertwiners within each connected component of the boundary graph.

In Appendix~\ref{AppDVD} we study the other 3 integrable vertex graphs with non-minimal faces. We do so to get a first understanding of what is excluded by our face criterion, but also because they provide simple examples of some interesting features:
diagrams non 3-link-connected yet integrable; diagrams with unbounded $l$ summations but which are nonetheless well approximated by the simplified model; presence of non-trivial $6j$ symbols and thus oscillation in the large spin limit.

\subsection{$DED$ foam}
For the vertex graph in Fig.~\ref{FigDED}, the associated spin foam amplitude is (here and in the following we omit the first, trivial step of listing the matrix elements, and use directly the graphical calculus)
\be \nn
\begin{split}
\label{figDED1}
W^{DED}(j_a,j'_a;i,t,i',t')&= d_id_{i'}d_td_{t'}
\begin{array}{c}
\psfrag{e}{$j^\prime_1$}
\psfrag{f}{$j^\prime_2$}
\psfrag{g}{$j^\prime_3$}
\psfrag{h}{$j^\prime_4$}
\psfrag{a}{$j_1$}
\psfrag{b}{$j_2$}
\psfrag{c}{$j_3$}
\psfrag{d}{$j_4$}
\psfrag{s}{$i^\prime$}
\psfrag{t}{$t^\prime$}
\psfrag{u}{$i$}
\psfrag{v}{$t$}
\psfrag{1}{$g_1$}
\psfrag{2}{$g_2$}
\psfrag{3}{$g_3$}
\psfrag{4}{$g_4$}
\psfrag{l}{$I$}
\includegraphics[scale=2]{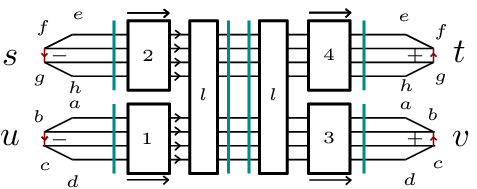}
\end{array}
\end{split}
\ee
Notice that, by choosing judiciously the integration to be removed, we can avoid the presence of booster functions with 8 legs. This choice further shows manifestly that the vertex graph reduces to two connected components, although this would be true with any other gauge-fixing chosen. Hence, the amplitude is again factorized, and no mixing of lower and upper spins occurs. 

Proceeding as before with \Ref{decaffeinato} and the normalization of the SU(2) generalized $\theta$ graphs we get
\be 
\begin{split}
W^{DED}(j_a,j_a';i,t,i',t') =d_id_{i'}d_td_{t'} & \sum_{k} d_k 
\begin{array}{c}
\psfrag{a}{$j_1$}
\psfrag{b}{$j_2$}
\psfrag{c}{$j_3$}
\psfrag{d}{$j_4$}
\psfrag{e}{$j_1$}
\psfrag{f}{$j_2$}
\psfrag{g}{$j_3$}
\psfrag{h}{$j_4$}
\psfrag{i}{$i$}
\psfrag{k}{$k$}
\includegraphics[width=2.2cm]{_images/B4.eps}
\end{array} 
\begin{array}{c}
\psfrag{a}{$j_1$}
\psfrag{b}{$j_2$}
\psfrag{c}{$j_3$}
\psfrag{d}{$j_4$}
\psfrag{e}{$j_1$}
\psfrag{f}{$j_2$}
\psfrag{g}{$j_3$}
\psfrag{h}{$j_4$}
\psfrag{i}{$t$}
\psfrag{k}{$k$}
\includegraphics[width=2.2cm]{_images/B4.eps}
\end{array} 
\\ \times & \sum_{k'} d_{k'}
\begin{array}{c}
\psfrag{a}{$j'_1$}
\psfrag{b}{$j'_2$}
\psfrag{c}{$j'_3$}
\psfrag{d}{$j'_4$}
\psfrag{e}{$j'_1$}
\psfrag{f}{$j'_1$}
\psfrag{g}{$j'_3$}
\psfrag{h}{$j'_4$}
\psfrag{i}{$i'$}
\psfrag{k}{$k'$}
\includegraphics[width=2.2cm]{_images/B4.eps}
\end{array}\begin{array}{c}
\psfrag{a}{$j'_1$}
\psfrag{b}{$j'_2$}
\psfrag{c}{$j'_3$}
\psfrag{d}{$j'_4$}
\psfrag{e}{$j'_1$}
\psfrag{f}{$j'_1$}
\psfrag{g}{$j'_3$}
\psfrag{h}{$j'_4$}
\psfrag{i}{$t'$}
\psfrag{k}{$k'$}
\includegraphics[width=2.2cm]{_images/B4.eps}
\end{array}.
\label{WDED}
\end{split}
\ee
Again only minimal spins enter, and the diagram is EPRLs-exact. The estimate \Ref{Puchta} gives the asymptotic behaviour
\be
W^{DED}(j_a,j'_a;i,t,i',t') \sim \f1{N^6}\f1{(4\pi)^8}\left[\f{6\pi}{(1+\g^2)}\right]^{6} \frac{ \d_{it}  \d_{i',t'} }{(\sum_a j_a)^3 (\sum_a j'_a)^{3} }.
\ee
This foam order $\l^2$ is thus suppressed with respect to the one of order $\l$ by three powers in the large spin limit.
As for correlations, the situation is identical to the previous one: the amplitude factorizes, and the only non-trivial correlations are within each connected component of the graph.

\subsection{$DLD$ foams}\label{SecDLD6}
Foams with an internal face give more interesting amplitudes. First of all, these foams turn out to contain sums over the magnetic $l$'s, hence the EPRL amplitudes differ from the simplified ones. Secondly, the presence of an internal face can couple lower and upper spins. This does not happen for all foams: the first three vertex graphs of Fig. \ref{DLDTutti} still give factorized amplitudes: to see this, notice that all spins of a connected part of the boundary go through the same internal edge. If we choose to remove the group integral associated with that edge on both vertices, we immediately see that those spins completely decouple from the rest of the diagram. 
The remaining four are all coupled. Among them, the simplest to evaluate is number 6: it has a six-valent node, whose integration can be removed by gauge-fixing, leaving only 3-valent and 4-valent strands behind. 
We study this one first and in more details, and report some properties of the others in the next Section. Unless otherwise stated, the reference value for the Immirzi parameter in the numerical calculations is $\g=6/5$.

The amplitude of $DLD_6$ is given by
\be \nn
\begin{split}
\label{DLDAmplitudeFoam}
W^{DLD_6}(j_a,j_a';i,t;i',t')&= d_id_{i'}d_td_{t'}
\begin{array}{c}
\psfrag{a}{$j_1$}
\psfrag{b}{$j_2$}
\psfrag{c}{$j_3$}
\psfrag{d}{$j_4$}
\psfrag{e}{$j^\prime_1$}
\psfrag{f}{$j^\prime_2$}
\psfrag{g}{$j^\prime_3$}
\psfrag{h}{$j^\prime_4$}
\psfrag{j}{$j_f$}
\psfrag{s}{$i^\prime$}
\psfrag{t}{$t^\prime$}
\psfrag{u}{$i$}
\psfrag{v}{$t$}
\psfrag{1}{$g_1$}
\psfrag{2}{$g_2$}
\psfrag{3}{$g_3$}
\psfrag{4}{$g_4$}
\psfrag{5}{$g_5$}
\psfrag{6}{$g_6$}
\psfrag{l}{$I$}
\includegraphics[scale=1.8]{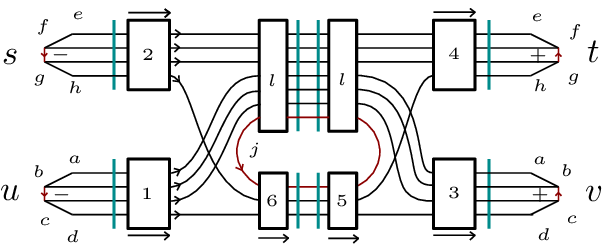}
\end{array}
\end{split}
\ee
where we used the internal SU(2) gauge invariance to route the faces within each boxes in a convenient way. Permuting spin labels or flipping the internal edges gives trivially related amplitudes.
We further use the freedom to remove one group integral by vertex to remove the highest valent ones, marked with $I$ in the picture.
The internal lines without the $Y$-map are magnetic $l$'s. However those that hit an $I$-box
are also projected by the $Y$-map to their minimal value. This leaves us with four free $l$'s only, those going through the $g_5$ and $g_6$ boxes. Splitting the boxes as in \Ref{decaffeinato} and using the orthogonality \Ref{Rule}, the $l$'s must be pairwise equal and  the SU(2) symbols reduce to generalized $\th$ graphs. The amplitude is then equivalent to
\be \label{DLD6}
\begin{split}
W^{DLD_6}(j_a,j_a';i,t;i',t' )
& =d_i d_t d_{i^\prime} d_{t^\prime} 
\sum_{l,l^\prime k, k^\prime} \f{ d_k d_{k^\prime} }{d_l d_{l^\prime}} \bigg[\sum_{j_f} d_{j_f}
\bigg( \begin{array}{c}
\psfrag{a}{$j_4$}
\psfrag{b}{$j^\prime_4$}
\psfrag{c}{$j_f$}
\psfrag{d}{$l$}
\psfrag{e}{$l^\prime$}
\psfrag{f}{$j_f$}
\includegraphics[width=1.7cm]{_images/B3.eps}
\end{array} \bigg)^2\ \bigg]
\\ & \hspace{1.cm} \times
\begin{array}{c}
\psfrag{a}{$j_1$}
\psfrag{b}{$j_2$}
\psfrag{c}{$j_3$}
\psfrag{d}{$j_4$}
\psfrag{e}{$j_1$}
\psfrag{f}{$j_2$}
\psfrag{g}{$j_3$}
\psfrag{h}{$l$}
\psfrag{i}{$i$}
\psfrag{k}{$k$}
\includegraphics[width=2.2cm]{_images/B4.eps}
\end{array}
\begin{array}{c}
\psfrag{a}{$j_1$}
\psfrag{b}{$j_2$}
\psfrag{c}{$j_3$}
\psfrag{d}{$j_4$}
\psfrag{e}{$j_1$}
\psfrag{f}{$j_2$}
\psfrag{g}{$j_3$}
\psfrag{h}{$l$}
\psfrag{i}{$t$}
\psfrag{k}{$k$}
\includegraphics[width=2.2cm]{_images/B4.eps}
\end{array}
\begin{array}{c}
\psfrag{a}{$j^\prime_1$}
\psfrag{b}{$j^\prime_2$}
\psfrag{c}{$j^\prime_3$}
\psfrag{d}{$j^\prime_4$}
\psfrag{e}{$j^\prime_1$}
\psfrag{f}{$j^\prime_2$}
\psfrag{g}{$j^\prime_3$}
\psfrag{h}{$l^\prime$}
\psfrag{i}{$i^\prime$}
\psfrag{k}{$k^\prime$}
\includegraphics[width=2.2cm]{_images/B4.eps}
\end{array}
\begin{array}{c}
\psfrag{a}{$j^\prime_1$}
\psfrag{b}{$j^\prime_2$}
\psfrag{c}{$j^\prime_3$}
\psfrag{d}{$j^\prime_4$}
\psfrag{e}{$j^\prime_1$}
\psfrag{f}{$j^\prime_2$}
\psfrag{g}{$j^\prime_3$}
\psfrag{h}{$l^\prime$}
\psfrag{i}{$t^\prime$}
\psfrag{k}{$k^\prime$}
\includegraphics[width=2.2cm]{_images/B4.eps}
\end{array}.
\end{split}
\ee
With respect to the previous diagrams, we have two new features: the first is 
a coupling of lower and upper labels, thanks to the internal face function
\be\label{C3def}
C_{l,l'}(j_4,j'_4) := \sum_{j_f={\max\{|j_4-j'_4,|l-l'|\}}}^{\min\{j_4+j_4'|,l+l'\}} d_{j_f} \Big( B_3^\g(j_4,j'_4,j_f;l,l',j_f) \Big)^2, 
\ee
where we made the extremes of the sum explicit with respect to \Ref{DLD6}.
The second is the summations over the magnetic-$l$ numbers: this amplitude differs then from the one computed with the simplified model.

\begin{figure}[H] 
\begin{center}
\begin{minipage}{0.4 \textwidth}
\centering
\vspace{0.75cm}
\includegraphics[width=7cm]{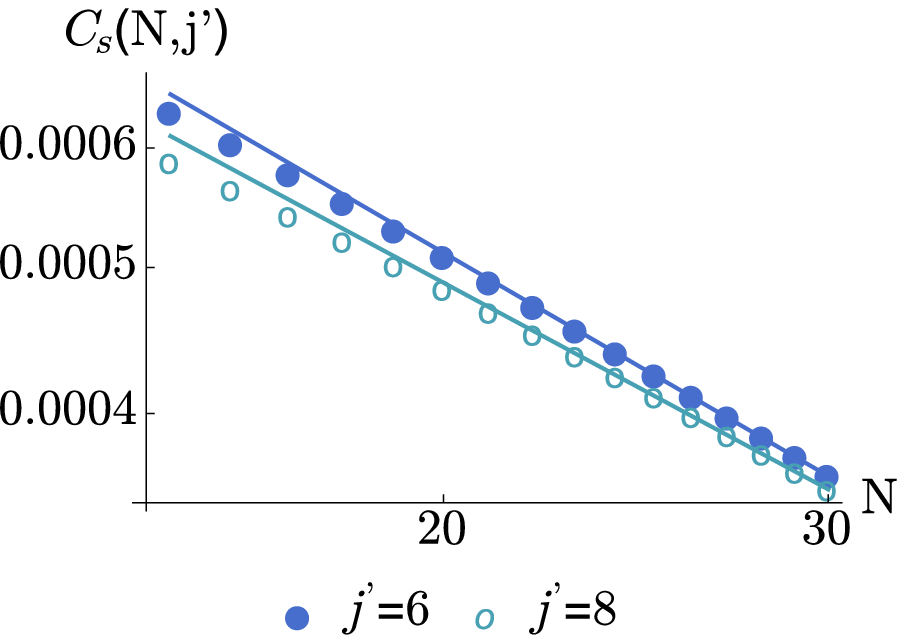}
\end{minipage}
\hspace{1cm} 
\begin{minipage}{0.4 \textwidth}
\centering
\includegraphics[width=7cm]{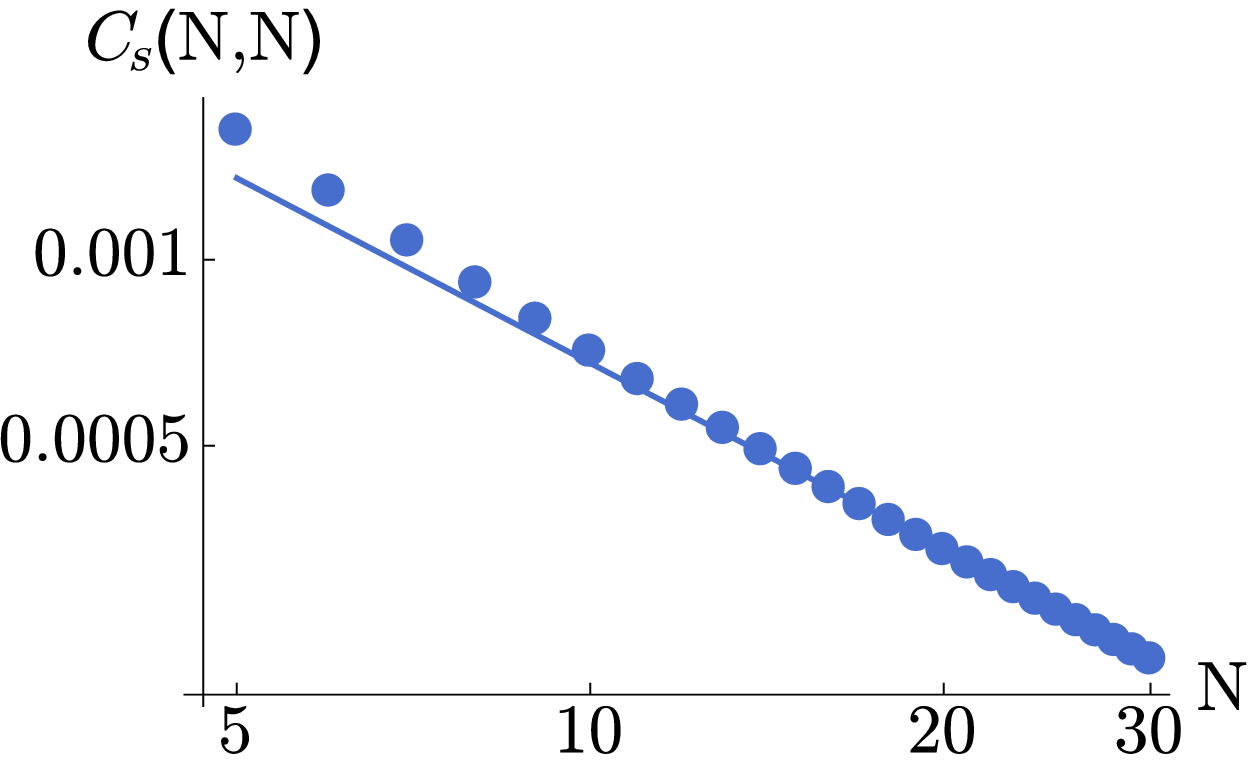}
\end{minipage}
\end{center}
\caption{\small{\emph{Numerical confirmation of the asymptotics \Ref{CsScalings} of the face correlation function  for the simplified model. The dots are the numerical data, the lines fits obtained assuming a power law $N^{-1}$.}  }
\label{FigCBscaling}}
\end{figure}

In order to study the large spin beahviour of \Ref{DLD6}, we begin within the simplified model: namely we cut the two $l$-summations to their first contribution: $l=j_4$, $l'=j_4'$, and define in this way the EPRLs amplitude $W_s^{DLD_6}(j_a,j_a';i,t;i',t' )$ and
its internal face correlation as 
\be\label{Cs}C_s(j,j') := C_{j,j'}(j,j').\ee
Using the inhomogeneous scaling $B^\g_3(Nj_1,Nj_2,j_3) \sim N^{-1}$, see \Ref{B3InHomo}, and the boundness of the summation in $j_f$, we can infer the power laws
\be\label{CsScalings}
C_s(Nj,j') \sim N^{-1}, \qquad C_s(Nj, Nj') \sim N^{-1}.
\ee
Both are confirmed with good accuracy by numerical analysis, see Fig.~\ref{FigCBscaling}.
Using this result and 
\Ref{BnLO} for the scaling of $B_4^\g$, we estimate the EPRLs contribution to \Ref{DLD6} to be 
\be\label{Ws6}
W_s^{DLD_6}(N j ,N j';i,t;i',t' ) \sim N^{-9}.
\ee
We further expect peakedness on equal intertwiners as before, $\d_{it}\d_{i't'}$.

\begin{figure}[H] 
\begin{center}
\begin{minipage}{0.4 \textwidth}
\centering
\vspace{0.4cm}
\includegraphics[width=7.cm]{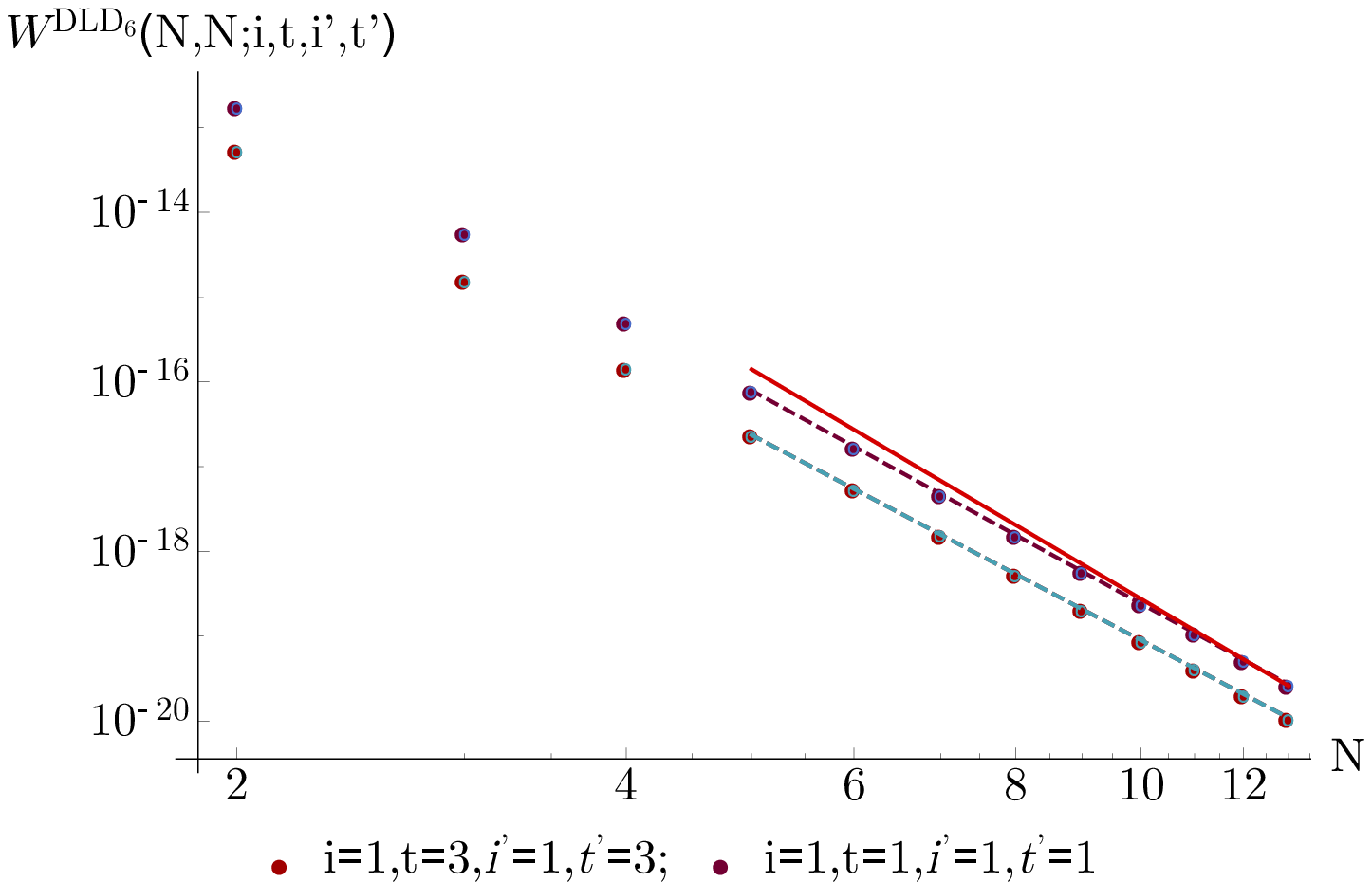}
\end{minipage}
\hspace{1cm} 
\begin{minipage}{0.4 \textwidth}
\centering
\includegraphics[width=6.cm]{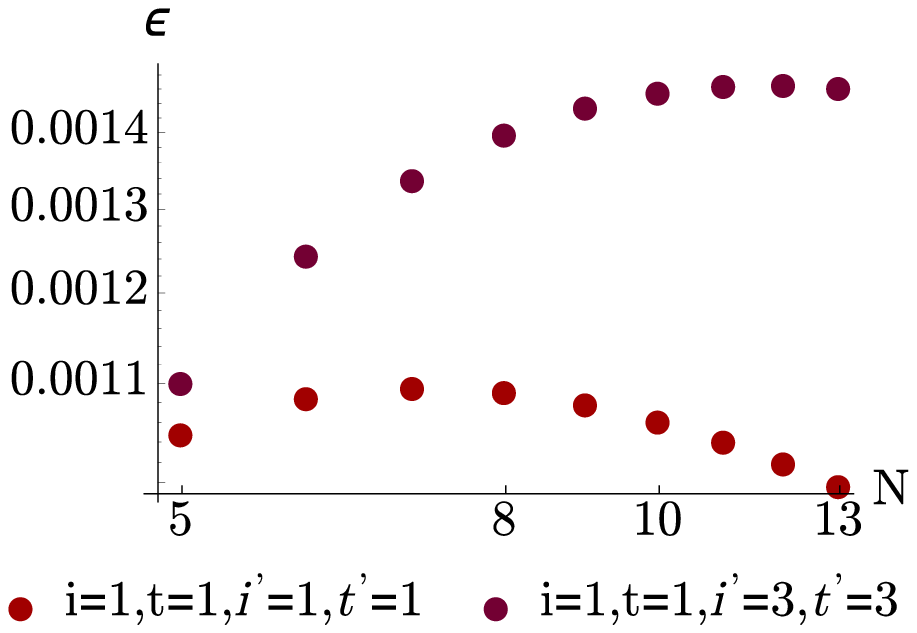}
\end{minipage}
\end{center}
\caption{\small{Left panel: \emph{Large spin scaling of $W^{DLD_6}(N,N;i,t;i',t' )$ in \Ref{DLD6} for equal spins and two different configurations of intertwiners. The dots are the numerical data, and the continuous lines the expected $N^{-9}$ decay derived in \Ref{W6}. The dashed lines are the actual fits at the spins we can access. Using the last 5 points, the fits give $N^{-8.4}$ and $N^{-8.1}$ respectively. The discrepancy can be imputed to the slow spins, at which already the indivudal booster function is not yet at asymptotic behaviour, see Fig.~\ref{B4asymp} in the Appendix. 
Data points for the simplified model are indistinguishable, the difference is in the per thousands and shown in the right panel.} 
Right panel: \emph{Numerical studies of the relative error 
between the EPRL model and the simplified one, $\eps:=|W^{DLD_6}-W^{DLD_6}_s|/W^{DLD_6}$. } }
\label{FigDLDscaling} }
\end{figure}

From the simplified to the full EPRL model there are only a finite number of extra terms: the sum over $l$ is bounded by $j_3$ and $k$, which in turn is bounded by $j_1$ and $j_2$; and similarly for $l'$. 
We thus expect the same scaling,
\be\label{W6}
W^{DLD_6}(Nj,Nj';i,t;i',t' ) \sim N^{-9}.
\ee
We performed numerical tests of both estimates \Ref{Ws6} and \Ref{W6} at spins of order 10, finding a qualitative agreement, see Fig.\ref{FigDLDscaling}.
This could be made sharper pushing to higher spins, since 
at such low spins, even the individual $B_4^\g$ is still not precisely at its asymptotic power law behaviour, see Fig.~\ref{B4asymp}.
However we lack at present the numerical power to do so, because Mathematica has instabilities in evaluating and integrating hypergeometric functions starting from around spins $j=30$.
Given the extremes of \Ref{C3def}, this puts our numerical limits at spins of order 10.\footnote{We can go beyond only for diagrams having only three-valent booster functions, since for these we can use a finite sum expression without hypergeometric functions \cite{Boosting}. }

This means that even if this foam is not EPRLs exact, the difference is very small. The simplified model captures the correct scaling, and also the spin correlations, as we will see in Section~\ref{spinCorr}. 

\subsection{$DLD$ foams: additional vertex graphs}

As we have seen, integrability and our face criterion do not single out a unique amplitude diagram for the $DLD$ complex. Next to $DLD_6$ studied in the previous Section, there are 3 additional types of vertex graphs leading to non-factorized amplitudes, those numbered 4,5 and 7. As before, these graphs give raise to different foams by considering permutations of spin labels, and swapping the routing through the internal edges. We consider here only one representative of these possible permutations. 

In order of numerical complexity, the next non-factorized amplitude is the one with vertex  graphs number 5, which only contains 4-stranded boosters. Proceeding as in the previous Section, we have
\begin{align}\nn
 W^{DLD_5}(j_a,j^\prime_a;i,t,i',t') & = 
 d_id_{i'}d_td_{t'}
\begin{array}{c}
\psfrag{a}{$j_1$}
\psfrag{b}{$j_2$}
\psfrag{c}{$j_3$}
\psfrag{d}{$j_4$}
\psfrag{e}{$j^\prime_1$}
\psfrag{f}{$j^\prime_2$}
\psfrag{g}{$j^\prime_3$}
\psfrag{h}{$j^\prime_4$}
\psfrag{j}{$j_f$}
\psfrag{s}{$i^\prime$}
\psfrag{t}{$t^\prime$}
\psfrag{u}{$i$}
\psfrag{v}{$t$}
\psfrag{1}{$g_1$}
\psfrag{2}{$g_2$}
\psfrag{3}{$g_3$}
\psfrag{4}{$g_4$}
\psfrag{5}{$g_5$}
\psfrag{6}{$g_6$}
\psfrag{l}{$I$}
\includegraphics[scale=1.8]{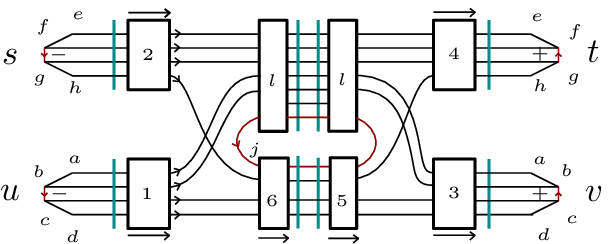}
\end{array}
 \\ \label{ADLD5}
 & =
 d_i d_t d_{i^\prime} d_{t^\prime}
\sum_{l_a,\bar l_a, l^{\prime}_4 \atop  k, k^{\prime},h} \frac{ d_k d_{k^{\prime}}  d_h } {d_{l_4} d_{l^{\prime}_4}}  
\bigg( \sum_{j_f} d_{j_f}
\begin{array}{c}
\psfrag{a}{$j_4$}
\psfrag{b}{$j_3$}
\psfrag{c}{$j^{\prime}_4$}
\psfrag{d}{$j_f$}
\psfrag{e}{$l_4$}
\psfrag{f}{$l_3$}
\psfrag{g}{$l^\prime_4$}
\psfrag{h}{$j_f$}
\psfrag{i}{$h$}
\psfrag{k}{$k$}
\includegraphics[width=2.2cm]{_images/B4.eps}
\end{array}
\begin{array}{c}
\psfrag{a}{$j_4$}
\psfrag{b}{$j_3$}
\psfrag{c}{$j^\prime_4$}
\psfrag{d}{$j_f$}
\psfrag{e}{$\bar l_4$}
\psfrag{f}{$\bar l_3$}
\psfrag{g}{$l^\prime_4$}
\psfrag{h}{$j_f$}
\psfrag{i}{$h$}
\psfrag{k}{$k^\prime$}
\includegraphics[width=2.2cm]{_images/B4.eps}
\end{array} \bigg)
\\ &   \hspace{2cm} \times
\begin{array}{c}
\psfrag{a}{$j_1$}
\psfrag{b}{$j_2$}
\psfrag{c}{$j_3$}
\psfrag{d}{$j_4$}
\psfrag{e}{$j_1$}
\psfrag{f}{$j_2$}
\psfrag{g}{$l_3$}
\psfrag{h}{$l_4$}
\psfrag{i}{$i$}
\psfrag{k}{$k$}
\includegraphics[width=2.2cm]{_images/B4.eps}
\end{array}
\begin{array}{c}
\psfrag{a}{$j_1$}
\psfrag{b}{$j_2$}
\psfrag{c}{$j_3$}
\psfrag{d}{$j_4$}
\psfrag{e}{$j_1$}
\psfrag{f}{$j_2$}
\psfrag{g}{$\bar l_3$}
\psfrag{h}{$\bar l_4$}
\psfrag{i}{$t$}
\psfrag{k}{$k$}
\includegraphics[width=2.2cm]{_images/B4.eps}
\end{array}
\begin{array}{c}
\psfrag{a}{$j^\prime_1$}
\psfrag{b}{$j^\prime_2$}
\psfrag{c}{$j^\prime_3$}
\psfrag{d}{$j^\prime_4$}
\psfrag{e}{$j^\prime_1$}
\psfrag{f}{$j^\prime_2$}
\psfrag{g}{$j^\prime_3$}
\psfrag{h}{$l^\prime_4$}
\psfrag{i}{$i^\prime$}
\psfrag{k}{$k^\prime$}
\includegraphics[width=2.2cm]{_images/B4.eps}
\end{array}
\begin{array}{c}
\psfrag{a}{$j^\prime_1$}
\psfrag{b}{$j^\prime_2$}
\psfrag{c}{$j^\prime_3$}
\psfrag{d}{$j^\prime_4$}
\psfrag{e}{$j^\prime_1$}
\psfrag{f}{$j^\prime_2$}
\psfrag{g}{$j^\prime_3$}
\psfrag{h}{$l^\prime_4$}
\psfrag{i}{$t^\prime$}
\psfrag{k}{$k^\prime$}
\includegraphics[width=2.2cm]{_images/B4.eps}
\end{array}\nn
\end{align}
There are five $l$ summations (the SU(2) symbol associated with the internal face identifies two initially different $l_4'$ and $\bar{l}_4'$), only one bounded. This EPRL amplitude differs thus  from the simplified model by an infinite number of terms, decreasing in magnitude with a power law in the $l$s.
The quantity in round brackets identifies the face correlation function coupling lower and upper spins. It is again a square of boosters, 4-stranded this time. 

As before, let us first estimate the scaling of the simplified model. 
The first thing to notice is the dependence of the face correlation function on the intertwiner labels. If we let the boundary spins grow, Clebsch-Gordan inequalities quickly reduce to zero the number of terms in the sum over the internal spin, unless we let the intertwiners also grow. Hence, the dominating contribution comes from large bulk intertwiners. But since in the simplified model the boundary booster functions (in the last line of \Ref{ADLD5}) are sharply peaked on equal intertwiners, this means that the whole amplitude scales differently for different values of the boundary intertwiners, with dominant contribution coming from the largest intertwiners. This is quite different from the previous foam considered, where the scaling was homogeneous in the intertwiners. 
To estimate the power law associated with this dominant scaling, we look 
at the inhomogeneous scaling of $B_4^\g$ with one small spin, see \Ref{B3InhomLim}. Then, since the number of terms to be summed over grows linearly with the spins, we estimate 
\be
\label{C4}
\sum_{j_f} d_{j_f} B_{4}^\g(N j_{\bar a},j_f; N j_{\bar a},j_f; N h, N k) B_{4}^\g(N j_{\bar a},j_f; N j_{\bar a},j_f; N h, N k') \sim N (N^{-5/2})^2 = N^{-4},
\ee
where the extremes of the summations are $$[{\rm max}\{N|j_4'-h|,N|j_4'-k|,N|j_4-k'|\}, \min\{ N(j_4'+h),N(j_4'+k),N(j_4+k') \}].$$
This function is hard to evaluate numerically, and we were able to compute it explicitly up to spins of order 10 only. At those value, a numerical fit gives a power law $N^{-3.2}$. This is in qualitative agreement with the estimate \Ref{C4}, since at the same values the inhomogeneous $B_4^\g$ only scales like $N^{-2.1}$ instead of $N^{-2.5}$. With this numerical support, we use \Ref{C4} to estimate the whole amplitude. The summations over the intertwiners are all cut down by the Gaussian peak in the boosters.
Putting \Ref{C4} together with the dimensional factors of the intertwiners and the homogeneous scalings of the boosters in the last line of the amplitude, and taking into account the fact that also the boundary intertwiners must be rescaled to get the dominant contribution, we get
for the simplified model
\be
\label{DLD5}
W^{DLD_5}_s (N j_a ,N j^\prime_a;Ni,Nt,Ni',Nt') \sim N^4 N N^{-4} (N^{-5/2})^4 = N^{-9}.
\ee
This is the same scaling estimated for the previous foam. At spins of order 10, a linear fit of the exact numerical data gives $N^{-8.1}$, same qualitative agreement as expected from \Ref{C4}. 

For the complete model, our results are more limited. The numerical evaluation is significantly harder because of the many $l$-summations, and out of reach of our current numerical power. 
This is unfortunate because unlike for the previous foam, the two scalings could differ this time. Indeed, the presence of unbounded summations of slow convergence (the booster functions now have a power-law decay in the $l$s, as opposed to the exponential one of the previous foam) means that the complete EPRL model could have a \emph{slower} decay. Consider in fact the following example: 
\be\label{pluto}
f(j) = \f1{j^x} \sum_{\D l=0}^{\infty} \f1{(j+\D l)^y} \sim \f1{j^{x+y-1}} \f1{y-1} 
\ee
where in the last step we approximated the sum with an integral. This suggests that an unbounded sum can slow by one power the large spin decay.
If this is the case, then this foam dominates and the overall expansion \Ref{sf} should be completed with an intermediate scaling, and spin correlations between the boundary dipoles would appear at a stronger power than $N^{-9}$. Since there are only two independent summations, the boosters decay and the example \Ref{pluto} suggest that the scaling would be at most two powers slower, that is $N^{-7}$.
\bigskip

The remaining two non-factorized amplitudes have a more complicated structure, with 5-stranded booster functions. Their expression is, respectively,
\be \nn
\begin{split}
 W^{DLD_4}(j_a,j^\prime_a;i,t,i',t') &=
 d_id_{i'}d_td_{t'}
\begin{array}{c}
\psfrag{a}{$j_1$}
\psfrag{b}{$j_2$}
\psfrag{c}{$j_3$}
\psfrag{d}{$j_4$}
\psfrag{e}{$j^\prime_1$}
\psfrag{f}{$j^\prime_2$}
\psfrag{g}{$j^\prime_3$}
\psfrag{h}{$j^\prime_4$}
\psfrag{j}{$j_f$}
\psfrag{s}{$i^\prime$}
\psfrag{t}{$t^\prime$}
\psfrag{u}{$i$}
\psfrag{v}{$t$}
\psfrag{1}{$g_1$}
\psfrag{2}{$g_2$}
\psfrag{3}{$g_3$}
\psfrag{4}{$g_4$}
\psfrag{5}{$g_5$}
\psfrag{6}{$g_6$}
\psfrag{l}{$I$}
\includegraphics[scale=1.8]{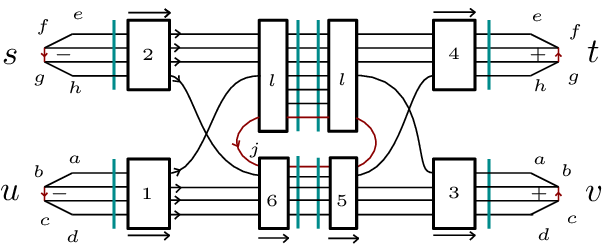}
\end{array}
 \\ & \hspace{-1cm}=
 d_i d_t d_{i^\prime} d_{t^\prime}
 \sum_{l_a,\bar l_a, l^{\prime}_a \atop k, k^{\prime},s,h,h^{\prime}} \frac{d_k d_{k^{\prime}}  d_s d_h d_h^\prime} {d_l d_{l^{\prime}}}  
\bigg(\sum_{j_f} d_{j_f}
\begin{array}{c}
\psfrag{a}{$j_4$}
\psfrag{b}{$j_3$}
\psfrag{c}{$j_2$}
\psfrag{d}{$j^{\prime}_4$}
\psfrag{e}{$j_f$}
\psfrag{f}{$l_4$}
\psfrag{g}{$l_3$}
\psfrag{h}{$l_2$}
\psfrag{i}{$l^\prime_4$}
\psfrag{l}{$j_f$}
\psfrag{u}{$j_1$}
\psfrag{v}{$k$}
\psfrag{s}{$h^\prime$}
\psfrag{t}{$h$}
\includegraphics[width=2.9cm]{_images/B5.eps}
\end{array}
\begin{array}{c}
\psfrag{a}{$j_4$}
\psfrag{b}{$j_3$}
\psfrag{c}{$j_2$}
\psfrag{d}{$j^{\prime}_4$}
\psfrag{e}{$j_f$}
\psfrag{f}{$l_4$}
\psfrag{g}{$\bar l_3$}
\psfrag{h}{$\bar l_2$}
\psfrag{i}{$l^\prime_4$}
\psfrag{l}{$j_f$}
\psfrag{u}{$j_1$}
\psfrag{v}{$s$}
\psfrag{s}{$h^\prime$}
\psfrag{t}{$h$}
\includegraphics[width=2.9cm]{_images/B5.eps}
\end{array}\bigg)
\\ & \hspace{2cm} \times
\begin{array}{c}
\psfrag{a}{$j_1$}
\psfrag{b}{$j_2$}
\psfrag{c}{$j_3$}
\psfrag{d}{$j_4$}
\psfrag{e}{$j_1$}
\psfrag{f}{$l_2$}
\psfrag{g}{$l_3$}
\psfrag{h}{$l_4$}
\psfrag{i}{$i$}
\psfrag{k}{$k$}
\includegraphics[width=2.2cm]{_images/B4.eps}
\end{array}
\begin{array}{c}
\psfrag{a}{$j_1$}
\psfrag{b}{$j_2$}
\psfrag{c}{$j_3$}
\psfrag{d}{$j_4$}
\psfrag{e}{$j_1$}
\psfrag{f}{$\bar l_2$}
\psfrag{g}{$\bar l_3$}
\psfrag{h}{$\bar l_4$}
\psfrag{i}{$t$}
\psfrag{k}{$k$}
\includegraphics[width=2.2cm]{_images/B4.eps}
\end{array}
\begin{array}{c}
\psfrag{a}{$j^\prime_1$}
\psfrag{b}{$j^\prime_2$}
\psfrag{c}{$j^\prime_3$}
\psfrag{d}{$j^\prime_4$}
\psfrag{e}{$j^\prime_1$}
\psfrag{f}{$j^\prime_2$}
\psfrag{g}{$j^\prime_3$}
\psfrag{h}{$l^\prime_4$}
\psfrag{i}{$i^\prime$}
\psfrag{k}{$k^\prime$}
\includegraphics[width=2.2cm]{_images/B4.eps}
\end{array}
\begin{array}{c}
\psfrag{a}{$j^\prime_1$}
\psfrag{b}{$j^\prime_2$}
\psfrag{c}{$j^\prime_3$}
\psfrag{d}{$j^\prime_4$}
\psfrag{e}{$j^\prime_1$}
\psfrag{f}{$j^\prime_2$}
\psfrag{g}{$j^\prime_3$}
\psfrag{h}{$l^\prime_4$}
\psfrag{i}{$t^\prime$}
\psfrag{k}{$k^\prime$}
\includegraphics[width=2.2cm]{_images/B4.eps}
\end{array}
\end{split}
\ee
and
\be \nn
\begin{split}
 W^{DLD_7}(j_a,j^\prime_a;i,t,i',t')  &=
 d_id_{i'}d_td_{t'}
\begin{array}{c}
\psfrag{a}{$j_1$}
\psfrag{b}{$j_2$}
\psfrag{c}{$j_3$}
\psfrag{d}{$j_4$}
\psfrag{e}{$j^\prime_1$}
\psfrag{f}{$j^\prime_2$}
\psfrag{g}{$j^\prime_3$}
\psfrag{h}{$j^\prime_4$}
\psfrag{j}{$j_f$}
\psfrag{s}{$i^\prime$}
\psfrag{t}{$t^\prime$}
\psfrag{u}{$i$}
\psfrag{v}{$t$}
\psfrag{1}{$g_1$}
\psfrag{2}{$g_2$}
\psfrag{3}{$g_3$}
\psfrag{4}{$g_4$}
\psfrag{5}{$g_5$}
\psfrag{6}{$g_6$}
\psfrag{l}{$I$}
\includegraphics[scale=1.8]{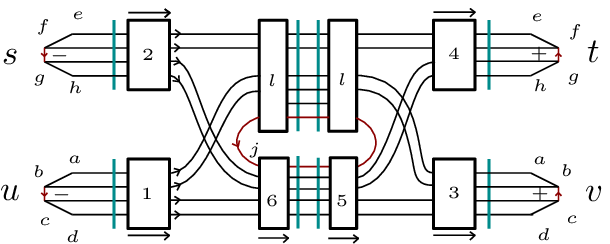}
\end{array}
 \\ & \hspace{-1cm}=
  d_i d_t d_{i^\prime} d_{t^\prime}
\sum_{l_a,\bar l_a, l^{\prime}_a, \bar l^{\prime}_a, \atop k, k^{\prime}} \frac{d_h d_h^\prime d_k d_{k^{\prime}}
} {d_{l_4} d_{l'_4}  } \bigg( \sum_{j_f} d_{j_f}
\begin{array}{c}
\psfrag{a}{$j_4$}
\psfrag{b}{$j_3$}
\psfrag{c}{$j^{\prime}_4$}
\psfrag{d}{$j^{\prime}_3$}
\psfrag{e}{$j_f$}
\psfrag{f}{$l_4$}
\psfrag{g}{$l_3$}
\psfrag{h}{$l^\prime_3$}
\psfrag{i}{$l^\prime_4$}
\psfrag{l}{$j_f$}
\psfrag{u}{$k$}
\psfrag{v}{$k^\prime$}
\psfrag{s}{$h^\prime$}
\psfrag{t}{$h$}
\includegraphics[width=2.9cm]{_images/B5.eps}
\end{array}
\begin{array}{c}
\psfrag{a}{$j_4$}
\psfrag{b}{$j_3$}
\psfrag{c}{$j^{\prime}_4$}
\psfrag{d}{$j^{\prime}_3$}
\psfrag{e}{$j_f$}
\psfrag{f}{$l_4$}
\psfrag{g}{$\bar l_3$}
\psfrag{h}{$ \bar l^\prime_3$}
\psfrag{i}{$l^\prime_4$}
\psfrag{l}{$j_f$}
\psfrag{u}{$k$}
\psfrag{v}{$k^\prime$}
\psfrag{s}{$h^\prime$}
\psfrag{t}{$h$}
\includegraphics[width=2.9cm]{_images/B5.eps}
\end{array}\bigg)
\\ & \hspace{1.5cm}\times
\begin{array}{c}
\psfrag{a}{$j_1$}
\psfrag{b}{$j_2$}
\psfrag{c}{$j_3$}
\psfrag{d}{$j_4$}
\psfrag{e}{$j_1$}
\psfrag{f}{$j_3$}
\psfrag{g}{$l_3$}
\psfrag{h}{$l_4$}
\psfrag{i}{$i$}
\psfrag{k}{$k$}
\includegraphics[width=2.2cm]{_images/B4.eps}
\end{array}
\begin{array}{c}
\psfrag{a}{$j_1$}
\psfrag{b}{$j_2$}
\psfrag{c}{$j_3$}
\psfrag{d}{$j_4$}
\psfrag{e}{$j_1$}
\psfrag{f}{$j_3$}
\psfrag{g}{$\bar l_3$}
\psfrag{h}{$\bar l_4$}
\psfrag{i}{$t$}
\psfrag{k}{$k$}
\includegraphics[width=2.2cm]{_images/B4.eps}
\end{array}
\begin{array}{c}
\psfrag{a}{$j^\prime_1$}
\psfrag{b}{$j^\prime_2$}
\psfrag{c}{$j^\prime_3$}
\psfrag{d}{$j^\prime_4$}
\psfrag{e}{$j^\prime_1$}
\psfrag{f}{$j^\prime_2$}
\psfrag{g}{$l'_3$}
\psfrag{h}{$l^\prime_4$}
\psfrag{i}{$i^\prime$}
\psfrag{k}{$k^\prime$}
\includegraphics[width=2.2cm]{_images/B4.eps}
\end{array}
\begin{array}{c}
\psfrag{a}{$j^\prime_1$}
\psfrag{b}{$j^\prime_2$}
\psfrag{c}{$j^\prime_3$}
\psfrag{d}{$j^\prime_4$}
\psfrag{e}{$j^\prime_1$}
\psfrag{f}{$j^\prime_2$}
\psfrag{g}{$\bar l^\prime_3$}
\psfrag{h}{$\bar l^\prime_4$}
\psfrag{i}{$t^\prime$}
\psfrag{k}{$k^\prime$}
\includegraphics[width=2.2cm]{_images/B4.eps}
\end{array}
\end{split}
\ee
The best we can do for these two foams is to give the large spin scaling of the simplified amplitude. Proceeding in the same way as before, suing this time the inhomogeneous scaling \Ref{B3InhomLim} of the 5-stranded booster, we obtain the following estimates:
\begin{align}
& W^{DLD_4}_s(N j_a ,N j^\prime_a;Ni,Nt,Ni',Nt')  \sim N^4 N^{3} N^{-6} (N^{-5/2})^4 = N^{-9},
\\ &
W_s^{DLD_7}(N j_a ,N j^\prime_a;Ni,Nt,Ni',Nt')   \sim N^4 N^{2} N^{-6} (N^{-5/2})^4  = N^{-10}.
\end{align}
The presence of unbounded summations means again that the amplitudes for the complete EPRL model could decay slower.

Finally, we have the three vertex graphs 1,2 and 3 of Fig. \ref{FigDLD}, corresponding to factorized $DLD$ amplitudes. Proceeding as above, we find that the amplitudes for the simplified model scale respectively as $N^{-6}$, $N^{-7}$, $N^{-11}$. As for the complete model, we did not invest numerical resources to study these diagrams, that give factorized amplitudes. Since the vertex graphs 1 and 2 are not 3-link-connected, we are not guaranteed that  the associated amplitudes are ill-defined. We present in Appendix~\ref{AppDVD} an example with a similar structure in which the amplitude is finite.
  
The results of our estimates and comparisons with the simplified model  are summarized in the following Table~\ref{Table}:
\begin{table}[H]
\begin{center}\begin{tabular}{ccccc}
{\bf Foam} & {\bf Factorization} & {\bf LOs scaling} & {\bf EPRLs=EPRL} & {\bf Numerics} \\ \hline 
$DVD$ & Y & $N^{-3}$ & Y & $\checkmark$ \\
$DED$ & Y & $N^{-6}$ & Y & $\checkmark$   \\
$DLD_1$ &Y & $N^{-6}$ & N & \\ 
$DLD_2$ &Y & $N^{-7}$ & N & \\ 
$DLD_3$ &Y & $N^{-11}$ & N &  \\
$DLD_4$ &N & $N^{-9}$ & N &  \\
$DLD_5$ &N & $N^{-9}$ & N &  \\
$DLD_6$ &N & $N^{-9}$ & N & $\checkmark$  \\
$DLD_7$ &N & $N^{-10}$ & N & 
\end{tabular}\end{center}
\caption{\label{Table} \small{\emph{Summary of scalings and properties of the foams. The column Factorization refers to whether the amplitude factorizes in two terms associated with each connected part of the boundary graph; LOs scaling gives the scaling of the simplified model; the next column marks the foams for which EPRL and EPRLs give the amplitude (so in particular the scaling is the same); Numerics lists those foams for which we have numerical evaluations. In the three cases considered, the numerics confirm the scaling of the simplified model. For the other foams, we notice in particular that $DLD_4$, $DLD_5$ and $DLD_7$ contain slowly converging unbounded summations over $l$s and could likely give slower fall offs. The summary of different scalings shows manifestly that already for the simplified EPRLs model the large spin scaling is not a simple function of $V,E$ and $F$.} }}
\end{table}

It goes without saying that we would have liked to perform numerical studies of (at least) the large spin scaling of all foams, especially for those like $DLD_4$, $DLD_5$ and $DLD_7$ that have non-factorized amplitudes and may have a slower power law decrease than the simplified model. This lays however beyond our current numerical means: Evaluating the booster functions is very slow and limited at spins of order 30, and the slow convergence of the sums over $l$ makes it impossible to deduce any meaningful behaviour from our data.

One thing that can be nonetheless fruitfully learned from the estimates for the EPRLs is that the power of the leading scaling is not a simple function of the number of vertices, edges and faces.
A priori, one has the following structure:
\begin{itemize} 
\item Number of vertices and their valence = Number and type of $SU(2)$ symbols;
\item Number of half-edges = Number of booster function;
\item Number of faces per half-edge = Valence of booster functions.
\end{itemize}
Using \Ref{Puchta} and the scalings of the SU(2) symbols one can use this structure to estimate the scaling of the amplitudes, at least for the simplified model. This works for simplicial foams, as explained in \cite{Boosting}.\footnote{The actual estimate there given does not take into account the removal of one integration per node and should accordingly be amended.} However for non-simplicial foams the situation is made more complicated by the large freedom in the routing of the faces and the valence of the edges. The routing in particular turns out to strongly influence the scaling, as shown in the Table above. As a particular example, the routing can lead to disconnected vertex graphs, which are always the dominant ones for given $V,E,F$: this is because gauge-invariance implies to remove one integral per connected component, and thus an additional booster power-law decay with it.
Only if the routing of the faces is rigidly fixed, one can hope to find easy rules to estimates the power-law decay without writing down the explicit form of the amplitude. For the complete EPRL model we have two possibilities: either the same scaling of the simplified model, or a slower scaling. The first situation occurs for simple foams with no unbounded $l$ summations, like $DVD$ and $DED$ above, or with fast convergent unbounded summations, like $DLD_6$ and the additional examples in Appendix \ref{AppDVD}. The second situation is however more general, and to estimate the difference one can look at how many independent unbounded $l$-summations are present, see discussion around \Ref{pluto} for an example. The slower decays impact also the analysis of the amplitude divergences associated to internal bubbles, like for the self-energy spin foam or the Pachner 1-5 move, see \cite{PietroInProgress}. The actual details on the behaviour of the complete EPRL model depend on the structure of the booster functions for non-minimal spins, and an improved analytic understanding of their large spin behaviour is certainly necessary if one wants to make progress in the evaluation of Lorentzian EPRL amplitudes. See \cite{pierreinprogress} for work in this direction.

\subsection{Higher vertices and Regge asymptotics}
\begin{figure}[H] 
\begin{center}
\begin{minipage}{0.4 \textwidth}
\centering
\vspace{0cm}
\includegraphics[width=1.8cm]{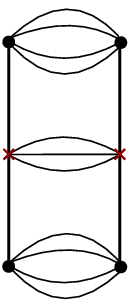}
\end{minipage}
\hspace{-1.5cm} 
\begin{minipage}{0.4 \textwidth}
\centering
\includegraphics[width=2.5cm]{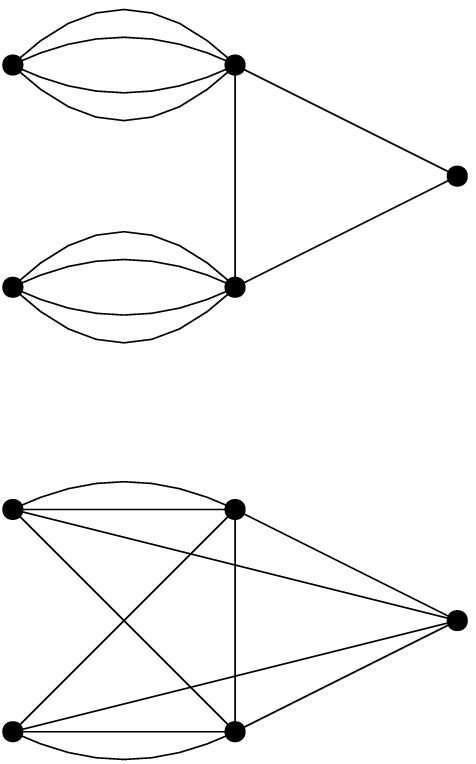}
\end{minipage}
\end{center}
\caption{\small{ \emph{An example of 2-complex with two pentavalent vertices, on the left, and two of its possible vertex graphs, on the right. The requirement of avoiding faces corresponding to non-minimal cycles excludes vertex graphs with the combinatorics of a $\{15j\}$ symbol.}}}
\label{v5} 
\end{figure}

The next foams in the vertex expansion have two vertices with three internal edges and, according to our criterion, three internal faces, see Fig.~\ref{v5}. The vertices are now 5-valent: one may wonder whether $\{15j\}$ symbols appear in the amplitudes, and allow us to make some contact with simplicial spin foams and their geometric interpretation. The answer is in the negative, due to the combinatorial structure of the faces: a  4-simplex vertex graph would require one of the faces to have two external links (one upper and one lower). But for this specific foam that would mean a non-minimal cycle, thus violating our face criterion.
$15j$ symbols will certainly appear with a high enough number of vertices, however it is hard to anticipate without a detailed analysis how much of their potential Regge-like behaviour will survive in the presence of so many non-simplicial symbols around them.

Although lacking a relation to Regge actions, the simple foams considered so far carry nonetheless non-trivial correlations, and we will look at some of them in the next Section.

\section{Spin Correlations} 
\label{spinCorr}
\begin{figure}[H] 
\begin{center}
\includegraphics[width=6.5cm]{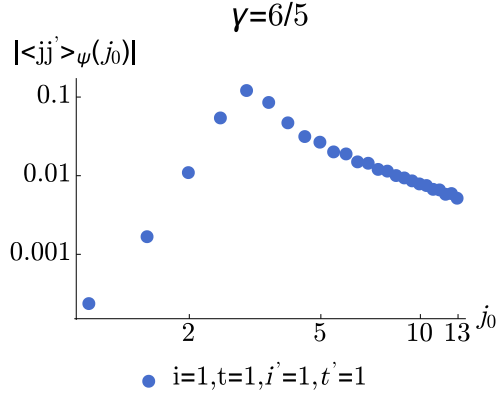}
\end{center}
\caption{\small{ \emph{ Correlations \Ref{corrCoer} as functions of the background spin $j_0$ introduced by the boundary state. The value of the peak and of the power-law tail depend on $\g$, here taken to be 6/5. The intertwiners labels are all equal to 1.} }
\label{Correlations2} }
\end{figure}

One of the most important properties of spin foam amplitudes is to introduce dynamical correlations among spin network states. These correlations have played an important role in the study of the 2-point function for simplicial foams \cite{Bianchi:2006uf}, and the relation with the graviton propagator in the linearized expansion (see e.g. \cite{Dittrich:2007wm}).\footnote{For recent alternative ideas based on entanglement to select physically correlated  states, see e.g. \cite{Perez:2014ura,Bianchi:2016hmk,Feller:2017ejs,Chirco:2017xjb}.}
For non-simplicial foams, we do not have a geometric interpretation of the correlations, but it is nonetheless important to study their structure if such amplitudes have to play a dynamical role in the theory. 
We focus here on the most interesting correlations encountered in our dipole-to-dipole transitions, those between spins associated with disconnected parts of the boundary graph. 
We have seen in the previous Section that such correlations appear only in the presence of an internal face and a routing mixing the boundary faces. 
The simplest foam for which this happens is  $DLD_6$, and in this Section we present a more detailed analysis of the spin correlations. Our analysis will however be only indicative in nature, for the following reason: we do not have sufficient computer power and time to evaluate the twelve sums over boundary spins and intertwiners. We resorted to a strong simplification provided by a `mini-superspace' model: we assume all lower and all upper spins to be equal, $j_a\equiv j$, $j_a'\equiv j'$, and study the amplitudes as probability distributions $W_{\rm mini}^{DLD_6}(j,j')$ in this two-dimensional space, at fixed $\g$ and intertwiners. 
This enormously reduces the contributions we need to sum over and makes numerical calculations an easy task. We hope that it still captures, at least qualitatively, the nature of the spin correlations of the true model. 
Using a Gaussian boundary state $\ket{\Psi} = \sum_{j,j'} \exp \{-(j-j_0)^2-(j'-j_0)^2\}\ket{j,j'}$ as in the general boundary framework, we evaluate
\be
\label{corrCoer}
\mean{j j'}_{\g,j_0} := \f{ \bra{W_{\rm mini}^{DLD_6}}\hat{J}_{l} \hat{J}_{l'} \ket{\Psi} } {\bra{W_{\rm mini}^{DLD_6}}\Psi\ra} -  \f{ \bra{W_{\rm mini}^{DLD_6}}\hat{J}_{l} \ket{\Psi} } {\bra{W_{\rm mini}^{DLD_6}}\Psi\ra} \f{ \bra{W_{\rm mini}^{DLD_6}}\hat{J}_{l'} \ket{\Psi} } {\bra{W_{\rm mini}^{DLD_6}}\Psi\ra}.
\ee
The results are reported in Fig.~\ref{Correlations2}.
We see a peak and a power-law large $j_0$ decay. These two features are qualitatively reminiscent of the simplicial spin correlations (see e.g. \cite{IoDan}), although the details differ, notably the power law and the width of the Gaussian.

\begin{figure}[h] 
\begin{center}
\includegraphics[width=6.5cm]{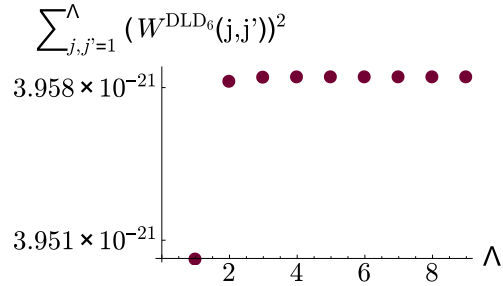}
\end{center}
\caption{\small{\emph{Normalizability of the minisuperspace amplitude. The plot shows the convergence of \Ref{converg} as a cut-off $\L$ on the summations is increased. Here we fixed all intertwiner labels to 1, and $\g=1.2$.}}}
\label{ConvSum} 
\end{figure}

\begin{figure}[H] 
\begin{center}
\begin{minipage}{0.4 \textwidth}
\centering
\vspace{1cm}
\includegraphics[width=9.5cm]{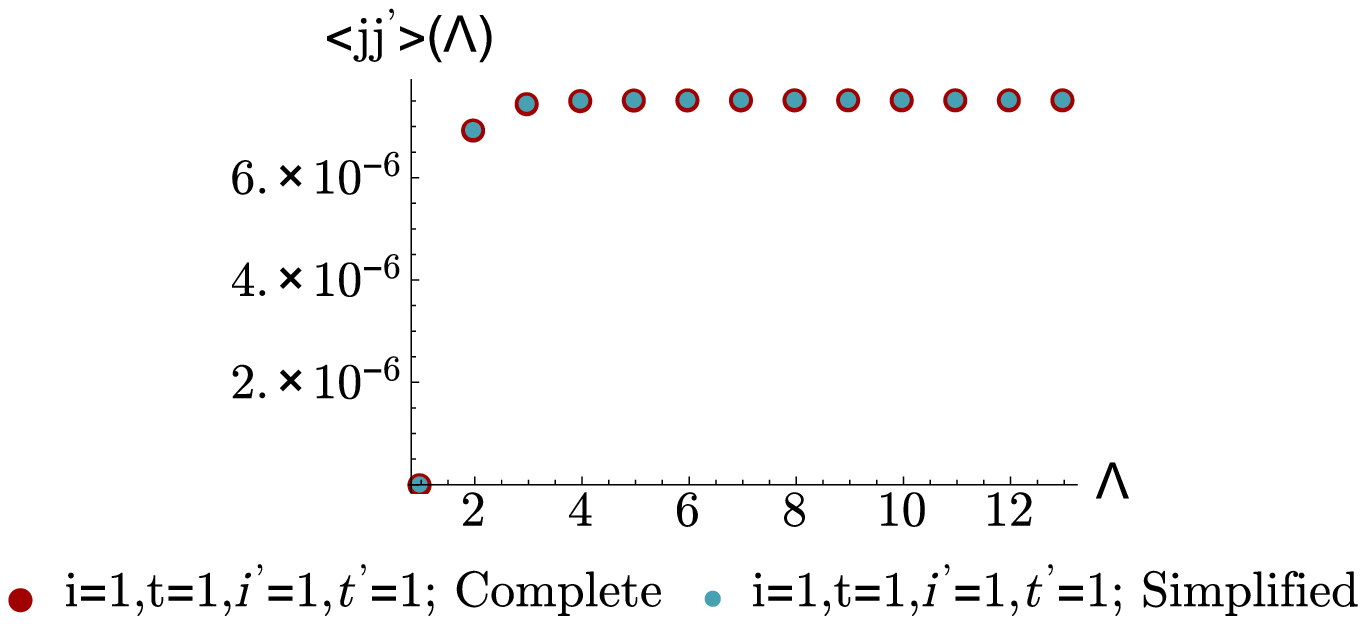}
\end{minipage}
\hspace{2cm} 
\begin{minipage}{0.4 \textwidth}
\centering
\includegraphics[width=6cm]{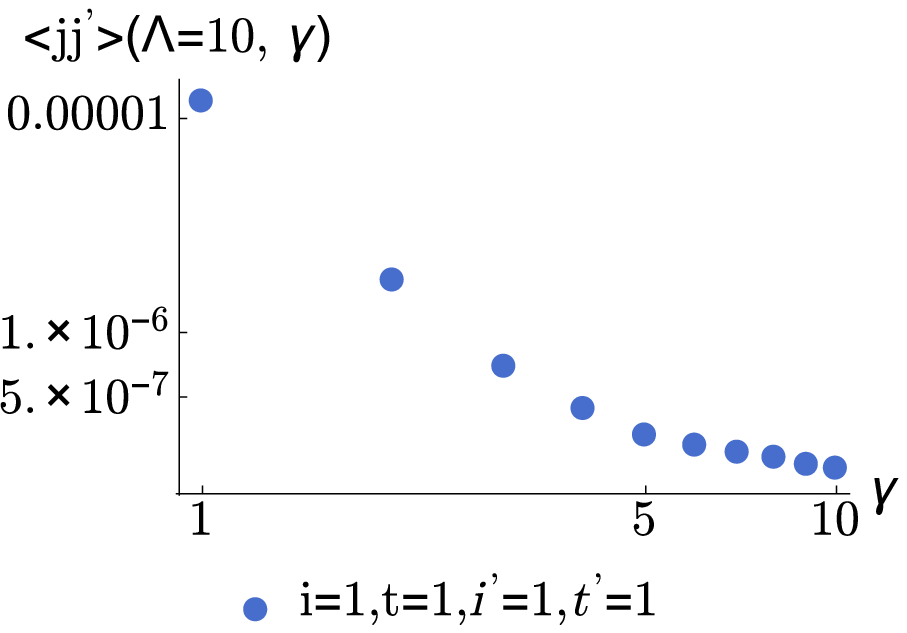}
\end{minipage}
\end{center}
\caption{\small{Left panel: \emph{Spin correlation \Ref{Corr1} for the complete and the simplified models, showing that convergence of the sum is achieved at small values of the cut-off $\Lambda$. The models give indistinguishable data points, the relative error being of order $10^{-4}$. 
The same fast convergence is seen for all values of $\g$ considered.
} 
Right panel: \emph{Spin correlation \Ref{Corr1} as a function of the Immirzi parameter, using the simplified model and cut-off $\Lambda=10$. 
} }
\label{Correlations} }
\end{figure}

An important consequence of the minisuperspace model is that the amplitude is normalizable:
\be
\bra{W_{\rm mini}^{DLD_6}}W_{\rm mini}^{DLD_6}\ra = \sum_{j,j'} W_{\rm mini}^{DLD_6}(j,j')^2 < \infty.
\label{converg}
\ee
Furthermore, convergence is achieved almost immediately, see Fig.~\ref{ConvSum}. 
Hence, we can define the correlations for this mini-superspace model also as in \Ref{PEV}, 
\be
\mean{j j'}_{DLD_6} :=\f{\sum_{j,j'}^\Lambda j j' W_{\rm mini}^{DLD_6}(j,j')^2}{\sum_{j,j'}^\Lambda W_{\rm mini}^{DLD_6}(j,j')^2 } 
-\f{\sum_{j,j'}^\Lambda j W_{\rm mini}^{DLD_6}(j,j')^2}{\sum_{j,j'}^\Lambda W_{\rm mini}^{DLD_6}(j,j')^2 } \f{ \sum_{j,j'}^\Lambda j' W_{\rm mini}^{DLD_6}(j,j')^2}{\sum_{j,j'}^\Lambda W_{\rm mini}^{DLD_6}(j,j')^2 },
\label{Corr1}
\ee
where, since the sums over $j$ and $j'$ are infinite, we put an homogeneous cut-off $\Lambda$ on them. The convergence in $\Lambda$ is very fast, see left panel of Fig.~\ref{Correlations}, as it was for the normalization alone. The same plot shows that the EPRL and simplified model give basically the same correlations: the difference is of order $10^{-4}$. Hence for this diagram the simplified model correctly captures both the scaling and the spin correlations.

Finally, we studied the dependence of \Ref{Corr1} on the Immirzi parameter $\g$, see left panel of Fig.~\ref{Correlations}. The correlations are positive, and decrease as $\g$ increases. 

\begin{figure}[H] 
\begin{center}
\includegraphics[width=8cm]{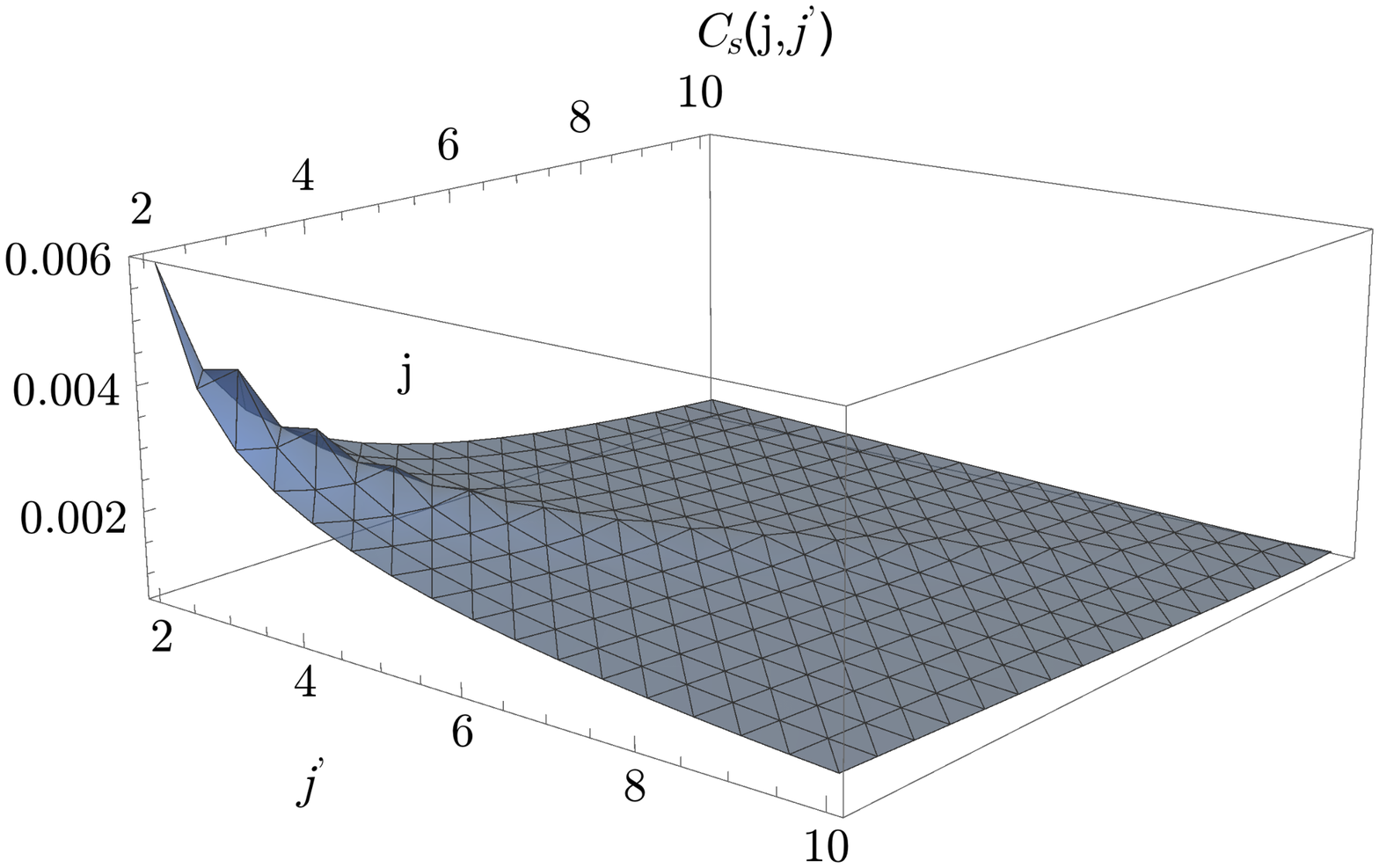} \\
\includegraphics[width=6cm]{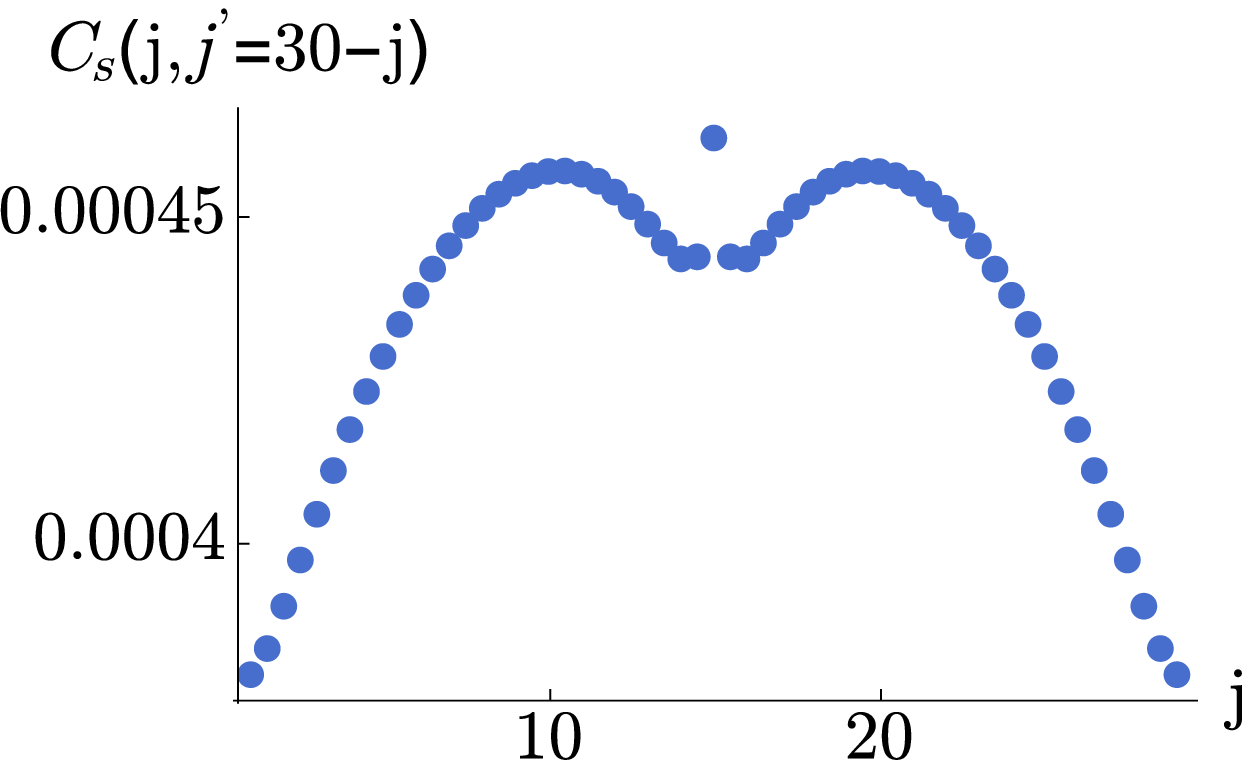} \hspace{1cm} 
\includegraphics[width=6cm]{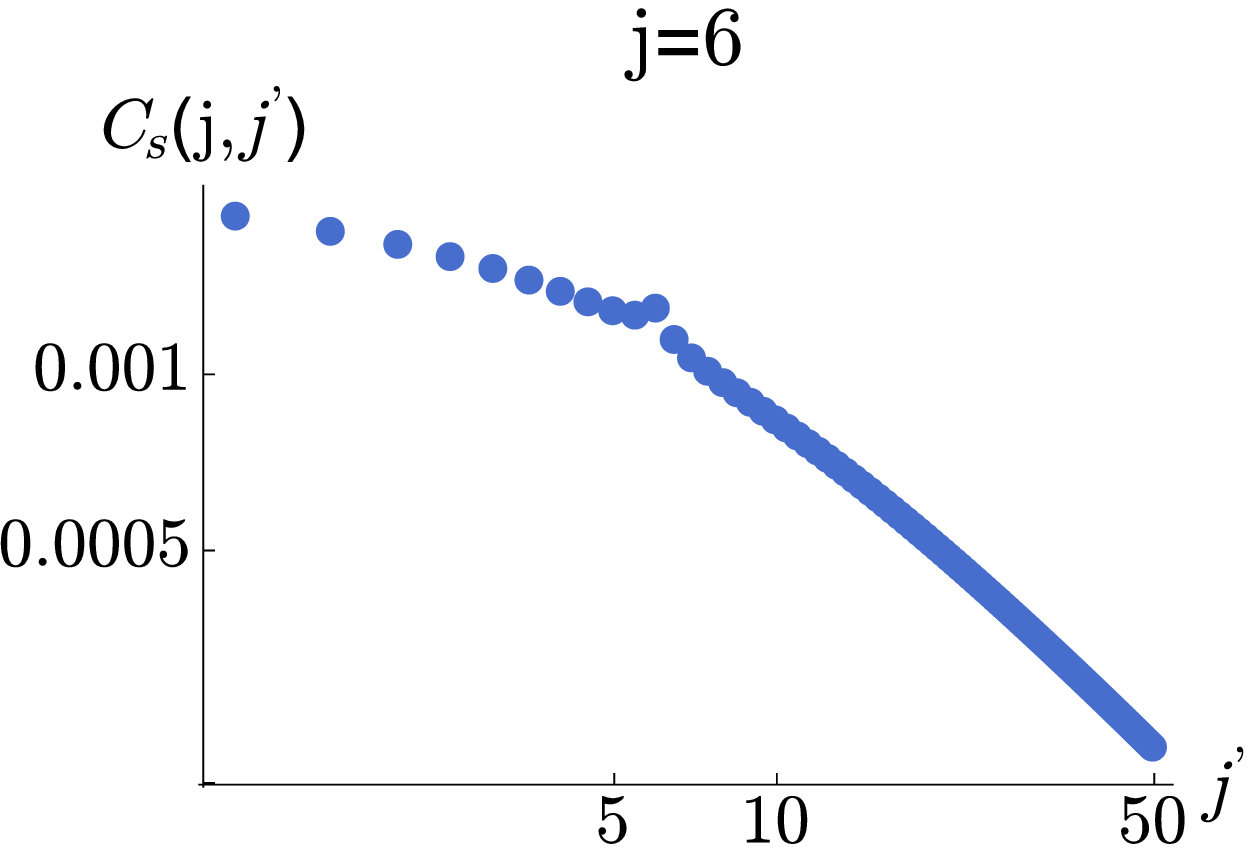}
\end{center}
\caption{\small{\emph{Numerical studies of the internal face correlation function, for the simplified model and $\g=6/5$.} Upper panel: \emph{3d plot of $C_s(j,j')$, showing a power-law decay superimposed with a peak for equal spins and symmetric sub-leading peaks.}
Lower panels: \emph{2d sections of $C_s(j,j')$, showing more details about the peaks. The peak in the fixed-$j$ cut may amusingly remind the reader of a resonance in particles scattering, like the famous Higgs-diphoton bump, however the background typically has opposite concavity. }}
\label{FigCs} }
\end{figure}

Let us now look at the origin of the correlations, namely the internal face function \Ref{C3def}
responsible for the coupling of upper and lower spins.
The main contribution is the one for the simplified model, \Ref{Cs}. As a distribution in the spins this function has an interesting behaviour, with a principal peak for equal spins and symmetric sub-leading peaks, see Fig.~\ref{FigCs}. 

\begin{figure}[H] 
\begin{center}
\includegraphics[width=7.5cm]{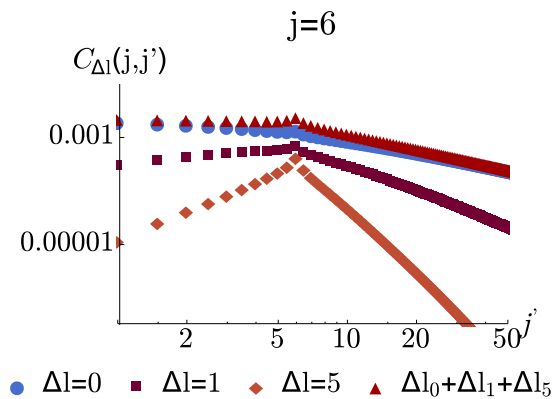}
\end{center}
\caption{\small{\emph{Peak enhancement for the cut-offed sums \Ref{giorgio}.}}
\label{FigC} }
\end{figure}

Adding the $l$ contributions of the complete EPRL model gives stronger correlations, although by a very small amount. For the internal face function, we can look at a cut-off summation,
\begin{align}\label{giorgio}
C_{\Delta l}(j,j') = \sum_{l=j}^{j+\Delta l} \sum_{l'=j'}^{j+\Delta l} C_{l,l'}(j,j'). 
\end{align}
Each new term added enhances the peak, as shown in Fig.~\ref{FigC}. However, the $l,l'$ to be summed over in \Ref{DLD6} are convoluted with booster functions, that has seen previously decrease as we increase the $l$ labels. This has the effect of damping the peak enhancement, and as anticipated above, the final correlations only differ by an order $10^{-4}$.

The resummed $C_{\Delta l}$ only gives a qualitative picture of the effect in the complete model, in particular it overestimate it since the $B_4$ convoluted over are also decaying functions in $\D l$. Indeed, the correct result can be seen by looking at the complete amplitude as a function of final (say all equal) spins at fixed initial (all equal) spins. That is the reason why, also for the correlation of the present foam and not only for the scaling behaviour, the simplified model offers a good approximation.

As a final comment, let us point out that the simplicity constraints play an important role in generating the correlations. Had 
Had we worked with SU(2) BF theory alone, in fact, the internal face function would simply be
\be
C_{BF} (j,j')= \sum_{j_f=|j-j'|}^{j+j'} d_{j_f}.
\ee
There is no more coupling of the spins in the summand, and the only almost-trivial correlations come from the extremes of the sum.
Hence these correlation are a good test of the dynamics of the EPRL model, within the limit of the simple foams and mini-superspace setup here considered.

\section{Conclusions}
Numerical calculations of spin foam amplitudes are notoriously difficult, especially for Lorentzian signature. In this paper we have considered amplitudes for a boundary graph with two 4-link dipoles, originally considered in \cite{Vidotto,VidottoLor} for cosmological applications. We have extended that analysis to foams with two vertices, and showed how the amplitudes can be computed explicitly using the method introduced in \cite{Boosting}. Our results show that the decoupling of the boundary dipoles present at one vertex \cite{Vidotto} is still present at two vertices with a single connecting edge. In order to couple the spins on the two dipole one needs an internal face. This is however not sufficient, as some foams with an internal face also can give factorized amplitudes, depending on the routing of the faces. For the simplest foam with a non-factorized amplitude we performed numerical studies of its scaling and spin correlations. The amplitude is suppressed by six powers in the large spin limit with respect to that with a single vertex, and the correlations between lower and upper spins, studied with a mini-superspace model, are positive and monotonically decreasing with $\g$. 
The origin of the correlations can be isolated in a specific function associated to the internal face, which features a non-trivial structure with multiple peaks. This study of correlations can be considered as a first step to understand the dynamics induced by the spin foam for the two dipoles, and it is relevant for the development of spin foam cosmology. Indeed, one of our main motivations to undertake the present analysis was to take a look at cosmological applications using spin foams \cite{Rovelli:2008aa,Vidotto,Borja:2011ha,Bianchi:2011ym,VidottoLor,Kisielowski:2012yv,Rennert:2013pfa,Bahr:2017eyi}, with the specific question of seeking evidence of the famous Big Bounce predicted by loop quantum cosmology \cite{Ashtekar:2006wn}. We are not there yet with our approach, partly because the path proved rich and complicated, partly because our current numerical power is significantly limited. We thus content ourselves for the time being with presenting the results obtained so far, and postpone a study of cosmological applications to future work.

For all foams considered, the spin sum over the internal face is convergent, and the amplitudes have a power law decay in the large spin limit. The power increases with the complexity of the foam, thus introducing a hierarchical scheme on top of the vertex coupling expansion. However, the exact scaling depends in a complicated way on the combinatorics of the foam, and cannot be easily established just looking at the number of vertices, edges and faces of the 2-complex. For instance a routing of faces leading to a disconnected vertex graphs always gives a dominant amplitude with respect to a connected routing, because of the extra group integrations to be removed.
These results leave us with a mixed feeling. Exposing the large spin hierarchy of the foams allows one to play around with expansions like \Ref{sf} as an arena to test different ideas on how the sum should be organized, or a continuum limit taken, and how different questions may be sensitive to different orders. On the other hand, the (relative) simplicity of the numerical evaluation of these non-simplicial spin foams has a payoff in the loose combinatorics of the faces. Even introducing a very strong criterion as we did, the proliferation is still important, and most unsatisfactorily, the scaling power law can not be a priori determined simply looking at the number of vertices, edges and faces, not even for the simplified EPRLs model.

In the course of presenting these results we also discussed detailed aspects of the evaluations, the guiding role that the simplified model can play, and what are our current numerical limits.
In particular, we were able to numerically evaluate only one -- and the simplest -- non-factorized foam for the complete EPRL model. 
Both types of results, on the scalings and the correlations, concern the very limited setting \Ref{sf} with at most two vertices and a single internal face, and it is necessary to push our analysis further before more meaningful conclusions can be drawn. 
The numerical evaluations are very costly, and we are furthermore limited by instabilities of hypergeometric functions in Mathematica to compute the booster functions at spins beyond 30. 
Our analysis tells us that we need to improve these instabilities and parallelize the code so to speed up the $l$ summations. Work is currently in progress in this direction, and these improvements will be crucial to testing the Regge asymptotic of the simplicial amplitude \cite{noiLor}.
Once this is achieved, the factorization method of \cite{Boosting} put the Lorentzian amplitudes on the same tractability level as SU(2) amplitudes. This is non-trivial progress, but far from saying that the hardest part of the work is done: it is actually at that point that one can face the true hard problem, namely the summations over internal $j$ spins for many-vertex foams. 
So from the numerical perspective our results are both good and bad: good because we were able to do explicit numerics at two vertices, bad because they showed how much more improvement is needed, in methods and numerical codes and/or approximation schemes.

Let us conclude with a comment on the dynamical meaning of the amplitudes here considered. 
Most of the literature on spin foams focuses on simplicial amplitudes, for their natural geometric interpretation and relation to Regge calculus. Nonetheless, non-simplicial amplitudes have been put forward as a mean to provide transition amplitudes to spin networks on arbitrary graphs \cite{KKL}. 
One can go out of the simplicial setting in two opposite directions: on the one,  considering simpler foams, with lower-valence vertices and simple boundary states. This is the set up proposed in  \cite{Vidotto,VidottoLor} for cosmological applications and considered here. On the other one, considering higher-valence vertices, which could be interesting from a coarse-graining perspective for instance, and has also been considered for cosmological applications \cite{Bahr:2017eyi}.
In the first case, the connection with Regge calculus and that geometric interpretation of the dynamics is lost, as we have seen in this paper no Regge actions enter the dynamics of the leading foams in the expansion. In the second case, the analysis of \cite{IoSU2asympt} (which is restricted to SU(2) theory, but already points out what the story is like for the EPRL model, see also \cite{BahrSteinhaus15} on this) shows that while it is possible in principle to get a Regge-like dynamics based on flat 4d polytopes instead of 4-simplices, the EPRL model as it is now allows for more general non-flat polytope configurations at the saddle point, corresponding to 3d data describing a class of conformal twisted geometries in which areas and angles match, but not the diagonal of the polygonal faces. 

The lack of a simple connection with Regge calculus, and of a simple rule to identify the large spin scaling of each diagram, 
are for us indications that more work is needed to assess the viability of non-simplicial EPRL amplitudes for quantum gravity.

\subsection*{Acknowledgements}
We are grateful to Pietro Don\`a and Marco Fanizza for many helpful discussions about numerical aspects and to Marcin Kisielowski and Thomas Krajewski about generalised spin foams, and to Francesca Vidotto for initial motivations and a reading of the manuscript. Simone thanks Daniele Oriti for discussions on spin foam expectation values.
\appendix

\setcounter{equation}{0}
\renewcommand{\theequation}{\Alph{section}.\arabic{equation}}

\section{$SU(2)$ and $SL(2,\C)$ graphical calculus \label{AppSU2}}
For the graphical calculus used to explicitly compute the amplitudes, we refer the reader to the monography \cite{Varshalovich}, whose conventions we use here. 

\subsection{SU(2) Symbols}
We represent Wigner's $3jm$ symbol as
\begin{equation}
\Wthree{j_1}{j_2}{j_3}{m_1}{m_2}{m_3} = \begin{array}{c}
\psfrag{a}{$j_{1}$}\psfrag{b}{$j_{2}$}\psfrag{c}{$j_{3}$}\psfrag{d}{$-$}
\includegraphics[width=1.5cm]{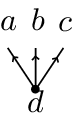}
\end{array}
=\begin{array}{c}
\psfrag{a}{$j_{3}$}\psfrag{b}{$j_{2}$}\psfrag{c}{$j_{1}$}\psfrag{d}{$+$}
\includegraphics[width=1.5cm]{_images/3JMPure.eps}
\end{array},
\end{equation} 
where the signs $\pm$ on the nodes keep track of the assignment of spins in the symbol, respectively anticlockwise/clockwise.

The contraction of two $3jm$ symbols via an intertwiner gives
\begin{equation}
\begin{split}
\begin{array}{c}\psfrag{a}{$j_{1}$}
\psfrag{b}{$j_{2}$}
\psfrag{c}{$j_{3}$}
\psfrag{d}{$j_{4}$}
\psfrag{i}{$i$}
\psfrag{e}{$-$}
\psfrag{f}{$-$}
\includegraphics[width=1.8cm]{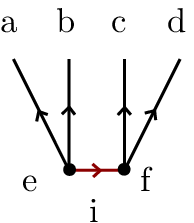}
\end{array}=\begin{array}{c}\psfrag{a}{$j_{1}$}
\psfrag{b}{$j_{2}$}
\psfrag{c}{$j_{3}$}
\psfrag{d}{$j_{4}$}
\psfrag{i}{$i$}
\psfrag{e}{$-$}
\psfrag{f}{$-$}
\includegraphics[width=1.8cm]{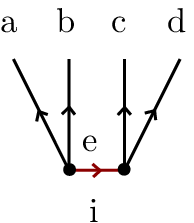}
\end{array}
 &= \sum_{m_{i}} (-1)^{i-m_{i}}  \Wthree{j_1}{j_2}{i}{m_1}{m_2}{m_{i}} \Wthree{i}{j_3}{j_4}{-m_{i}}{m_3}{m_4} \\ &= :
\Wfour{j_1}{j_2}{j_3}{j_4}{m_1}{m_2}{m_3}{m_4}{i}
\label{CG4}
\end{split}
\end{equation} 
which we used to define a $4jm$ symbol with four external legs. From now on we forget to use dots for nodes and we discard signes assuming they are always minus if not otherwise stated. These symbols satisfy orthogonality properties such as
\be
\sum_{m_1,m_2} \Wthree{j_1}{j_2}{j_3}{m_1}{m_2}{m_3} \Wthree{j_1}{j_2}{l_3}{m_1}{m_2}{n_3} = \f{\d_{j_{3}l_{3}}\d_{m_3n_3}}{d_{j_3}},  \label{orto3j}
\ee
\be
 \sum_{m_a} \Wfour{j_1}{j_2}{j_3}{j_4}{m_1}{m_2}{m_3}{m_4}{j_{12}} \Wfour{j_1}{j_2}{j_3}{l_4}{m_1}{m_2}{m_3}{n_4}{l_{12}} = 
 \f{\d_{j_{12}l_{12}}}{d_{j_{12}}} \f{\d_{j_{4}l_{4}} \d_{m_4n_4}}{d_{j_{4}}}.\label{orto4j}
\ee
Unlike the $3jm$ symbol, the $4jm$ symbol we defined is not normalized.\footnote{A normalised $4jm$-symbol is obtained multiplying the right-hand side of \Ref{CG4} by $\sqrt{d_{j_{12}}}$. We chose the non-normalised convention because it is the one that corresponds to a 4-valent node in the SU(2) graphical calculus, and because it is more convenient to work with in order to reconstruct the invariants $\{nj\}$-symbols associated to graphs.}

In the graphical notation of the $4jm$ symbol we distinguish one spin, $j_{12}$, which corresponds to an intertwiner label in the recoupling channel $12$.
Graphically, tracing over the free magnetic indices the orthogonality relations \Ref{orto3j} and \Ref{orto4j} we obtain the evaluation of the $\th$ graph and its generalized version with 4 links, 
\begin{equation}
\begin{array}{c}\psfrag{a}{$j_{1}$}
\psfrag{b}{$j_{2}$}
\psfrag{c}{$j_{3}$}
\includegraphics[width=1.8cm]{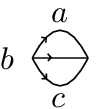}
\end{array}=1;
\hspace{1cm}
\begin{array}{c}\psfrag{a}{$j_{1}$}
\psfrag{b}{$j_{2}$}
\psfrag{c}{$j_{3}$}
\psfrag{d}{$j_{4}$}
\psfrag{i}{$i$}
\psfrag{k}{$k$}
\includegraphics[width=2.8cm]{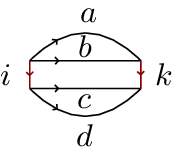}
\end{array}=\frac{\d_{i,k}}{d_i}.
\label{thetaeq}
\end{equation} 
Above and everywhere we implicitly assumed that the Clebsch-Gordan inequalities are satisfied, else the evaluations vanish.

For the simple foams considered in this paper, the rule that we needed the most is the orthogonality relation (\ref{orto3j}), namely
\be 
\begin{array}{c}
\psfrag{a}{$j_1$}
\psfrag{b}{$j_2$}
\psfrag{c}{$j_3$}
\psfrag{d}{$j_4$}
\psfrag{e}{$d_{j_1}^{-1}$}
\psfrag{=}{$=$}
\includegraphics[width=7.5cm]{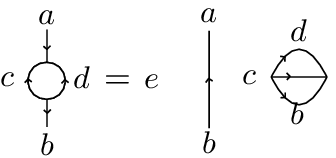}
\end{array}\label{Rule}
\ee
Other two useful rules are the node sign change, equation (6) in Section 8.5 of \cite{Varshalovich}, and the $6j$ graph identity at the bottom of page 429. 

Inverting the orientation of an internal line with spin $j$ gives a $(-1)^{2j}$ phase. Inverting the orientation of an external line gives a parity map $\eps=i\s_2.$ Depending on whether we are acting on a ket or a bra we get 
\begin{equation}
\begin{split}
&D^{(j)}_{mn} (i \sigma_2) =\epsilon^{(j)}_{mn} =
\begin{array}{c}
\includegraphics[width=1.8cm]{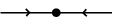}
\end{array}
=
(-1)^{j+m}\delta_{m,-n} \\ \\
&{{\epsilon}^{-1}}^{(j)mn}=\epsilon^{(j) nm}
\begin{array}{c}
\includegraphics[width=1.8cm]{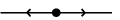}
\end{array}
=
(-1)^{j-m}\delta_{m,-n}
\end{split}
\end{equation} 
So for instance
\begin{equation}
\begin{array}{c}\psfrag{a}{$j_{1}$}
\psfrag{b}{$j_{2}$}
\psfrag{c}{$j_{3}$}
\psfrag{d}{$j_{4}$}
\psfrag{i}{$i$}
\psfrag{e}{$-$}
\includegraphics[width=1.5cm]{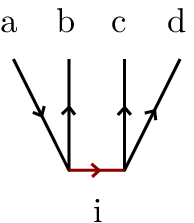}
\end{array}=(-1)^{j_1-m_1}\Wfour{j_1}{j_2}{j_3}{j_4}{-m_1}{m_2}{m_3}{m_4}{i}\,.
\end{equation}

Since $\sum_l j_l\in\N$ at a node, reversing all external lines has no effect:
\begin{equation}
\begin{array}{c}
\psfrag{a}{$j_{1}$}
\psfrag{b}{$j_{2}$}
\psfrag{c}{$j_{3}$}
\psfrag{d}{}
\includegraphics[width=1.5cm]{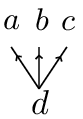}
\end{array}
=\begin{array}{c}
\psfrag{a}{$j_{1}$}
\psfrag{b}{$j_{2}$}
\psfrag{c}{$j_{3}$}
\psfrag{d}{}
\includegraphics[width=1.5cm]{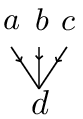}
\end{array}\,,\qquad
\begin{array}{c}
\psfrag{a}{$j_{1}$}
\psfrag{b}{$j_{2}$}
\psfrag{c}{$j_{3}$}
\psfrag{d}{$j_{4}$}
\psfrag{i}{$i$}
\psfrag{e}{}
\includegraphics[width=1.8cm]{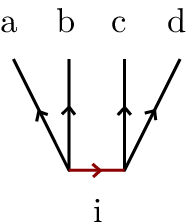}
\end{array}=
\begin{array}{c}
\psfrag{a}{$j_{1}$}
\psfrag{b}{$j_{2}$}
\psfrag{c}{$j_{3}$}
\psfrag{d}{$j_{4}$}
\psfrag{i}{$i$}
\psfrag{e}{}
\includegraphics[width=1.8cm]{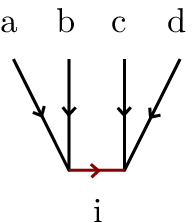}
\end{array}\,.\label{keep}
\end{equation} 
Changing the orientation of the virtual line gives a phase in agreement with the rule for internal lines:
\begin{equation}
\begin{split}
\begin{array}{c}
\psfrag{a}{$j_{1}$}
\psfrag{b}{$j_{2}$}
\psfrag{c}{$j_{3}$}
\psfrag{d}{$j$}
\psfrag{i}{$i$}
\psfrag{e}{$+$}
\psfrag{s}{$-$}
\includegraphics[width=1.8 cm]{_images/4JM1out.eps}
\end{array}  
&=\sum_{m_{i}}
\Wthree{j_1}{j_2}{i}{m_1}{m_2}{m_{i}}(-1)^{i-m_{i}}\Wthree{i}{j_3}{j}{-m_{i}}{m_3}{m}
\\&=\sum_{m_{i}}
\Wthree{j_1}{j_2}{j_{12}}{m_1}{m_2}{-m_{i}}(-1)^{i+m_{i}}\Wthree{i}{j_3}{j}{m_{i}}{m_3}{m}\\ 
&=\sum_{m_{i}}\,(-1)^{2(i-m_{i})}\,\Wthree{j_1}{j_2}{i}{m_1}{m_2}{-m_{i}}(-1)^{i+m_{i}}\Wthree{i}{j_3}{j}{m_{i}}{m_3}{m} \\
& =(-1)^{2i}
\begin{array}{c}
\psfrag{a}{$j_{1}$}
\psfrag{b}{$j_{2}$}
\psfrag{c}{$j_{3}$}
\psfrag{d}{$j$}
\psfrag{i}{$i$}
\psfrag{e}{$+$}
\psfrag{s}{$-$}
\includegraphics[width=1.8cm]{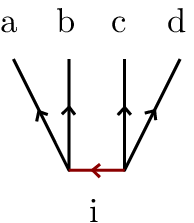}
\end{array}.
\end{split}
\nonumber
\end{equation} 

\subsection{Boxes and Integrations}
We refer the reader to \cite{BarrettLorAsymp} for a more detailed description of graphical calculus for $\SL(2,\C)$. Here we just need two elements to complement the SU(2) calculus. The $\SL(2,\C)$ integration boxes, and the booster functions. We represent every half-edge of a foam with a box, every face with a strand and the $Y$ map with a blue line orthogonal to the strands. If e.g. four faces touch an half edge the corresponding box will have four strands. The $Y$ map projects the `magnetic spins' to their minimal value, which we always denote with a $j$ in this paper. `Magnetic spins' allowed to vary freely above their minimal value are always labeled with a $l$. Hence for a 4-stranded half-edge,
\be
\begin{array}{c}\label{A10}
\psfrag{a}{$g$}
\includegraphics[width=2cm]{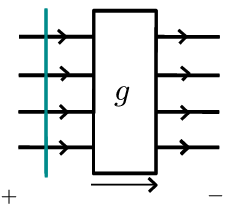}
\end{array}
=\int_{SL(2,\C)} dg \bigotimes_{a=1}^4 \, D^{(\gamma j_a,j_a)}_{j_a m_al_a n_a}(g).
\ee
The arrow under the box is needed to keep track of whether the group element to be integrated over is $g$ or $g^{-1}$. For instance, following the same convention as before for outgoing and ingoing lines, we have
\be
\begin{array}{c}
\psfrag{a}{$j_1$}
\psfrag{b}{$j_2$}
\psfrag{c}{$l_1$}
\psfrag{d}{$l_2$}
\psfrag{g}{$g$}
\psfrag{h}{$h$}
\includegraphics[width=3.5 cm]{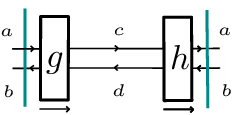}
\end{array}=
\int dh\,dg\sum_{l_1,r_1,l_2,r_2}D^{\g j_1,j_1}_{j_1m_1l_1r_1}(g)D^{\g j_1,j_1}_{l_1r_1j_1n_1}(h^{-1})D^{\g j_2,j_2}_{j_2m_2l_2r_2}(g^{-1})D^{\g j_2,j_2}_{l_2r_2j_2n_2}(h)\,.
\ee
Following \cite{Boosting}, the $\SL(2,\C)$ integrals can be split in two SU(2) Wigner's $njm$ symbols summed over a booster function, for instance
\begin{equation}
\begin{array}{c}
\psfrag{a}{$\{j_a\}$}
\psfrag{b}{$\{l_a\}$}
\psfrag{i}{$i$}
\psfrag{k}{$k$}
\psfrag{=}{$=$}
\psfrag{e}{$d_i$}
\psfrag{+}{$+$}
\psfrag{-}{$-$}
\psfrag{g}{$g$}
\psfrag{h}{$d_k$}
\psfrag{s}{$\sum_{i,k}$}
\includegraphics[width=13.0cm]{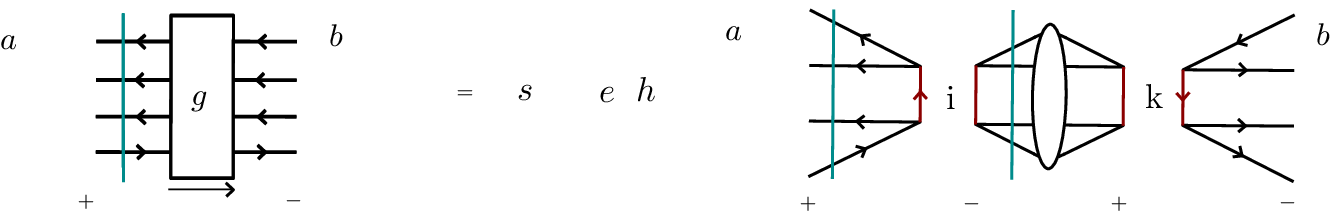} 
\end{array}\,.\label{decaffeinato}
\end{equation}
The graphical notation for the booster function is the one introduced in the main text. With this choice the signs of the nodes are alternating. We use this property also in the main text, to avoid keeping track of the signs on all diagrams: we assign + or - to the boundary states, then the signs of the internal nodes are implicitly assigned according to this alternate rule.

The booster functions do not depend on the strand orientations. To see this, notice that in a box with $n$ lines we can have two possible configurations:
\be
\begin{array}{c}
\psfrag{j}{$j$}
\psfrag{l}{$l$}
\includegraphics[width=0.8cm]{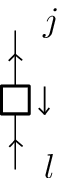}
\end{array}
= D^{\g j,j}_{lnjm}(g^{-1})
\qquad
\begin{array}{c}
\psfrag{j}{$j$}
\psfrag{l}{$l$}
\includegraphics[width=0.8cm]{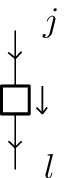}
\end{array}= D^{\g j,j}_{jmln} (g),
\ee
with $g \in SL(2,\C)$. On the first case the decomposition reads \cite{Boosting}
\be
\begin{split}
D^{\g j,j}_{lnjm} (g^{-1}) &= \overline{D}^{\g j,j}_{jmln} (g)=(-1)^{j-l+m-n}D^{\g j,j}_{j-ml-n} (g) \\ &=(-1)^{j-l+m-n}D^j_{-mp}(u)d^{\g j,j}_{jlp}(r)D^l_{p-n}(v),
\end{split}
\ee
while in the second case 
\be
\begin{split}
D^{\g j,j}_{jmln} (g) &=D^j_{mp}(u)d^{\g j,j}_{jlp}(r)D^l_{pn}(v)
\end{split}.
\ee
In both cases we can split the $SL(2,\C)$ group element without any phase depending on magnetic index $p$, thus the booster functions do not get any arrow. 

\section{Booster functions: numerical results}
\label{AppBooster}

\begin{figure}[H] 
\begin{center}
\includegraphics[width=6.5cm]{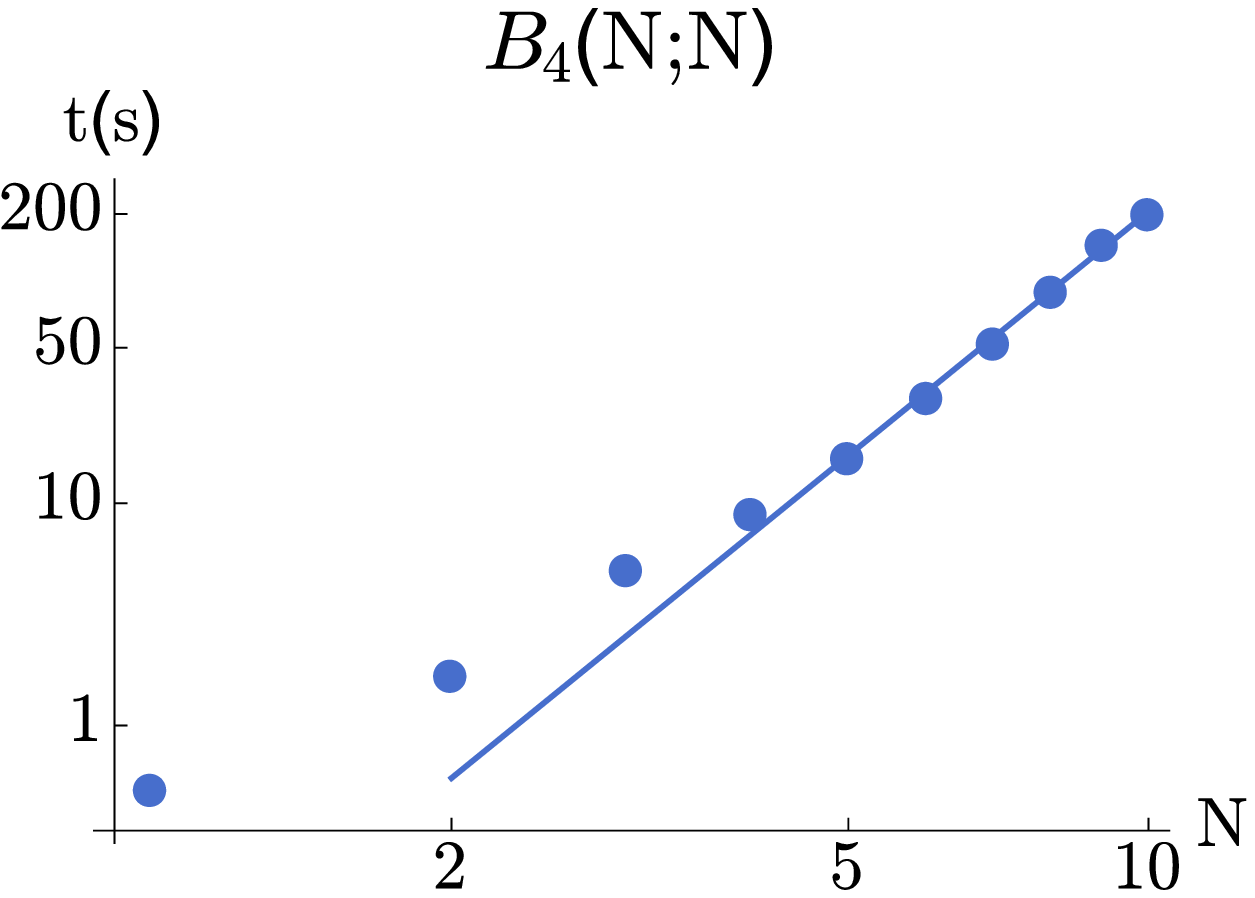}
\hspace{1cm}
\includegraphics[width=6.5cm]{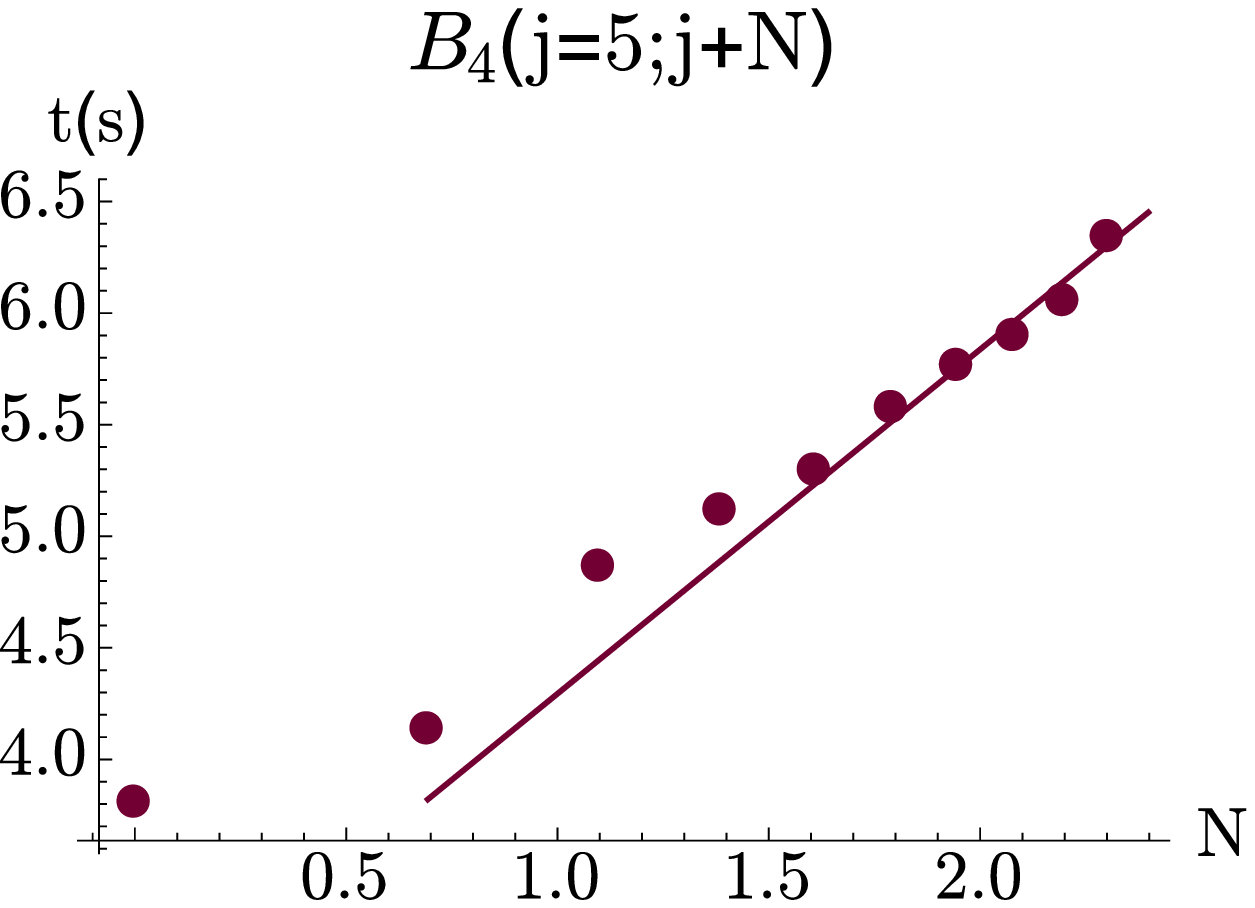}
\end{center}
\caption{\small{Left panel: \emph{Time (in seconds) to evaluate a $B^\g_4(N j, Nj)$ for minimal and equal spins for all possible intertwiners, the important contribution is given by numerical integration while, at low spins, $4j$ symbol calculation time is negligible using \cite{Johansson:2015cca} } Right panel: \emph{ Calculation time for $B^\g_4(j, j+N \Delta l)$, for all intertwiners and with a homogeneous rescaling of magnetic spins. } }
\label{FigboosterTiming} }
\end{figure}

In this Appendix we collect numerical results for the booster functions \Ref{Bsn}. For $n=3$ there is a fast and exact formula in terms of finite sums, see \cite{Boosting}. Using this we can go up to spins of order a hundred within seconds. For larger $n$ we can do much less. A formula based on the finite sums for presented in \cite{Boosting}, and in spite of promising properties, it presents numerical instabilities due to subtractions of ratios of large numbers. For this paper we then used the basic definition \Ref{Bsn} with the boost matrix elements expressed in terms of hypergeometric functions. This gives reliable evaluations with Mathematica, although slow, up to spins of order 30 where numerical instabilities appear. 
Fig.~\ref{FigboosterTiming} gives an idea of the evaluation times.

\begin{figure}[H]
\centering
\includegraphics[width=10cm]{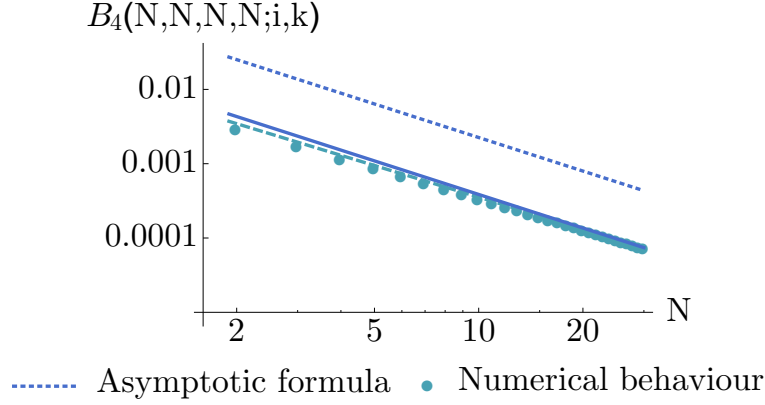} 
\caption{ \small{\emph{Homogeneous large spin scaling of $B_4^\g$, with equal intertwiners $i=k=1$ and $\g=6/5$.  Dots are the data points, and the continuous line is a numerical fit with power $N^{-3/2}$. The dashed line is a numerical fit with a free power, which gives a scaling $N^{-1.4}$ using the last 5 points. This is compatible with the $N^{-3/2}$, and suggests that the asymptotics is reached at spins of order 50 to 100, as it is for the $B_3^\g$, see \cite{Boosting}. 
Finally, the dotted line is the analytic formula \Ref{Puchta}, which captures the right scaling but not the numerical factor. Further analysis \cite{Boosting}  shows that also the $\g$ and $j_i$ dependences are well captured, but not so well the intertwiner one which is a Gaussian rather than a Kronecker delta.}}
\label{B4asymp}}
\end{figure}

Spins of order 30 are enough to see convincingly the $N^{-3/2}$ scaling, although it is still a few per cents away with a fit, see Fig.~\ref{B4asymp}. Experience from the 3-stranded boosters shows in fact that a 1\% fit needs spins of order 60. The same figure also shows that the analytic estimate \Ref{Puchta} is off by a numerical factor of order 1. Further studies show also that the decay with unequal intertwiners has the same power law, hence the Kronecker delta reported in \Ref{Puchta} should be more precisely be replaced by a Gaussian.

\begin{figure}[H] 
\begin{center}
\includegraphics[width=7cm]{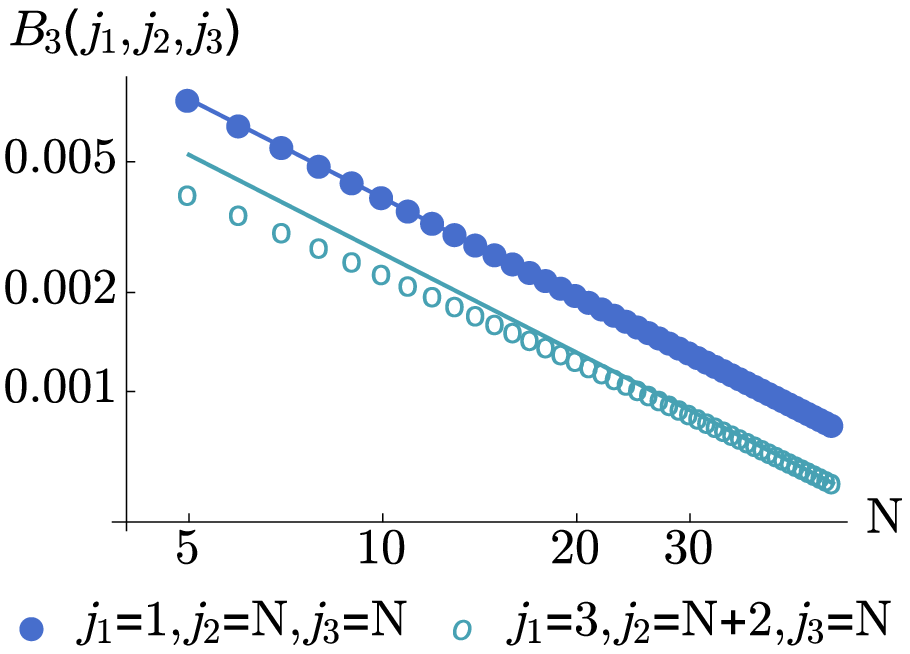} 
\hspace{1cm} 
\includegraphics[width=7cm]{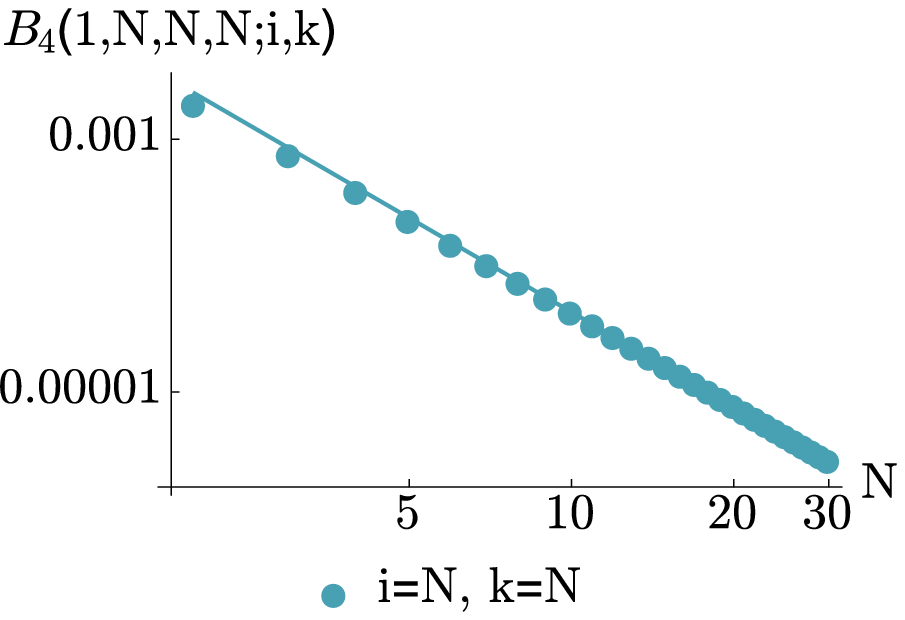}
\includegraphics[width=7cm]{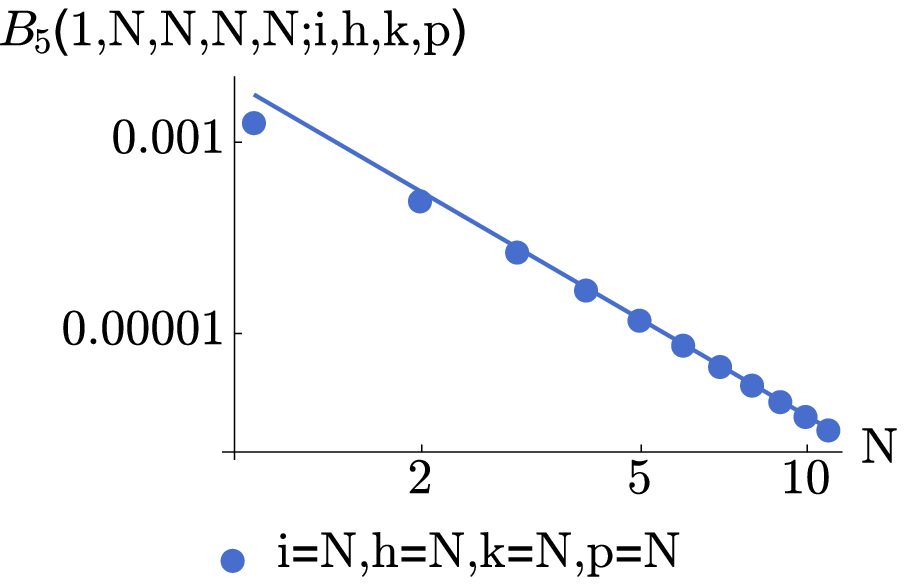}
\end{center}
\caption{\small{\emph{Inhomogeneous large spin scalings of booster functions with a small spin, see \Ref{BInhomo} in the main text. The dots are data points, the lines numerical fits. We find a power law $N^{-1}$ for $B_3^\g$ (top left), $N^{-5/2}$ for  $B_4^\g$ with intertwiners equal and large (top right), $N^{-7/2}$ for  $B_5^\g$ with both intertwiners equal and large (bottom).  } }
\label{B3InHomo} }
\end{figure}

Next, we report in Fig.~\ref{B3InHomo} the studies of the inhomogeneous scalings, with a single small spin, and all other large.

\begin{figure}[H]
\centering
\includegraphics[width=7cm]{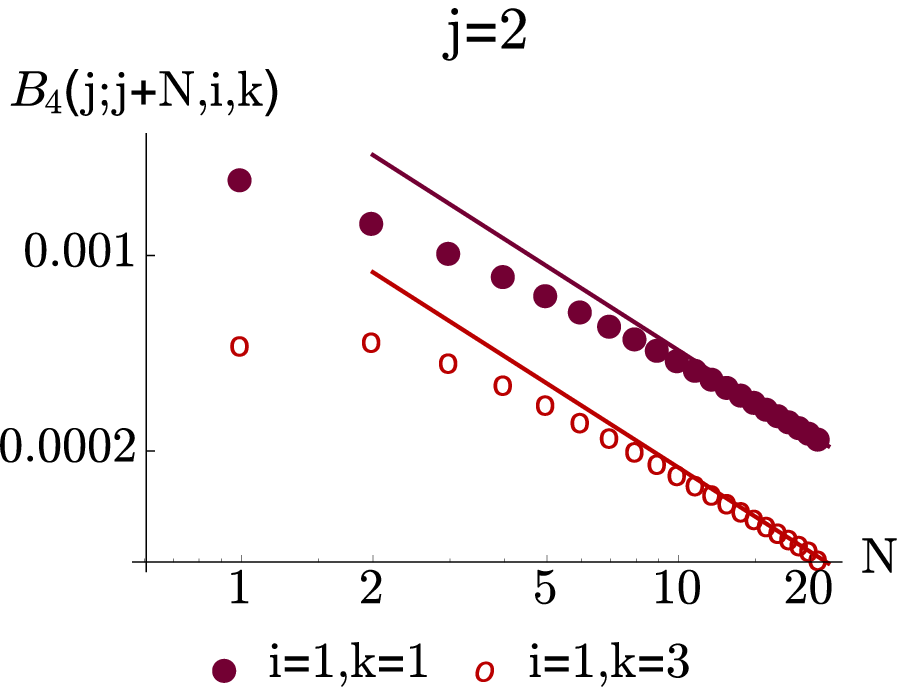} \hspace{1cm} 
\includegraphics[width=7cm]{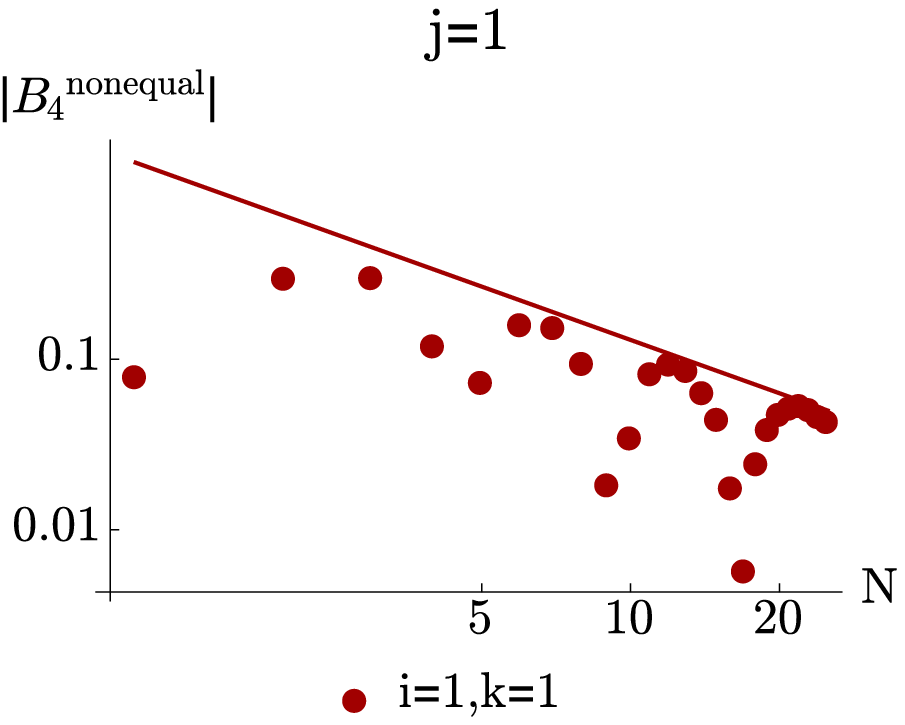} 
\caption{ \small{\emph{Non-minimal spin decays, homgeneous.} Left panel: \emph{Equal spins, equal and non-equal intertwiners. Numerical fits give a power law $N^{-1}$ in both cases.} Right panel: \emph{ Configuration with non-equal spins, $B_4^{nonequal}=B_4(j,j,j+1,j+2; j+N,j+N,j+1+N,j+2+N;i,k)$, showing oscillatory behaviour with the same $N^{-1}$ power law.}}
\label{FigB4NonMinimal} }
\end{figure}

Another important property of the booster functions is the way they decay as we increase the $l$ `magnetic numbers', starting from their minimal value $j$. This can be either a power law or an exponential decay. Fig.~\ref{FigB4NonMinimal} shows non-oscillating and oscillating power laws obtained increasing homogeneously all $l$'s of $B_4^\g$ for different spin configurations.

\begin{figure}[H] 
\begin{center}
\begin{minipage}{0.4 \textwidth}
\centering
\vspace{0.75cm}
\includegraphics[width=7cm]{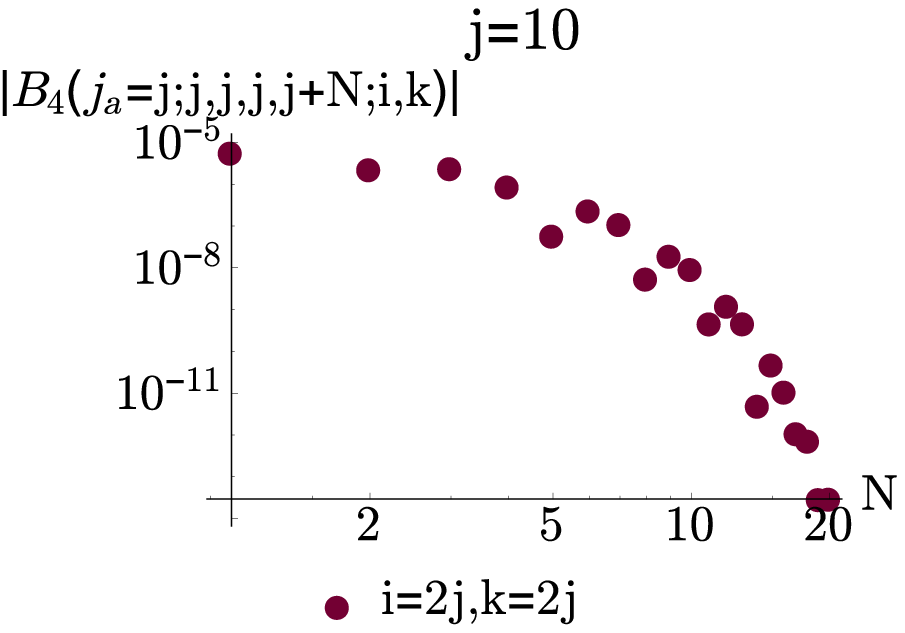}
\end{minipage}
\hspace{1cm} 
\begin{minipage}{0.4 \textwidth}
\centering
\includegraphics[width=7cm]{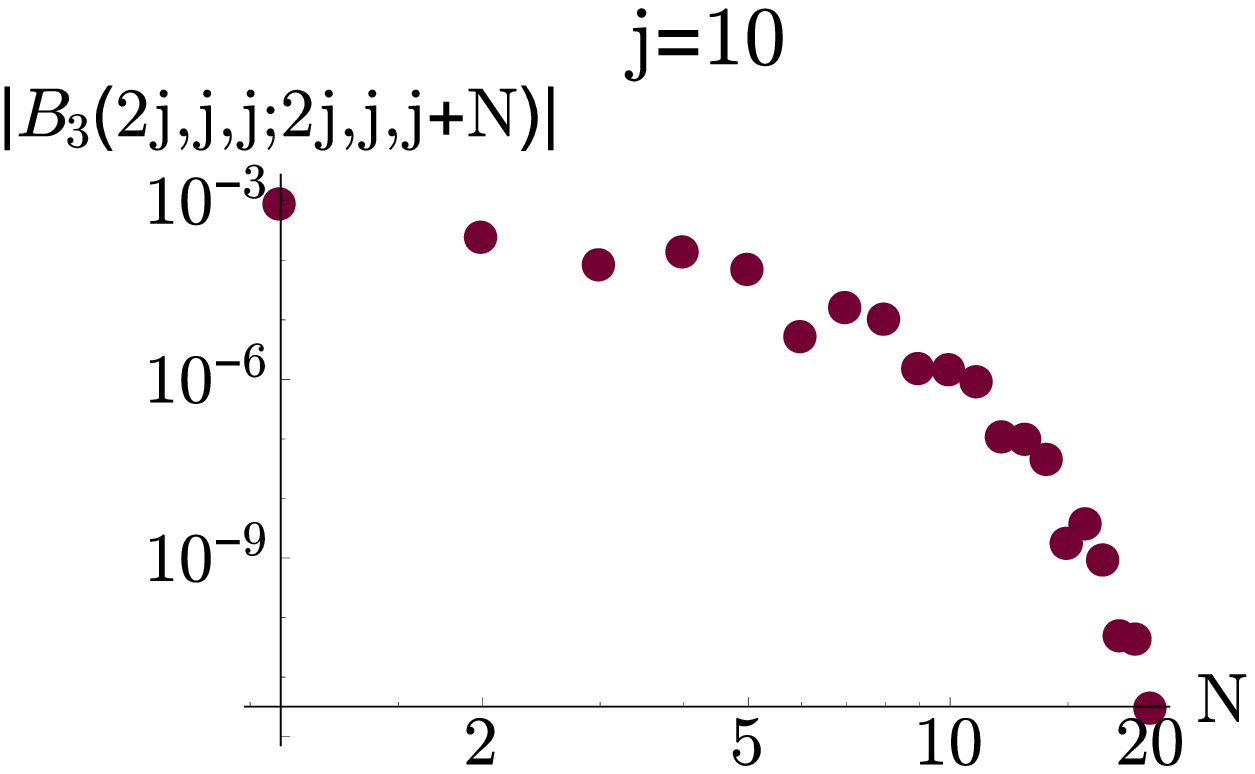}
\end{minipage}
\end{center}
\caption{\small{Left panel: \emph{Decaying at j=10 for a $B_4$ while increasing one $l$. The difference between the first two points ($N=0$ and $N=1$) is one order of magnitude while between the first and the last point nine orders. } Right panel: \emph{ Decaying at j=10 for a $B_3$ while increasing one $l$. } }
\label{FigDecB} }
\end{figure}

The final Fig.~\ref{FigDecB} shows that allowing a single $l$ to grow large with respect to its minimal value give an exponential decay.

\section{Extended graph analysis}
\label{AppDVD}
In the main body of the paper, we used two different types of arguments to reduce the number of vertex graphs to be studied. First, the existence of non-integrable Lorentzian spin networks; second, our criterion to define the spin foam faces as minimal cycles. In this Appendix we relax the second argument, and show what it would look like to take into account all possible graphs which are a priori integrable. Accordingly, we only exclude graphs containing uni-valent and bi-valent nodes, as well as tadpoles. We keep a priori also non-3-link-connected graphs: as explained in the main text, 3-link connectivity is only a sufficient criterion, and indeed we will see below examples of integrable non-3-link connected graphs.

\begin{figure}[ht]   
\centering      
\includegraphics[width=2cm]{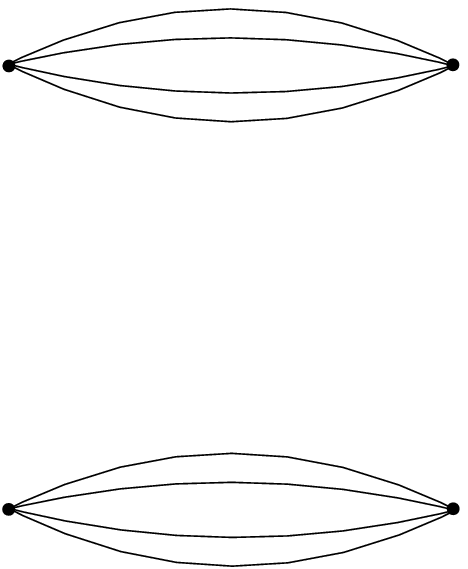}\quad\quad
\includegraphics[width=2.6cm]{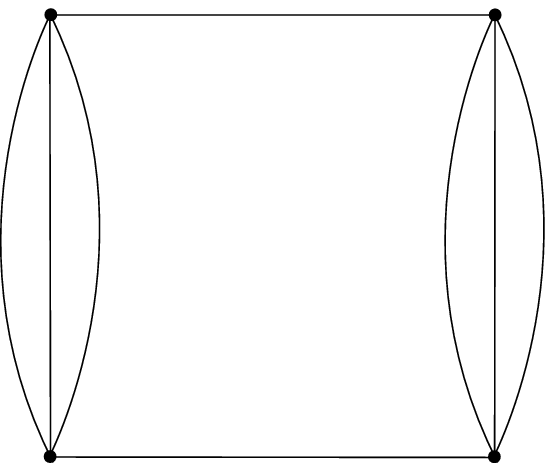}\quad\quad
\includegraphics[width=2.4cm]{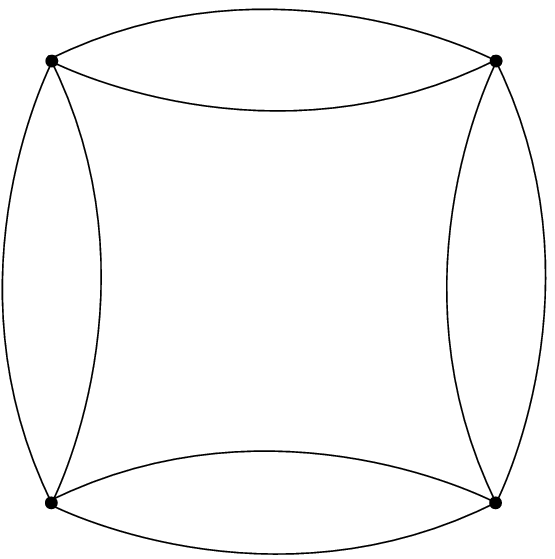}\quad\quad
\includegraphics[width=2.6cm]{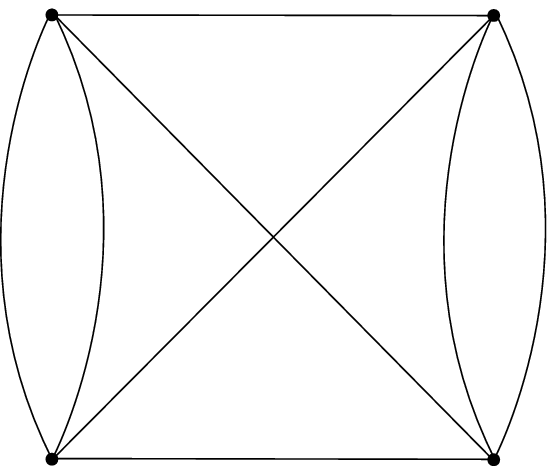}
\caption{\emph{\small{Integrable graphs for a 4-valent vertex with 8 faces. The second is not 3-link-connected, yet it turns out to be integrable. The last one is the only one to contain a non-trivial symbol, a $6j$. When applied to the DVD foam, all but the first violate the minimal-cycle faces criterion; all non-minimal-cycle faces in this case lead to `face-rigid' amplitudes, meaning one face has two boundary links thus imposing equality of those boundary spins. }}}
\label{DVDTutti}
\end{figure}

\begin{figure}[H]   
\centering      
\includegraphics[width=7.5cm]{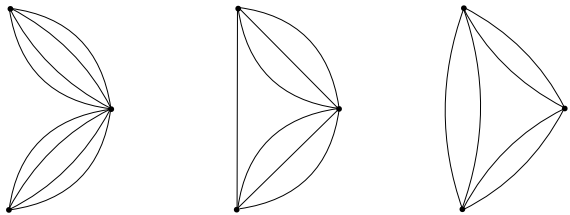}
\caption{\small{\emph{Integrable graphs for a 3-valent vertex with 8 faces. When applied to $DED$ the last two give `face-rigid' amplitudes.}}}
\label{DEDTutti}
\end{figure}

\begin{figure}[H]   
\centering      
\includegraphics[width=11cm]{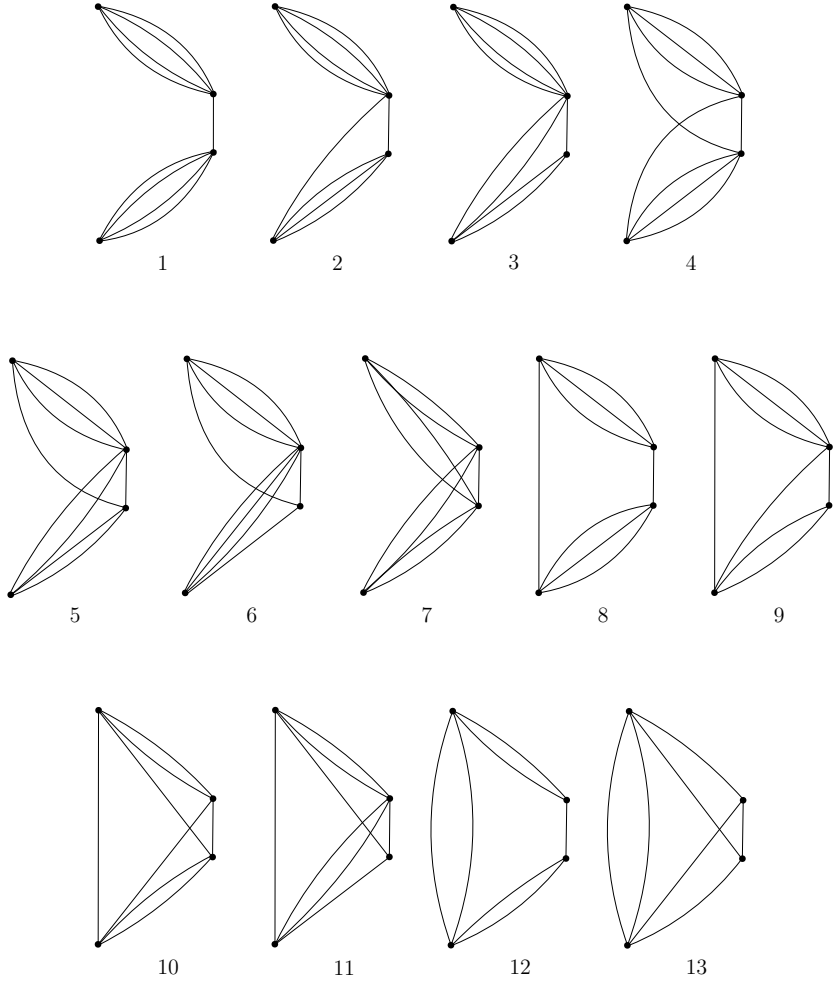}
\caption{\emph{Integrable graphs for a 4-valent vertex with 9 faces. Number 1,2 and 8 are not 3-link-connected. The simplified amplitude is finite, but we do not know whether the complete EPRL amplitude is finite or not. }} 
\label{DLDTutti}
\end{figure}

Allowing for faces corresponding to non-minimal cycles, we have 4 possible vertex graphs for $DVD$, see Fig.~\ref{DVDTutti}, 3 for $DED$, see Fig.~\ref{DEDTutti}, and 13 for $DLD$, see Fig.~\ref{DLDTutti}.
We have listed only the topologically distinct vertex graphs; to list the associated spin foams one has to take into account also label permutations and alternative but topologically equivalent routings (for these simple graphs, these can be captured by rotating the graph by $\pi/2$ and by flipping it vertically).
Many of these diagrams lead to `face-rigid' amplitudes for the dipole transitions: these can be easily identified as those with a link connecting the top and bottom nodes. It is also easy to identify those that would lead to factorized amplitudes, by using the criterion explained in the main text: if all the spins of a single boundary are linked to a unique node of the vertex graph then decoupling can be immediately seen choosing to remove the group integration at that node. This is the case for the first graph of \Ref{DEDTutti} and for the first three graphs of \Ref{DLDTutti}.

The admissible graph analysis shown in these figures was performed by hand, using rules analogue to those explained in details in \cite{Kisielowski:2012yv}. For more general graphs it is highly recommendable to switch to an automatic evaluation on Mathematica. 

\subsection{Evaluations of the `face-rigid' DVD foams}

We report here the explicit evaluation of the three additional vertex graphs of Fig.~\ref{DVDTutti}.
These calculations, although not directly relevant to the main goal of the paper, allow us to highlight some properties of the EPRL model useful for more general evaluations.
First, the fact the non-3-link connected graphs can have integrable Lorentzian amplitudes (in $DVD_2$). 
Second, the fact that even in the presence of unbounded `magnetic $l$'s' summations the simplified model can still provide precise leading order approximations of the EPRL model, if the sums quickly converge (in $DVD_2$ and $DVD_3$). Third, it allows us also to see a diagram in which a non-trivial $6j$ symbol appears (in $DVD_4$). 

\begin{figure}[h] 
\begin{center}
\begin{minipage}{0.4 \textwidth}
\centering
\vspace{-1.cm}
\includegraphics[width=7cm]{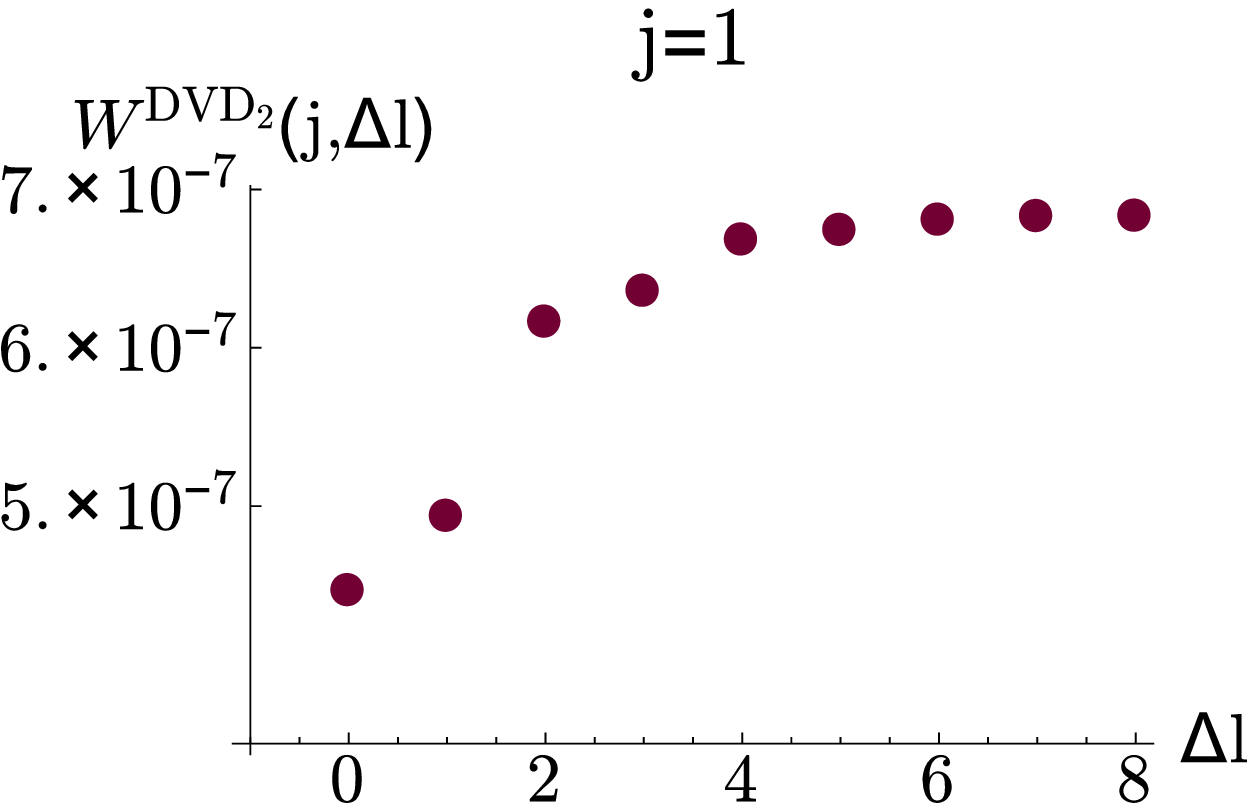}
\end{minipage}
\hspace{1cm} 
\begin{minipage}{0.4 \textwidth}
\centering
\includegraphics[width=7cm]{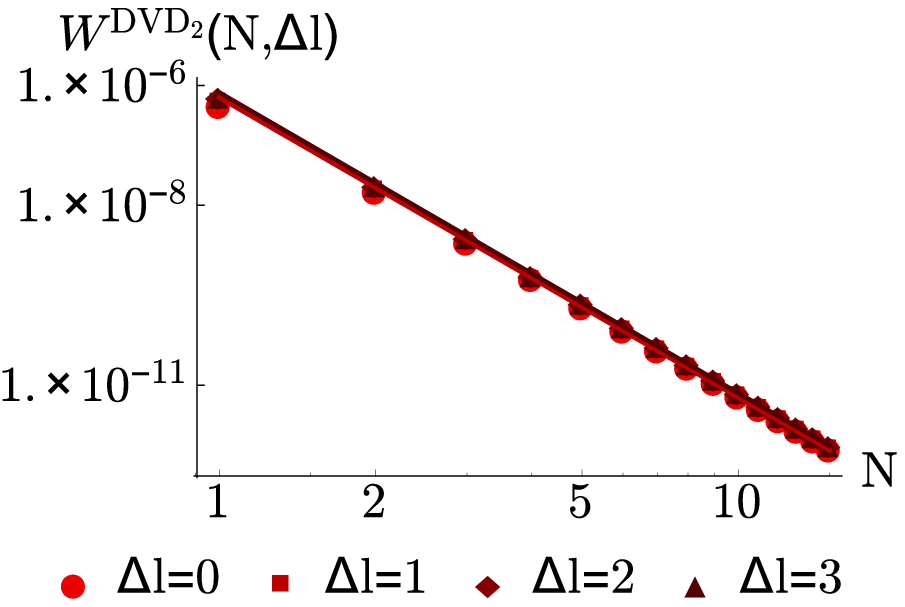}
\end{minipage}
\end{center}
\caption{\small{Left panel: \emph{Convergence of the sums over $l$'s in \Ref{DVD2} with an homogeneous cut off $\Delta l$. Boundary spins and intertwiner fixed to 1, $\g=6/5$. The relative error between the first and the last point is of 35$\%$ while between the seventh and the last one is 0.03$\%$ } Right panel: \emph{Scaling of $W^{DVD_2}$ for different values of the cut off with all boundary's intertwiners fixed to 1. The roughly 10\% differences are squashed by the scale of the $y$-axis. The continuous line is a numerical fit, which gives $N^{-5.01}$ for all of them, a value compatible with the EPRLs analytic estimate at this low spins at which the boosters have not yet reached asymptotic scaling. }
}
\label{DVDConv2} }
\end{figure}

With reference to Fig.~\ref{DVDTutti}, the second vertex graph for DVD gives the following amplitude:
\begin{equation} 
\label{DVD2}
\begin{split}
W^{DVD_2}(j_a,j_a';i,t,i',t')=didi'dt & \sum_{k',l_a} dk'(-1)^{2(k'+t')} \frac{\delta_{j'_3,l_4}}{d_{j'_3}}\delta_{j_1,j'_1}\delta_{j_2,j'_2}\delta_{j_3,j'_3}
\\ &
\begin{array}{c}
\psfrag{a}{$j_1$}
\psfrag{b}{$j_2$}
\psfrag{c}{$j_3$}
\psfrag{d}{$j_4$}
\psfrag{e}{$j_1$}
\psfrag{f}{$j_2$}
\psfrag{g}{$j_3$}
\psfrag{h}{$l_4$}
\psfrag{i}{$t$}
\psfrag{k}{$t'$}
\includegraphics[width=2.2cm]{_images/B4.eps}
\end{array}
\begin{array}{c}
\psfrag{a}{$j_1$}
\psfrag{b}{$j_2$}
\psfrag{c}{$j_3$}
\psfrag{d}{$j_4$}
\psfrag{e}{$l_1$}
\psfrag{f}{$l_2$}
\psfrag{g}{$l_3$}
\psfrag{h}{$l_4$}
\psfrag{i}{$i$}
\psfrag{k}{$k'$}
\includegraphics[width=2.2cm]{_images/B4.eps}
\end{array}
\begin{array}{c}
\psfrag{a}{$j'_1$}
\psfrag{b}{$j'_2$}
\psfrag{c}{$j'_3$}
\psfrag{d}{$j'_4$}
\psfrag{e}{$j'_1$}
\psfrag{f}{$l_3$}
\psfrag{g}{$l_2$}
\psfrag{h}{$l_1$}
\psfrag{i}{$i'$}
\psfrag{k}{$k'$}
\includegraphics[width=2.2cm]{_images/B4.eps}
\end{array}
\end{split}
\end{equation}
We see in the first line above the three `face-rigid' conditions imposing equal spins between the lower and the upper dipole. The amplitude is not 3-link-connected, thus convergence of the $l$-summations is not guaranteed. Of the 4 $l$ summations, 3 are unbounded due to triangular inequalities, and further the $l$'s enter in 3 or 4 slots of the (last two) booster functions, whose decay is a slow power-law. Nonetheless, the summations quickly converge, and the amplitude is finite, see Fig.~\ref{DVDConv2}. The quick convergence of the summations indicates also that the simplified model can give a good scaling estimate, which is also confirmed numerically. Using \Ref{BnLO} we estimate the large spin limit of the EPRLs amplitude to scale as $N^{-11/2}$, and the numerical fit on the right panel of Fig.~\ref{DVDConv2} shows that already at spins of order 10 the power approaches this value.

\begin{figure}[h] 
\begin{center}
\begin{minipage}{0.4 \textwidth}
\centering
\vspace{-1.cm}
\includegraphics[width=7cm]{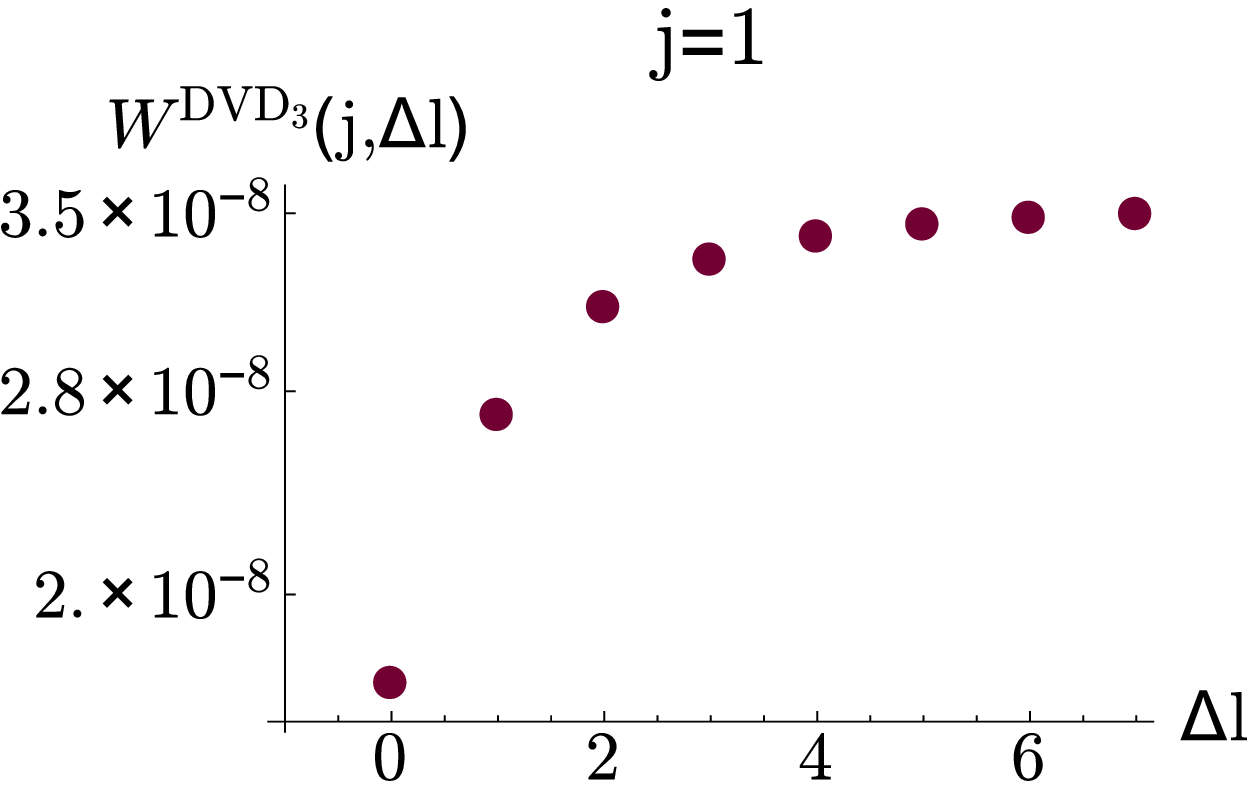}
\end{minipage}
\hspace{1cm} 
\begin{minipage}{0.4 \textwidth}
\centering
\includegraphics[width=7cm]{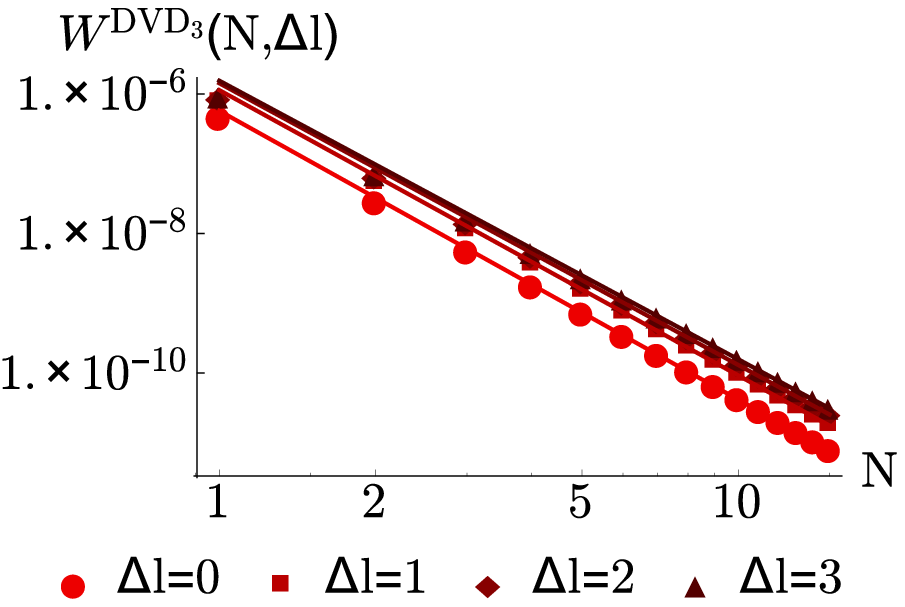}
\end{minipage}
\end{center}
\caption{\small{Left panel: \emph{Convergence of the sums over $l$'s in \Ref{DVD3} with an homogeneous cut off $\Delta l$. Boundary spins and intertwiner fixed to 1, $\g=6/5$. The relative error between the first and the last point is of 50$\%$ while between the seventh and the last one is 0.4$\%$. } Right panel: \emph{Scaling of \Ref{DVD3} for different values of the cut off with all boundary's intertwiners fixed to 1. The convergence is now slower and we can tell the different cut-offs apart. The numerical fits in the continuous lines give $N^{-4.1}$, again compatible with the EPRLs estimates.}
}
\label{DVDConv1} }
\end{figure}

We have a similar story for $DVD_3$, whose unbounded $l$ summations also quickly converge, making the EPRLs scaling coincide with the complete evaluations. The amplitude reads
\begin{equation}
\label{DVD3}
W^{DVD_3}(j_a,j_a';i,t,i',t')=didi'dt\sum_{l_a}\delta_{j_1,j'_1}\delta_{j_2,j'_2}
\begin{array}{c}
\psfrag{a}{$j_1$}
\psfrag{b}{$j_2$}
\psfrag{c}{$j_3$}
\psfrag{d}{$j_4$}
\psfrag{e}{$j_1$}
\psfrag{f}{$j_2$}
\psfrag{g}{$l_3$}
\psfrag{h}{$l_4$}
\psfrag{i}{$t$}
\psfrag{k}{$t'$}
\includegraphics[width=2.2cm]{_images/B4.eps}
\end{array}
\begin{array}{c}
\psfrag{a}{$j_1$}
\psfrag{b}{$j_2$}
\psfrag{c}{$j_3$}
\psfrag{d}{$j_4$}
\psfrag{e}{$l_1$}
\psfrag{f}{$l_2$}
\psfrag{g}{$l_3$}
\psfrag{h}{$l_4$}
\psfrag{i}{$i$}
\psfrag{k}{$t'$}
\includegraphics[width=2.2cm]{_images/B4.eps}
\end{array}
\begin{array}{c}
\psfrag{a}{$j'_1$}
\psfrag{b}{$j'_2$}
\psfrag{c}{$j'_3$}
\psfrag{d}{$j'_4$}
\psfrag{e}{$j'_1$}
\psfrag{f}{$j'_2$}
\psfrag{g}{$l_2$}
\psfrag{h}{$l_1$}
\psfrag{i}{$i'$}
\psfrag{k}{$t'$}
\includegraphics[width=2.2cm]{_images/B4.eps}
\end{array}
\end{equation}
with only two $l$ summations unbounded. The simplified scaling is $N^{-9/2}$, and the numerical studies are reported in Fig.~\ref{DVDConv1}.

These results show that the EPRLs captures well the scaling of the complete model, but with a caveat: it could happen that the quick convergence of the $l$ summations, exposed in the numerics above at small spins (left panels of Figs. \ref{DVDConv2} and \ref{DVDConv1}, is not preserved at high spins. This could result in larger contributions and slower decays. The fact that the numerical fits (in the right panels of the same figures) are unchanged shows that this should be true up to $j$ of order 10, but we do not have stronger evidence beyond that.

\begin{figure}[h] 
\begin{center}
\includegraphics[width=7cm]{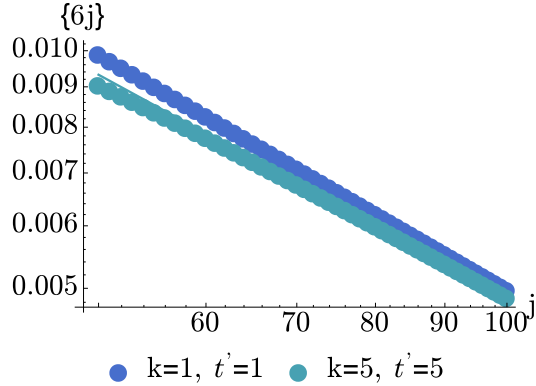}
\end{center}
\caption{\small{ \emph{Scaling of a $\{6j\}$ symbol with 4 large entries. The dots are data points, the continuous line a $N^{-1 }$ fit.}}}
\label{6JPlot}
\end{figure}

Finally, $DVD_4$ carries a $\{6j\}$ symbol due to the crossing of two of the faces:  
\begin{equation}
\begin{split}
W^{DVD_4}(j_a,j_a';i,t,i',t')=didi'dt &\sum_{k,l_a} dk(-1)^{2t'+3l_4+j'_1}\delta_{j_1,j'_1}\delta_{j_2,j'_2}\delta_{j_3,j'_3}  \\ &
\begin{array}{c}
\psfrag{a}{$j_1$}
\psfrag{b}{$j_2$}
\psfrag{c}{$j_3$}
\psfrag{d}{$j_4$}
\psfrag{e}{$l_1$}
\psfrag{f}{$l_2$}
\psfrag{g}{$j_3$}
\psfrag{h}{$l_4$}
\psfrag{i}{$i$}
\psfrag{k}{$k$}
\includegraphics[width=2.2cm]{_images/B4.eps}
\end{array}
\begin{array}{c}
\psfrag{a}{$j_1$}
\psfrag{b}{$j_2$}
\psfrag{c}{$j_3$}
\psfrag{d}{$j_4$}
\psfrag{e}{$j_1$}
\psfrag{f}{$j_2$}
\psfrag{g}{$l_3$}
\psfrag{h}{$l_4$}
\psfrag{i}{$t$}
\psfrag{k}{$t'$}
\includegraphics[width=2.2cm]{_images/B4.eps}
\end{array}
\begin{array}{c}
\psfrag{a}{$j'_1$}
\psfrag{b}{$j'_2$}
\psfrag{c}{$j'_3$}
\psfrag{d}{$j'_4$}
\psfrag{e}{$j'_1$}
\psfrag{f}{$l_3$}
\psfrag{g}{$l_2$}
\psfrag{h}{$l_1$}
\psfrag{i}{$i'$}
\psfrag{k}{$k$}
\includegraphics[width=2.2cm]{_images/B4.eps}
\end{array}
\left\{\begin{matrix} k & l_3 & j'_4\\ t' & l_3 & l_4\end{matrix}\right\}
\end{split}
\end{equation}
For the simplified model with boundary intertwiners fixed, the involved $\{6j\}$ symbol has only four large entries, and scales like $N^{-1}$, see Fig.~\ref{6JPlot}. 
The estimate for the simplified amplitude gives
\be
W^{DVD_4}_s(Nj,Nj;i,t,i',t') \sim (N^{-3/2})^3 N^{-1} = N^{-11/2}.
\ee

The properties of the DVD foams studied in this Appendix are summarized in the following Table~\ref{tableDVD}:
\begin{table}[h]
\begin{center}\begin{tabular}{ccccc}
{\bf Foam} & {\bf Factorization} & {\bf LOs scaling} & {\bf EPRLs=EPRL} & {\bf Face-rigidity} \\ \hline 
$DVD_1$ &Y & $N^{-3}$ & Y & \\
$DVD_2$ &N & $N^{-10/2}$ & N & 3-face rigid  \\ 
$DVD_3$ &N & $N^{-9/2}$ & N & 2-face rigid \\ 
$DVD_4$ &N & $ N^{-11/2}$ & N & 3-face rigid
\end{tabular}
\caption{\label{TableDVD} \small{\emph{Summary of scaling and some properties of the DVD foams.}}
\label{tableDVD}}\end{center}
\end{table}

\providecommand{\href}[2]{#2}\begingroup\raggedright\endgroup


\end{document}